\documentclass{article} 
\usepackage{iclr2027_conference,times}

\usepackage{amsmath,amsfonts,bm}

\newcommand{\mbf}[1]{\mathbf{#1}}
\newcommand{\mbb}[1]{\mathbb{#1}}

\newcommand{\mcal}[1]{\mathcal{#1}}

\def\eqref#1{equation~\ref{#1}}

\def\1{\bm{1}}

\def\rva{{\mathbf{a}}}
\def\rvb{{\mathbf{b}}}
\def\rvc{{\mathbf{c}}}

\def\rvo{{\mathbf{o}}}
\def\rvp{{\mathbf{p}}}
\def\rvq{{\mathbf{q}}}
\def\rvr{{\mathbf{r}}}
\def\rvs{{\mathbf{s}}}

\def\rvt{{\mathbf{t}}}

\def\rvx{{\mathbf{x}}}

\def\rvz{{\mathbf{z}}}

\DeclareMathAlphabet{\mathsfit}{\encodingdefault}{\sfdefault}{m}{sl}
\SetMathAlphabet{\mathsfit}{bold}{\encodingdefault}{\sfdefault}{bx}{n}

\usepackage{bm}  

\newcommand\scalemath[2]{\scalebox{#1}{\mbox{\ensuremath{\displaystyle #2}}}}

\def\x{{\mathbf{x}}}

\def\z{{\mathbf{z}}}

\newcommand{\pluseq}{\mathrel{+}=}
\usepackage{siunitx} 
\usepackage{amsmath}

\usepackage{hyperref}
\usepackage{url}
\usepackage{amssymb}
\usepackage{wrapfig,tikz}
\usepackage{threeparttable}
\usepackage{siunitx}
\usepackage{booktabs}
\usepackage{nicematrix} 
\usepackage{multirow}
\usepackage[table]{xcolor}
\usepackage{enumitem}

\usepackage{etoc}
\etocdepthtag.toc{mtchapter}
\etocsettagdepth{mtchapter}{subsection}
\etocsettagdepth{mtappendix}{none}

\usepackage{framed} 
\definecolor{shadecolor}{rgb}{0.94, 0.97, 1.0}

\definecolor{cite_color}{HTML}{114083}
\definecolor{link_color}{RGB}{153, 0,0}  
\definecolor{url_color}{RGB}{153, 102,  0}
\definecolor{emp_color}{RGB}{0,0,255}
\definecolor{aliceblue}{rgb}{0.94, 0.97, 1.0}
\hypersetup{
 colorlinks,
 citecolor=cite_color,
 linkcolor=link_color,
 urlcolor=url_color}

\newcommand{\ourM}{\texttt{GPCRLMD}\xspace}
\newcommand{\ourVAE}{AAT-VAE\xspace}

\newcommand{\isometric}{dimension-preserving\xspace}

\renewcommand{\cite}{\citep}
\usepackage[font=small]{caption}

\setcitestyle{numbers,square,comma}

\makeatother
\usepackage[algo2e,ruled,vlined]{algorithm2e}

\usepackage[capitalize,noabbrev]{cleveref}

\title{All-Atom GPCR-Ligand Dynamics Simulation via a Residual  Latent Flow Model}
\author{
Jiying Zhang$^1$ 
\And
Shuhao Zhang$^1$
\And
Pierre Vandergheynst$^1$
\And
Patrick Barth $^1$\\
\texttt{patrick.barth@epfl.ch} \\
$^1$ École Polytechnique Fédérale de Lausanne (EPFL), Lausanne, Switzerland
}

\iclrfinalcopy 
\begin{document}

\maketitle

\begin{abstract}
G-protein-coupled receptors (GPCRs), which are targeted by over one-third of approved drugs, undergo intricate conformational transitions to transduce signals. While Molecular Dynamics (MD) is essential for elucidating this transduction process, particularly within ligand-bound complexes, conventional all-atom MD simulation is computationally prohibitive. In this paper, we introduce \ourM, a deep generative framework for efficient all-atom GPCR-ligand simulation. \ourM employs an {Atom-Anchored Thermal Variational Autoencoder (\ourVAE)} to first map the complex into a regularized \isometric latent space, maintaining geometric topology via physics-informed constraints. Within this latent space, a Residual Latent Flow samples evolution trajectories, which are subsequently decoded back to atomic coordinates. By capturing temporal dynamics via relative displacements anchored to the initial structure, this residual mechanism effectively decouples static topology from dynamic fluctuations. Experimental results demonstrate that \ourM achieves state-of-the-art performance in {GPCR-ligand dynamics simulation}, faithfully reproducing ensemble observables and critical ligand-receptor interactions.
\end{abstract}

\section{Introduction}

\begin{wrapfigure}{R}{0.4\linewidth}
    \centering
            
    \includegraphics[width=1.0\linewidth]{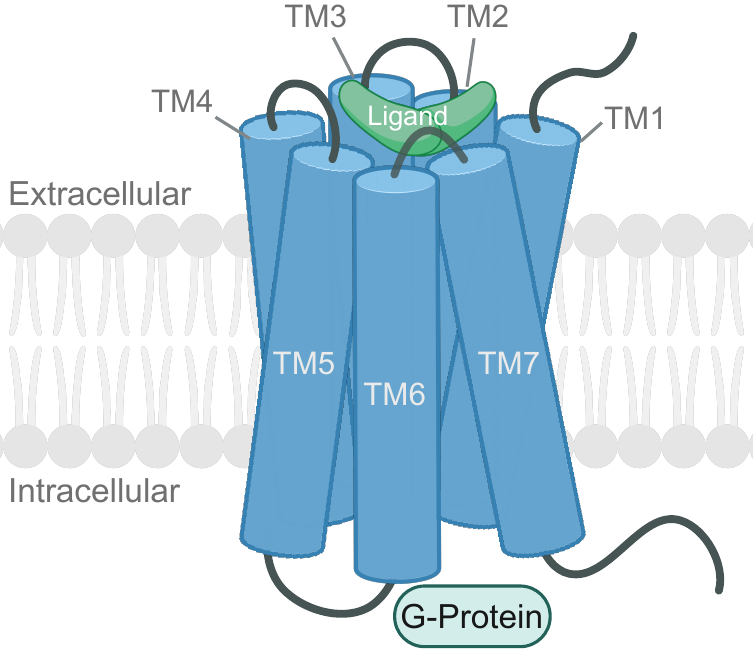}
    \vspace{-4mm}
    \caption{Schematic structure of a ligand-bound GPCR. A GPCR contains seven transmembrane helices (TM1-TM7). It senses ligands and transfers this signal into the cell.}
\vspace{-4mm}
    \label{fig:gpcr}
\end{wrapfigure}

Comprising over 800 distinct members, G protein-coupled receptors (GPCRs) form an essential class of seven-transmembrane proteins (Figure \ref{fig:gpcr}) that modulate a vast array of biological functions~\cite{hauser2017trends,aranda2025large,conflitti2025functional}. Their ubiquity and accessibility have made them the cornerstone of modern medicine, with more than one-third of all marketed drugs targeting this protein 
family~\cite{lorente2025gpcr}.  Understanding the mechanism of these therapeutics requires characterizing GPCR-ligand complexes, where the dynamic coupling between the ligand and the receptor shapes the conformational landscape necessary for signal transduction~\cite{kapla2021can,aranda2025large}.

Molecular Dynamics (MD) simulations serve as a critical tool for elucidating the kinetic and thermodynamic mechanisms governing biomolecular function~\cite{karplus2002molecular,torrens2020molecular,frenkel2023understanding}. This approach is particularly indispensable for investigating GPCR-ligand complexes~\cite{ciancetta2015advances,latorraca2017gpcr}, as the receptor adopts distinct conformational ensembles dictated by the functional profile of the bound ligand~\cite{zhang2024g,conflitti2025functional}. However, conventional MD is limited by its sampling efficiency~\cite{lindorff2011fast}. Relying on step-wise numerical integration to propagate dynamics, adequately exploring the vast, high-dimensional conformational landscape of protein-ligand interactions entails prohibitive computational costs~\cite{newport2019memprotmd,rodriguez2020gpcrmd,siebenmorgen2024misato}. This often renders the generation of comprehensive statistical ensembles a significant challenge, even with enhanced sampling methods~\cite{roessner2025unveiling,li2025enhanced}.

\begin{wrapfigure}{L}{0.56\linewidth}
\vspace{-2mm}
    \centering
    \includegraphics[width=0.980\linewidth]{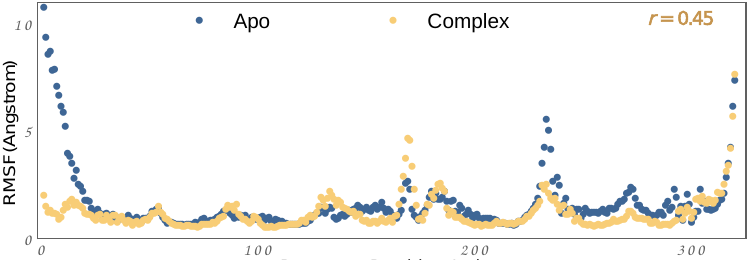}
    \vspace{-2mm}
    \caption{Comparison of Root Mean Square Fluctuation (RMSF) between the apo (ligand-free) and complex (ligand-bound) states. The receptor exhibits significantly different fluctuation profiles when bound to the ligand (Pearson correlation $r=0.45$), demonstrating that ligand binding substantially modulates receptor dynamics. Results are shown by residue index for PDB ID $\mathrm{6LW5}$, utilizing trajectory data from ~\citet{aranda2025large}.}
    \label{fig:apo_complex}
    \vspace{-3mm}
\end{wrapfigure}

Deep generative learning has emerged as a transformative paradigm for accelerating molecular simulations, offering a data-driven shortcut to bypass the high cost of physical integration steps~\cite{feng2025biomd,jing2024generative,lewis2025scalable}. Within the GPCR domain specifically, recent studies have successfully leveraged these models to capture the diverse conformational ensembles of the receptor itself~\cite{sengar2025generative,sengar2025beyond}. However, modeling the dynamic evolution of the GPCR-ligand complex remains a critical unmet need. As shown in Figure \ref{fig:apo_complex}, the presence of a ligand induces distinct fluctuation patterns compared to the apo state. The introduction of the ligand significantly increases the system's complexity~\cite{aranda2025large,conflitti2025functional}, creating a coupled dynamic challenge that existing generative frameworks have yet to address efficiently.

In this work, we address this challenge by introducing \ourM, a deep generative framework designed for the trajectory forecasting setting—predicting the temporal evolution of the GPCR-ligand system given an initial conformational state~\cite{jing2024generative,feng2025biomd}. \ourM uniquely integrates an Atom Anchored Thermal Variational Autoencoder to transform the atomic coordinates of the GPCR-ligand complex into a regularized, \isometric latent space. Unlike traditional dimensionality reduction that sacrifices structural detail, this approach preserves full spatial resolution while mapping the molecular distribution onto a tractable manifold governed by physics-informed constraints. Furthermore, by leveraging attention mechanisms to fuse receptor-ligand features, our model can capture the critical non-covalent interactions driving allostery. Subsequently, we employ a Residual Latent Flow model to capture temporal dynamics. By explicitly learning the relative displacement field anchored to the initial structure, our model effectively decouples static topology from dynamic fluctuations, enabling the efficient and parallel generation of future states. Experimental results demonstrate that \ourM generates biophysically realistic and structurally diverse molecular motions that are consistent with known ensemble and kinetic properties.

In summary, our main contributions are highlighted as follows:
\textbf{Benchmark Curation:} We establish a comprehensive benchmark for GPCR-ligand complex dynamics derived from the GPCRMD dataset~\cite{aranda2025large}. Unlike existing protocols that focus on protein-only systems, we evaluate model performance in capturing the coupled dynamics of the receptor-ligand interface.

\textbf{Novel Framework:} We propose \ourM, a deep generative framework for all-atom GPCR-ligand simulation. It comprises: (1) an Atom-Anchored Thermal VAE that maps complexes into a \isometric latent space via an atom-centered prior; and (2) Residual Latent Flow Matching. By anchoring latent displacements to the initial Cartesian frame, we establish an explicit mathematical linkage between physical and generative spaces. This decouples static equilibrium from stochastic dynamics, forcing the network to focus purely on transition probabilities rather than reconstructing covalent geometries.

\textbf{State-of-the-Art Performance:} We comprehensively evaluate \ourM on both ensemble and trajectory prediction benchmarks. The results demonstrate that our method achieves state-of-the-art performance across key fidelity and diversity metrics. Notably, \ourM reproduces the biologically critical transmembrane helix motions characteristic of GPCR dynamics, ensuring high structural plausibility. Demonstrating broader applicability, we extend our framework to general protein-ligand systems and achieve state-of-the-art results on the MISATO~\citep{siebenmorgen2024misato} dataset.
\section{Related work}

\paragraph{Molecular Dynamics Simulation with Generative Models.} Generative model-based molecular dynamics (MD) simulation methods~\cite{dao2025deep} have attracted significant attention, as they can efficiently produce trajectories or ensembles compared to conventional MD simulations~\cite{han2024geometric, str2str,wu2023diffmd}. In particular, protein-centric approaches have achieved impressive performance in capturing conformational ensembles of the receptor itself at varying resolutions~\cite{str2str,lewis2025scalable,jingAlphaFoldMeetsFlow2024,jing2024generative,lu2025aligning,sengar2025generative,sengar2025beyond,shen2025simultaneous}. However, these methods generally ignore non-protein components and thus cannot be directly applied to protein-ligand complexes. Additionally, while methods like NeuralMD~\cite{liu2023group} and UnbindingFlow~\cite{li2025enhanced} focus on simulating ligand dynamics, they fail to capture fine-grained protein fluctuations. Although BioMD~\cite{feng2025biomd} recently attempted to address all-atom protein-ligand simulation, it is constrained by the Cartesian coordinate representation and lacks the necessary GPCR domain knowledge. As a result, it fails to achieve satisfactory performance on GPCR-ligand dynamics (Table \ref{tab:gpcr_ligand_emsemble_biomd}). Therefore, all-atom GPCR-ligand complex simulation remains an open challenge. 

\vspace{-2mm}

\paragraph{Structure Generation in Latent Space.} \citet{mansoor2024protein} demonstrate that Variational Autoencoders (VAEs) offer a more suitable representation for generating plausible ensembles compared to Cartesian space. Furthermore, several studies~\cite{geffner2025proteina,xu2023geometric,samaddar2025efficient} indicate that latent spaces capture continuous and smooth structural information, thereby enhancing the robustness of generative models in producing diverse and physically valid biomolecular structures~\cite{kong2024full,kong2025unimomo}. While recent approaches~\cite{sengar2025generative,sengar2025beyond} have leveraged this to generate GPCR conformation ensembles, modeling receptor-ligand complexes within latent space remains largely under-explored. In this work, we bridge this gap by anchoring latent displacements directly to the initial Cartesian frame. To the best of our knowledge, no existing work in the molecular domain has established such an explicit mathematical linkage between physical Cartesian space and the generative latent space.

\vspace{-2mm}

\paragraph{GPCR Dynamic Simulations.} G-protein coupled receptors (GPCRs) are pivotal in mediating numerous physiological processes~\cite{zhang2024g,conflitti2025functional} by transducing chemical and physical signals, such as hormones, light, and pressure~\cite{y2008gq}. GPCR signaling is driven by complex, ligand-induced conformational transitions~\cite{zhang2024g,hilger2018structure,y2008gq}. While databases like GPCRMD~\cite{aranda2025large} provide reference trajectories, they are limited to specific pre-simulated pairs. Furthermore, recent generative surrogates~\cite{lopez2023gpcr,sengar2025generative,sengar2025beyond} lack generalizability, being restricted to single-system training that requires re-optimization for each new target. Additionally, their reliance on amino acid-based featurization inherently limits their ability to generalize across diverse GPCR-small molecule systems. Consequently, developing a unified framework capable of efficient, zero-shot trajectory synthesis for unseen GPCR-ligand complexes remains a significant open challenge.

\vspace{-2mm}
\section{Preliminaries}
\vspace{-3mm}
\paragraph{Notation.} The trajectory of the complex is denoted  as $ \mcal{X}_T =[\x_0, \x_1,\cdots, \x_{T-1}]$, where $\x_t=[\x_t^{\mcal{P}} ,\x_t^{\ell}] \in \mbb{R}^{N\times 3}$ represents the Cartesian coordinates of GPCR $\mcal{P}$ and ligand  $\ell$ at time step $t$.
The corresponding latent trajectory is denoted as $\mathcal{Z}_T = [\mathbf{z}_0, \mathbf{z}_1, \dots, \mathbf{z}_{T-1}]$.

\vspace{-2mm}
\subsection{Molecular Dynamics}
\vspace{-1mm}
Molecular Dynamics (MD) is an \textit{in silico} technique that simulates the motion of atoms and molecules under near-physiological conditions. In classical MD, the time evolution of the system is governed by Newton’s equations of motion,
$m_i \frac{d^2 \mathbf{r}_i}{dt^2} = \mathbf{F}_i$,
where $m_i$ and $\mathbf{r}_i$ denote the mass and position of atom $i$, and $\mathbf{F}_i = -\nabla U(\mathbf{r}_i)$ is the force derived from the potential energy $U$. This links atomic interactions and system energetics with dynamical behavior.  
In practice, MD employs numerical integration schemes such as the Verlet algorithm~\cite{verlet1967} or stochastic approaches like Langevin dynamics to propagate trajectories that approximate the Boltzmann distribution, thereby enabling the estimation of thermodynamic and kinetic properties. Despite its utility in revealing atomistic mechanisms, MD remains computationally demanding, restricting accessible timescales and system sizes.

\subsection{Variational Autoencoder (VAE)}
A variational autoencoder (VAE)~\cite{kingma2019introduction} is a probabilistic generative model that learns to map the original data distribution into a latent space, which is typically assumed to follow a standard Gaussian prior $\mathcal{N}(\mbf{0},\mbf{I})$. Unlike a deterministic autoencoder, the VAE introduces a variational posterior $q_\theta(\rvz \mid \rvx)$ to approximate the true but intractable posterior $p(\rvz \mid \rvx)$. 
From a statistical perspective, training a VAE is equivalent to maximizing the evidence lower bound (ELBO) of the marginal likelihood $p(\x)$:
$
    \scalemath{0.89}{\log p(\x)} 
    \scalemath{0.89}{\geq \mathbb{E}_{q_\theta(\rvz \mid \x)}\big[\log p_\phi(\x \mid \rvz)\big] 
    - \mathrm{KL}\big(q_\theta(\rvz \mid \rvx)\,\Vert\,p(\rvz)\big)} ,$
where $p(\rvz) = \mathcal{N}(\mbf{0},\mbf{I})$ is the prior over the latent variable $\rvz$, $p_\phi(\x \mid \z)$ is the likelihood (decoder), which is usually parameterized as Gaussian distribution, and $\mathrm{KL}(\cdot \,\Vert\, \cdot)$ denotes the Kullback–Leibler divergence. 

\subsection{Flow Matching}
Flow matching~\cite{lipman2023flow} is a class of generative models that learns continuous transformations between probability distributions by parameterizing the dynamics of an ordinary differential equation (ODE). 
Formally, given a flow $\phi_\tau: \mathbb{R}^d \times [0,1] \to \mathbb{R}^d$, its dynamics are described by
$
    \frac{d \phi_\tau(\x)}{d\tau} = v(\phi_\tau(\x), \tau),$
where $v: \mathbb{R}^d \times [0,1] \to \mathbb{R}^d$ is a time-dependent vector field (or velocity). The goal of flow matching is to learn a parameterized vector field $v_\theta$ such that the induced flow $\phi_\tau$ transports samples from a simple base distribution (e.g., $\mathcal{N}(\mbf{0},\mbf I)$) to the target data distribution.

\section{Our Method: \ourM}
\begin{figure*}
    \vspace{-4mm}
    \centering
    \includegraphics[width=1.0\linewidth]{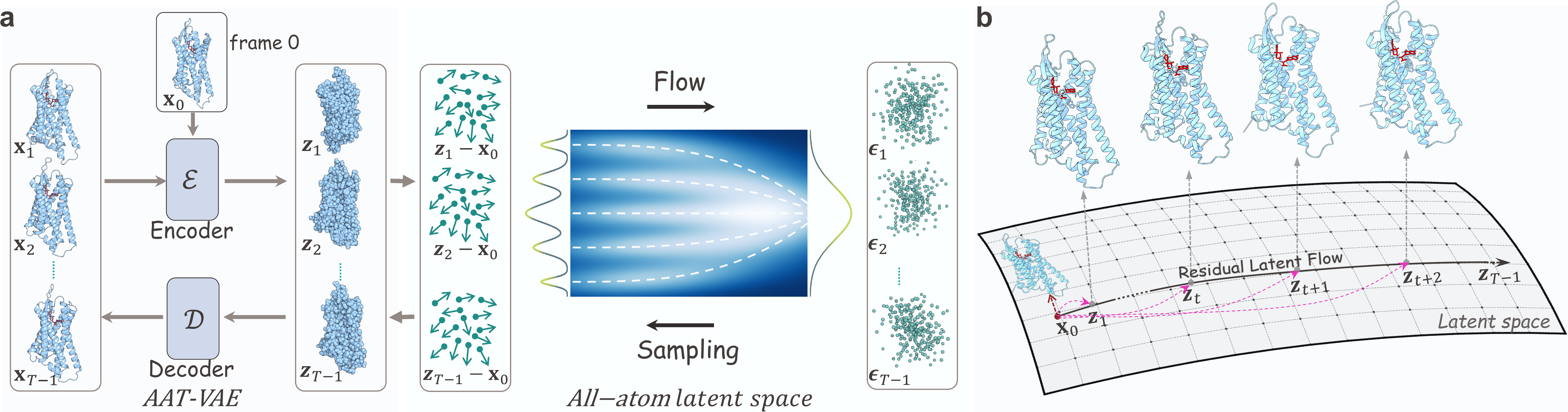}
    \vspace{-5mm}
    \caption{
    Pipeline of \ourM. \textbf{a) Framework Overview.} The complete pipeline maps the GPCR-ligand trajectory $\{\mbf{x}_i\}_{i=1}^{T-1}$  into a \isometric latent space via \textbf{AAT-VAE}, enabling the Flow Matching network to learn dynamics on a smooth, structure-preserving manifold. \textbf{b) Residual Latent Flow.} To effectively decouple the static equilibrium structure from stochastic dynamic fluctuations, the flow models the relative evolution $\mathbf{r}_t = \mathbf{z}_t - \mathbf{x}_0$ anchored to the initial frame $\mathbf{x}_0$. The detailed network architecture can be found in App. Figure~\ref{fig:architecture_detail}.
    }
    \label{fig:framework}
    \vspace{-4mm}
\end{figure*}

Direct modeling of molecular conformations in Cartesian coordinate space often suffers from inherent instability, as even small perturbations can lead to unphysical distortions or violations of steric constraints. 
To address this, we introduce a latent representation governed by a \textbf{physics-inspired atom-anchored prior}, constructing a structured and regularized manifold for generative modeling. 
Unlike standard approaches that compress diverse geometries into a generic isotropic Gaussian space~\cite{kingma2019introduction}, our design utilizes a \textit{\isometric} latent space anchored to atomic coordinates. This provides a stable foundation for probabilistic learning, enabling the flow model to capture smooth conformational transitions while avoiding the irregularities characteristic of raw Cartesian coordinates. 
A decoder subsequently maps these refined latent variables back to precise three-dimensional positions, ensuring physical plausibility. This design aligns with recent geometric representation learning advancements~\cite{kong2024full,mansoor2024protein,geffner2025proteina,kong2025unimomo}, demonstrating the effectiveness of decoupling representation robustness from generative dynamics.

\subsection{All-atom Dynamic Modeling via Atom-Anchored Thermal VAE}
\label{sec:harmonic_vae}

Unlike standard autoencoders that provide deterministic point estimates, a Variational Autoencoder (VAE) introduces a probabilistic latent space to capture the intrinsic uncertainty and conformational entropy of molecular systems. 
However, the standard Gaussian prior $\mathcal{N}(\mathbf{0}, \mathbf{I})$ used in conventional VAEs is structurally agnostic, forcing spatially distinct atoms to collapse toward a featureless origin. Although recent work considers coarse-grain geometric constraints in the latent space~\cite{kong2024full}, this ignores the fundamental all-atom geometric nature of molecules, especially in the dynamics simulation. 
To overcome this, we propose an Atom Anchored Thermal VAE (AAT-VAE), which imposes a physically grounded constraint on the all-atom \isometric latent space.
\vspace{-2mm}
\paragraph{All Atom Anchored Thermal Prior.}
In molecular dynamics (MD), atomic motions are often modeled as fluctuations around equilibrium positions governed by local force fields. Inspired by this, we replace the standard Gaussian prior with an \textbf{data-dependent atom-centered prior} that respects the instantaneous geometry of the molecule.
Because the generative dynamics in our framework are handled subsequently by the flow matching model, the VAE is tasked with probabilistically regularizing the representation space rather than acting as a standalone zero-shot generator. Thus, for each atom $i$ with coordinate $\mathbf{x}_i$, we formulate the variational autoencoding objective with a spatially-anchored regularization:
\begin{align}
    \mathcal{J}(\mathbf{x}_i) 
    = \mathbb{E}_{q_\theta(\mathbf{z}_i \mid \mathbf{x}_i, \mathbf{c}_i)}\big[\log p_\phi(\mathbf{x}_i \mid \mathbf{z}_i, \mathbf{c}_i)\big]  - \mathrm{KL}\big(q_\theta(\mathbf{z}_i \mid \mathbf{x}_i, \mathbf{c}_i)\,\Vert\, \mathcal{N}(\mathbf{x}_i, \sigma_p^2 \mathbf{I}) \big),
    \label{eq:vae_ac_elbo}
\end{align}

where $\mathbf{c}_i$ denotes conditional context (e.g., residue type, atom type, transmembrane helix indices). 
Crucially, the prior $p(\mathbf{z}_i) = \mathcal{N}(\mathbf{x}_i, \sigma_p^2 \mathbf{I})$ is a harmonic regularization term (The sensitivity analysis for $\sigma$ is shown in App. Table.~\ref{tab:sigma_sensitivity}.). This choice is motivated by three key physical considerations:

\vspace{-2mm}
\begin{enumerate}[leftmargin=*, nosep]
    \item \textbf{Harmonic Approximation:} The log-probability of our prior $p(\rvz_i)$ corresponds to a quadratic penalty term $\|\mathbf{z}_i - \mathbf{x}_i\|^2$. This mimics the harmonic approximation of local potential energy wells used in normal mode analysis~\cite{hinsen2005normal,zhou2024harmonic}, keeping latent codes near their atomic centers. 
    \item \textbf{Thermal Fluctuation:} Atoms in MD simulations naturally exhibit thermal noise around mean positions~\cite{hoyt2010fluctuations}, quantified by Root-Mean-Square Fluctuation (RMSF)~\cite{frenkel2023understanding,karplus2002molecular,dong2018structural}. This formulation naturally aligns the probabilistic latent noise with the intrinsic stochasticity of atomic vibrations, treating randomness as a physical property.
    \item \textbf{Geometric Preservation:} By anchoring the KL divergence to $\mathbf{x}_i$, we prevent the \textit{posterior collapse} often seen in VAEs (see App.~\ref{appsec:Posterior} for quantitative evidence) and ensure the latent space retains the explicit topology of the protein-ligand complex. This creates a compact, informative manifold that filters high-frequency noise while preserving essential structural motifs.
\end{enumerate}

\vspace{-2mm}
\paragraph{Training Objectives.}
We input all frames from one trajectory $\mathcal{X}_T$ into the encoder $q_\theta$. Implementation-wise, the encoder is built upon the Diffusion Transformer (DiT) architecture (details in Sec.~\ref{sub:model_architecture}), which employs attention layers to deeply fuse the protein and ligand features, ensuring that the latent codes capture the coupled allosteric effects.
 
The encoder $q_\theta$ output the corresponding latent coordinates $\mathcal{Z} = \{\mathbf{z}_0, \dots, \mathbf{z}_{T-1}\}$, which are subsequently reconstructed by the decoder $p_\phi$ as $\hat{\mathcal{X}}_T$. The total VAE objective aggregates the reconstruction error and the regularization:
\begin{align}\label{eq:HPVAE_loss}
\scalemath{0.8}{
    \mathcal{L}_{vae} }= \scalemath{0.89}{\frac{1}{TN}\sum_{t=0}^{T-1} \sum_{i=1}^{N} \Big( \|\hat{\mathbf{x}}_{t,i} - \mathbf{x}_{t,i}\|_2^2 }  \scalemath{0.89}{+ \lambda_{\text{KL}} \cdot \mathrm{KL}\big(\mathcal{N}(\mathbf{z}_{t,i}; \mu_\theta, \sigma_\theta) \,\Vert\, \mathcal{N}(\mathbf{x}_{t,i}, \sigma \mathbf{I})\big) \Big) + \mathcal{L}_{aux}}
\end{align}
where $\mathbf{z}_{t,i}$ is sampled from the encoder distribution. The reconstruction term corresponds to the negative log-likelihood under an isotropic Gaussian decoder. In practice, we use weighted rigid aligned MSE (details in App. \ref{appsec:weighedMSE}) to ensure that the MSE loss remains invariant to global rotational and translational differences.
The KL term explicitly penalizes deviations from the input geometry, effectively learning a \textit{denoised} and \textit{distinct} representation of the trajectory. Here, $\lambda_{\text{KL}}$ is a trade-off hyperparameter.
$\mathcal{L}_{aux}$ represents auxiliary geometric consistency losses (e.g., bond lengths, angles), detailed in App.~\ref{appsec:auxiliary}. To further enhance physical fidelity during long-trajectory sampling, we extend the framework to an Energy-Guided \ourVAE, as detailed in App.~\ref{appsec:Energy_HPVAE}.

\subsection{Residual Latent Flow Matching}
\label{sec:latent_flow}
 \vspace{-2mm}
Different from previous latent diffusion or flow approaches that compress molecular structures into low-dimensional semantic spaces~\citep{sengar2025generative, sengar2025beyond}, our method performs flow matching in a \textit{\isometric} all-atom-level latent space. This design preserves the original topological information to the greatest extent while enhancing the robustness of the generative process. 

Crucially, we propose a residual generative paradigm to model the evolution of dynamics. Standard flow matching typically integrates an ODE trajectory from uninformative Gaussian noise $\mathcal{N}(\mathbf{0}, \mathbf{I})$ to the data distribution~\cite{lipman2023flow}. However, MD simulation is intrinsically an evolution process governing how a system propagates from an initial state $\mathbf{x}_0$, rather than emerging from chaos. 
Leveraging the property that our latent coordinates maintain the same dimensionality as the original input (dimension is $3$), we explicitly leverage the initial frame from Cartesian space as a \textit{structural prior}. 
We define the flow on the \textit{relative displacement vector} between the current frame $t$ and the initial frame $\mathbf{x}_0$  in the latent space. To the best of our knowledge, no existing
work in the molecular domain has established such an explicit \textit{mathematical linkage} between physical
Cartesian space and the generative latent space.
This formulation effectively \textit{decouples} the static equilibrium structure from the stochastic dynamic fluctuations. By shifting the learning target to the displacement field, the model focuses its capacity on learning the transition probability density and the complex force fields required to cross energy barriers, rather than reconstructing the trivial covalent geometries from scratch (Figure~\ref{fig:framework}b). 
Formally, we define the flow on the relative latent coordinate space. Let $\mathbf{r}_{t,i} = \mathbf{z}_{t,i} - \mathbf{x}_{0,i}$ denote the \textbf{residual vector}. The flow is defined as:
\begin{align}
    \scalemath{0.89}{\frac{d\phi_\tau(\mathbf{r}_{t,i})}{d\tau} } = \scalemath{0.89}{v(\phi_\tau(\mathbf{r}_{t,i}), \tau),} \quad
    \scalemath{0.9}{
    \phi_0(\mathbf{r}_{t,i}) = \epsilon \sim \mathcal{N}(\mathbf{0}, \mathbf{I}),} \quad  
    \scalemath{0.9}{\phi_1(\mathbf{r}_{t,i}) = \mathbf{r}_{t,i} = \mathbf{z}_{t,i} - \mathbf{x}_{0,i}} \notag
\end{align}
where $\tau \in [0,1]$ represents the flow time. We employ the Rectified Flow framework~\cite{lipman2023flow,liu2023flow,hertrich2025relation} to learn straight paths between the noise and the target displacement. The intermediate state is interpolated as $\mathbf{r}_{t,i}^\tau := \tau\mathbf{r}_{t,i}^1 + (1-\tau) \epsilon$. The target velocity is simply $\mathbf{r}_{t,i}^1 - \mathbf{r}_{t,i}^0 = \frac{\mathbf{r}_{t,i}^1 - \mathbf{r}_{t,i}^\tau}{1-\tau}$. We parameterize a neural network $v_\theta$ to regress the clean data $\mathbf{r}_{t,i}^1$~\cite{stark2023harmonic}, yielding the velocity form $v_\theta(\mathbf{r}_{t,i}^\tau, \mathbf{c}_i, \tau) = \frac{\mathbf{r}_\theta(\mathbf{r}_{t,i}^\tau, \mathbf{c}_i, \tau) - \mathbf{r}_{t,i}^\tau}{1-\tau}$. 
The training objective in the latent space is implemented as:
\begin{align}
\scalemath{0.8}{
    \mathcal{L}_{fm} = \frac{1}{N(T-1)}\sum_{t=1}^{T-1}\sum_{i=1}^N \| \mathbf{r}_\theta(\mathbf{r}_{t,i}^\tau, \mathbf{c}_i, \tau) - \mathbf{r}_{t,i}^{1} \|_2^2}
\end{align}
This residual formulation offers significant advantages over absolute coordinate prediction. First, it aligns well with the physics of conformational ensembles, where dynamics can be viewed as ``pushing" a representation away from a reference state. Second, it improves generative robustness; since the flow learns perturbations around a valid physical structure ($\mathbf{x}_0$), the generated latent representations are less likely to drift outside the valid region of the VAE manifold, ensuring high fidelity in the decoded structures.

\vspace{-3mm}
\paragraph{Sampling.} Given the first frame, the flow model will generate the relative latent coordinate $\mcal{R}^{1}=[\rvr_{1}^{1}, \rvr_{1}^{2},\cdots, \rvr_{T-1}^{1}]$ of the trajectory simultaneously via an ODE-solver, taking the Euler method as an example,
\begin{align}
\scalemath{0.9}{
    \hat{\mcal{R}}^{\tau + d\tau}= \hat{\mcal{R}}^\tau + v_\theta(\hat{\mcal{R}}^{\tau}, \mcal{C}, \tau) d\tau,\; 
    \hat{\mcal{R}}^0 =[\bm{\epsilon}_1,\bm{\epsilon}_2,\cdots,\bm{\epsilon}_{T-1}] }
\end{align}
where $\mcal{C} = \{\rvc_i\}_{i=0}^{T-1}$ denotes all the conditional information.
After completing the iteration, we input the latent trajectory $\hat{\mcal{Z}}^{1}_{1:T-1}=\hat{\mcal{R}}^{1} + \x_0$ into the decoder $\mcal{D}_{\phi}$, obtaining the whole trajectory in the Cartesian space, i.e. $\hat{\mcal{X}}_{1:T}=\mcal{D}_\phi (\hat{\mcal{Z}}^1_{1:T-1})$.

\subsection{Model Architecture}\label{sub:model_architecture}
 \vspace{-2mm}
 
We develop an Asymmetric Spatio-Temporal AutoEncoder to bridge the training-inference gap. Specifically, we omit temporal modules in the encoder (aligning with inference inputs) and utilize a hierarchical spatial-temporal module within the decoder to strongly enhance temporal relevance modeling.
The core backbone of our architecture is built upon the Diffusion Transformer (DiT)~\cite{peebles2023scalable}. The detailed schematic is provided in App. Figure~\ref{fig:architecture_detail}.

\vspace{-2mm}
\paragraph{Structural Conditioning and Domain Embedding.}
Inspired by ~\cite{feng2025biomd},
the initial conformation ($\mathbf{x}_0$) is processed by a {Graph Transformer} to extract SE(3)-invariant single ($\mathbf{s}_i$) and pair ($\mathbf{z}_{ij}$) representations to establish a structural prior (alg. \ref{alg:main_loop}).
To integrate biological priors, we explicitly inject GPCR-specific domain knowledge: (1) Transmembrane (TM) Topology: The TM helix index (TM1–TM7) of each residue is embedded as a learnable vector. (2) Pharmacological Classification: The functional category of the ligand (e.g., agonist, antagonist) is similarly embedded. These domain features are concatenated with the token-level single representations—where a token corresponds to a receptor residue or a ligand atom—serving as the conditioning context for the encoder, decoder, and velocity network.

\begin{table*}[ht]\centering
\vspace{-5mm}
\caption{Statistical metrics on the MD ensemble benchmark of the GPCRMD test set (sequence similarity $<$ 50\%), where the median across all test targets is reported. The runtime is reported as \textbf{GPU second} required per sample, averaged over all test targets.}
\vspace{-3mm}
\label{tab:gpcr_ligand_emsemble_biomd}
\small
\setlength{\tabcolsep}{10pt} 
\scalebox{0.8}{
\begin{NiceTabular}{cl|cc|cc|cc}
    \CodeBefore
        \columncolor{blue!5}{3-8}
    \Body
    \toprule[1.5pt]
    \Block[]{2-2}{Metrics / Methods} & & \Block[]{1-2}{100 $ns$} & & \Block[]{1-2}{200 $ns$} & & \Block[]{1-2}{500 $ns$}  \\
    \cmidrule(lr){3-4} \cmidrule(lr){5-6} \cmidrule(lr){7-8}
    & & BioMD  & \ourM & BioMD  & \ourM & BioMD  & \ourM \\
    \midrule
    \Block[]{3-1}{Predicting\\flexibility}
    & Pairwise RMSD $r$ $\uparrow$     & 0.45 & \textbf{0.67} & 0.30 & \textbf{0.62} & 0.60 & \textbf{0.64} \\
    & Global RMSF $r$ $\uparrow$       & 0.69 & \textbf{0.76} & 0.65 & \textbf{0.74} & 0.57 & \textbf{0.75} \\
    & Per-target RMSF $r$  $\uparrow$  & 0.77 & \textbf{0.81} & 0.71 & \textbf{0.79} & 0.60 & \textbf{0.79} \\
    \midrule
    \Block[]{6-1}{Distributional\\accuracy}
    & Root mean $\mathcal{W}_2$-dist. $\downarrow$ & 2.81 & \textbf{2.40} & 2.87 & \textbf{2.53} & \textbf{2.83} & 2.92 \\
    & $\hookrightarrow$ Trans. contrib. $\downarrow$ & 2.43 & \textbf{2.24} & 2.49 & \textbf{2.34} & \textbf{2.51} & 2.68 \\
    & $\hookrightarrow$ Var. contrib. $\downarrow$   & 1.41 & \textbf{0.87} & 1.38 & \textbf{0.97} & 1.34 & \textbf{1.10} \\
    & MD PCA $\mathcal{W}_2$-dist. $\downarrow$    & 1.30 & \textbf{1.16} & 1.29 & \textbf{1.16} & 1.24 & \textbf{1.10} \\
    & Joint PCA $\mathcal{W}_2$-dist. $\downarrow$ & 2.17 & \textbf{1.92} & 2.22 & \textbf{2.02} & \textbf{2.19} & 2.41 \\
    & \% PC-sim $>0.5$ $\uparrow$        & 7.83 & \textbf{14.78} & 1.74 & \textbf{3.48} & 1.04 & \textbf{1.74} \\
    \midrule
    \Block[]{4-1}{Ensemble\\observables}
    & Weak contacts $J$ $\uparrow$       & 0.11 & \textbf{0.52} & 0.12 & \textbf{0.51} & 0.16 & \textbf{0.48} \\
    & Transient contacts $J$ $\uparrow$ & 0.26 & \textbf{0.30} & 0.24 & \textbf{0.27} & 0.20 & \textbf{0.22} \\
    & Exposed residue $J$ $\uparrow$    & 0.30 & \textbf{0.54} & 0.29 & \textbf{0.55} & 0.29 & \textbf{0.53} \\
    & Exposed MI matrix $\rho$ $\uparrow$ & 0.17 & \textbf{0.27} & 0.15 & \textbf{0.27} & 0.15 & \textbf{0.24} \\
    \midrule

    \Block[]{1-1}{Runtime\textsuperscript{*}}
    & GPU sec. per sample  & 10.0 & \textbf{2.0} & 20.0 & \textbf{4.0} & 50.0 & \textbf{10.0} \\
\bottomrule[1.5pt]
\end{NiceTabular}

\vspace{-6mm}
}
\par 
    \vspace{1pt}
    \parbox{1.0\linewidth}{
        \scriptsize 
        \textbf{Note:} \textsuperscript{*}\ourM achieves faster inference speeds because its latent velocity network has significantly fewer parameters than BioMD's Cartesian velocity network ($1.6$M vs.\ $78$M). 
A traditional 500 ns MD simulation requires over 48 hours to complete.

    }
 \vspace{-6mm}   
\end{table*}
\vspace{-2mm}
\paragraph{Spatio-Temporal AAT-VAE.} To efficiently model high-resolution dynamics, we employ a hierarchical interaction mechanism inspired by AlphaFold3~\cite{abramson2024accurate}.
(i) \textbf{Spatial Encoder:} As detailed in alg. \ref{alg:encoder}, we utilize an \texttt{Atom Attention Encoder} to process the multi-frame coordinates. By treating the receptor and ligand as a contiguous sequence of atoms, the sequence-local all-atom attention naturally captures the joint distribution of the complex. This ensures that receptor and ligand structural information are deeply fused in the resulting atom-level latent variables $\mathbf{q}_l$ (subsequently projected to $(\vec{\mu}_l, \vec{\sigma}_l)$).
(ii) \textbf{Temporal Decoder:} As detailed in alg. \ref{alg:Decoder}, the sampled atomic latents $\mathbf{q}_l \sim \mathcal{N}(\vec{\mu}_l, \vec{\sigma}_l)$ are first aggregated into tokens to capture global temporal dependencies via a \texttt{Token Temporal Attention Decoder} (attention maps are visualized in App. Figure \ref{fig:attnmap}). Subsequently, an \texttt{Atom Temporal Attention Decoder} reconstructs the fine-grained coordinates $\mathbf{x}_t$ by processing the atomic latents $\mathbf{q}_l$ conditioned on these temporally updated tokens. This hierarchical design ensures that global dynamics (token-level) effectively guide local atomic interactions (atom-level) during reconstruction.

\vspace{-2mm}
\paragraph{Latent Velocity Network.}
The velocity network $v_\theta$, responsible for the flow matching process, is distinct from the VAE. It accepts the noisy residual latent coordinates $\mathbf{r}^{\tau}$, and the conditional single and token representations $\mathbf{s}_{i}$ and $ \rva_i$ from the structural conditioner. It utilizes an atomic DiT-based architecture with cross-attention to the conditioned structural tokens to predict the flow velocity in the all-atom latent space. To better utilize the token information from the encoder and bridge the gap between training and inference, we creatively proposed a Stochastic Token Propagator, which actively transfers the structural context from the initial frame tokens ($\mbf{a}_0$) into the temporal domain by
mixing them with time-scaled diffusion noise, as shown in alg. \ref{alg:latent_velocity_network},

 \vspace{-2mm}
 \section{Experimental Results}

\begin{figure*}[ht]
    \centering
    \vspace{-5mm}
    \includegraphics[width=0.99\linewidth]{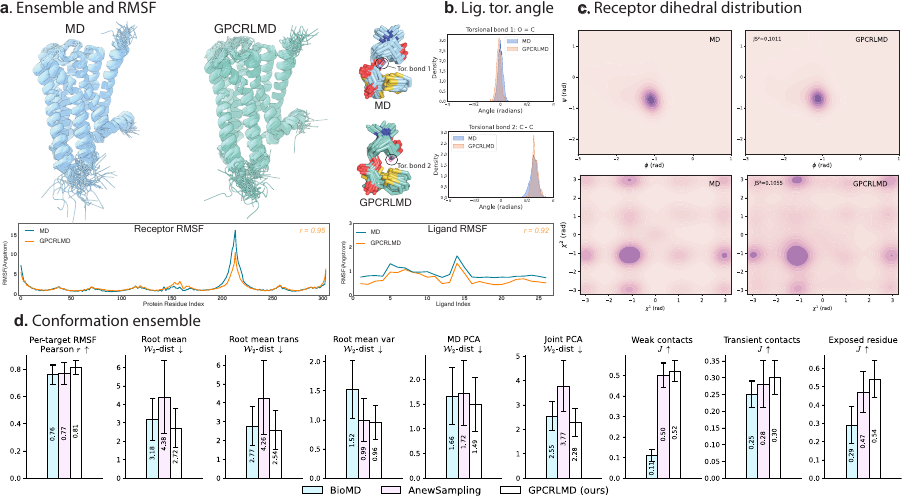}
    \vspace{-2mm}
    \caption{
    Evaluation of ensemble fidelity and structural validity. \textbf{a)} Receptor and ligand RMSF profiles compared to ground-truth MD (PDBID $\mathrm{5DSG}$). \textbf{b)} Ligand torsional angle distributions ( $\mathrm{5DSG}$). \textbf{c)} Coupled backbone ($\phi$-$\psi$) and side-chain ($\chi^1$-$\chi^2$) dihedral distributions ( $\mathrm{5DSG}$). \textbf{d)} Conformation ensemble benchmarks against AnewSampling and BioMD. Comparison with baselines including Str2str~\cite{str2str}, AlphaFlow\cite{jingAlphaFoldMeetsFlow2024} and MDgen~\cite{jing2024generative}, see App. Table \ref{tab:gpc_emsemble_bioemu}.  Metrics are averaged across the test set (250 conformations per sample).}
    \vspace{-3mm}
    \label{fig:emsemble_results}
\end{figure*}

\begin{figure*}[ht]
    \centering
    \vspace{-7mm}
    \includegraphics[width=0.99\linewidth]{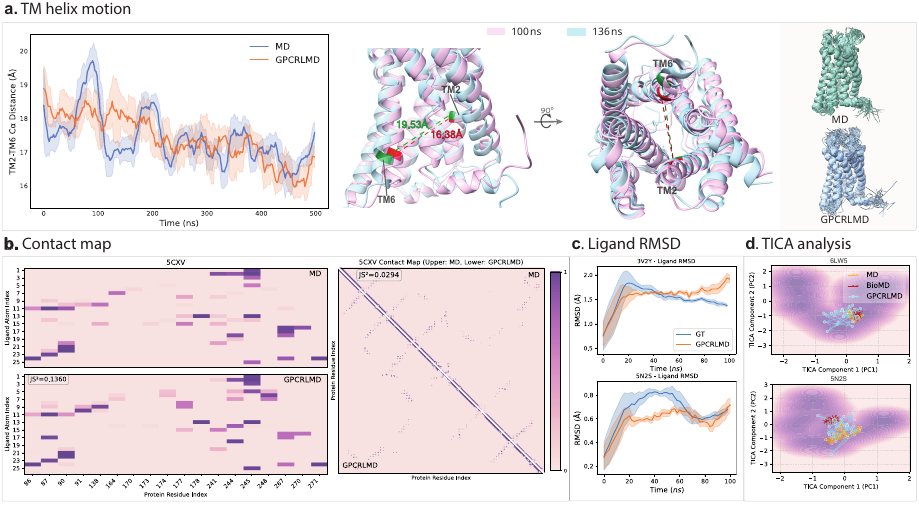}
    \vspace{-2mm}
    \caption{
    Evaluation of complex dynamics and interaction stability. \textbf{a)} Transmembrane (TM) helix motion. Time-evolution of the critical TM2-TM6 distance (PDBID: $\mathrm{6LW5}$) alongside a structural overlay of the conformational transition. More examples in Figure~\ref{fig:tm_distance}. \textbf{b)} Dynamic contact maps. Receptor-ligand and receptor-receptor interaction frequencies calculated using a dual-cutoff metric (details in Appendix~\ref{appsec:dual_contactmap_calculate}). \textbf{c)} Ligand stability. Time-dependent Ligand RMSD relative to the initial frame ($t=0$). \textbf{d)} Free Energy Landscape (TICA). Projection of generated and reference trajectories onto the backbone energy surface, visualizing conformational space coverage. More examples and results are shown in Figure \ref{fig:tica_bb_500ns}, \ref{fig:tica_sidechain_500ns} and Table~\ref{tab:tica_js_divergence}.
    }
        \vspace{-5mm}
    \label{fig:dynamic_results}
\end{figure*}
 We primarily evaluate \ourM on the GPCRMD dataset. To further demonstrate its generalization capabilities, we extend our framework to \textbf{MISATO}~\citep{siebenmorgen2024misato}, a comprehensive protein-ligand MD dataset. As detailed in App. Table~\ref{tab:misato_results}, \ourM achieves state-of-the-art performance on this benchmark.

 \vspace{-2mm}
\paragraph{Data Curation:} 
We constructed our dataset from the GPCRMD database~\cite{rodriguez2020gpcrmd}, leveraging the extensive GPCR dynamics sampled by \citet{aranda2025large}, supplemented by additional test trajectories generated in-house. The raw dataset comprises 703 molecular dynamics trajectories (detailed in Table~\ref{tab:data_summary}), with each replica spanning 500~ns (2,500 frames). To ensure rigorous generalization and prevent data leakage, we implemented a strict split based on receptor sequence identity ($<50\%$), resulting in a partition of 531:57:115 trajectories for training, validation, and testing, respectively. Notably, over 95\% of the test samples feature ligand scaffolds distinct from those in the training set. Crucially, this training partition constitutes a massive-scale dataset of over 1.32 million distinct conformational states. Unlike static protein datasets, this dense temporal sampling allows our model to learn the continuous dynamical manifold and rare allosteric transitions, rather than merely memorizing discrete crystal structures.
\textbf{Training Protocol and Strided Sampling:} To efficiently capture long-timescale dynamics within memory constraints, we employ a strided window sampling strategy. During training, we sample sub-trajectories of $50$ frames with a temporal stride of $10$ frames. This effectively covers a $100 \text{ns}$ physical timescale per input sample ($50 \times 10 \times 0.2 \text{ns}$), enabling the model to learn temporal correlations across significant biological windows while maintaining computational tractability.
\textbf{Evaluation Metrics:} For quantitative assessment, we adopt the rigorous benchmarking criteria proposed by~\citet{jingAlphaFoldMeetsFlow2024}, focusing on predicting
flexibility, distributional consistency, and ensemble validity. Please refer to App.~\ref{appsec:main_metrics} for details.

\vspace{-2mm}
\subsection{Ensemble Fidelity and Structural Validity} \label{sub:ensemble_eval} 
\vspace{-3mm}
We first assess whether \ourM captures the correct ensemble distribution of the receptor-ligand complex. Because \ourM is, to the best of our knowledge, the first generative framework specifically tailored for GPCR-ligand dynamics, we benchmark it against leading methods from related general domains: BioMD~\cite{feng2025biomd} (a general protein-ligand dynamics model, retrained on our dataset) and AnewSampling~\cite{wang2026learning} (a recent SOTA protein-ligand ensemble generation method, evaluated via its official \href{https://anewbt-mind.healthybaike.com/}{API}). Comparisons with BioEmu~\cite{lewis2025scalable} (a SOTA protein ensemble generator, using official checkpoints) are provided in App.~\ref{appsec:bioemu}.

\vspace{-3mm}
\paragraph{Statistical Benchmarks and Flexibility.} As summarized in Table~\ref{tab:gpcr_ligand_emsemble_biomd}, \ourM achieves state-of-the-art performance across most metrics for $100$ns--$500$ns simulations. It significantly surpasses the BioMD baseline in \textit{Predicting Flexibility} and \textit{Ensemble Observables}. We also evaluate our conformational ensemble against AnewSampling~\cite{wang2026learning}; the results in Figure~\ref{fig:emsemble_results}d demonstrate that \ourM consistently outperforms AnewSampling across all primary metrics. This quantitative success is visually mirrored in the residue-wise RMSF profiles for both the receptor and ligand (Figure~\ref{fig:emsemble_results}a; App. Figures~\ref{fig:rmsf_protein} and \ref{fig:rmsf_ligand}). The profiles exhibit high similarity to ground-truth MD: \ourM correctly maintains structural rigidity in the transmembrane helices while reproducing high diversity in the flexible loops, and it accurately captures the fluctuation patterns of the ligand.

\vspace{-2mm}
\paragraph{Structural Validity (Dihedrals and Torsions).}To ensure physical plausibility, we analyze the coupled distribution of backbone ($\phi$-$\psi$) and side-chain ($\chi^1$-$\chi^2$) dihedral angles (Figure~\ref{fig:emsemble_results}c), as well as the ligand torsional angle distributions (Figure~\ref{fig:emsemble_results}b). \ourM maintains high distributional similarity with ground-truth (GT) MD across both protein and ligand degrees of freedom, indicating that the model respects local chemical geometry. 
\vspace{-2mm}
\subsection{Complex Dynamics and Interaction Stability}\label{sub:dynamics_eval}
\vspace{-2mm}
We further evaluate the model's ability to simulate time-evolving biological mechanisms.

\vspace{-3mm}
\paragraph{Capturing Critical State Transitions.} The relative movement between Transmembrane Helix 2 (TM2) and TM6 is a hallmark of GPCR activation~\cite{aranda2025large}. As shown in Figure~\ref{fig:dynamic_results}a, \ourM accurately reproduces the distance fluctuations and opening/closing trends of the TM helices. This confirms the model learns the underlying allosteric mechanism rather than just memorizing static coordinates.

\vspace{-3mm}
\paragraph{Dynamic Contact Stability.} Non-covalent interactions are the driving force of complex stability. We compute the dynamic contact map using a dual-cutoff metric to measure interaction frequency over time. As shown in Figure~\ref{fig:dynamic_results}b. \ourM faithfully reproduces the specific contact patterns observed in the GT MD, maintaining critical receptor-ligand interactions throughout the trajectory.

\vspace{-3mm}
\paragraph{Ligand Stability and Energy Landscape.}We examine the ligand's dynamic stability via RMSD profiles (Figure~\ref{fig:dynamic_results}c). The generated ligands exhibit bounded RMSD fluctuations consistent with stable binding. 
Finally, to assess global dynamics, we project the trajectories onto the Free Energy Landscape (FEL) using TICA. As shown in Figure~\ref{fig:dynamic_results}d, the \ourM trajectory consistently maps to valid low-energy basins defined by the reference replica MD. This demonstrates that the model recovers essential metastable states and effectively samples diverse conformations, robustly maintaining physical validity.

\begin{wraptable}{R}{0.65\textwidth}
\centering
\vspace{-4mm}
\caption{Ablation study comparing \ourM with four variants.
}
\label{tab:ablation_study}
\small
\vspace{-3mm}
\setlength{\tabcolsep}{3pt}
\resizebox{0.65\textwidth}{!}{
\begin{NiceTabular}{l|ccccc}
    \CodeBefore
        \rowcolor{blue!5}{2}
    \Body
    \toprule[1.5pt]
    \RowStyle{\bfseries}
    Methods & \Block{1-1}{Global\\RMSF $r$ } & \Block{1-1}{Per-target\\RMSF $r$ } & \Block{1-1}{Root mean\\$\mathcal{W}_2$-dist } & \Block{1-1}{Joint PCA\\$\mathcal{W}_2$-dist} & \Block{1-1}{Transient \\contacts $J$} \\
    \midrule
    \ourM (w.all)   & \textbf{0.76} & \textbf{0.81} & \textbf{2.40} & \textbf{1.92} & \textbf{0.30} \\
        w/o Residual Vector & 0.71          & 0.76          & 4.00          & 3.53          & 0.19 \\
    w/o AAT Prior (Std. Gauss)  & 0.75 & 0.75 & 3.57 & 3.18 & 0.13 \\
    w/o Latent Flow     & 0.71          & 0.78          & 2.41          & 1.97          & 0.29 \\
    w/o Time Dep. AAT Prior &  0.72 & 0.78 & 2.42 & 1.95 & 0.28 \\

    w/o VAE (Deterministic AE)     & 0.63          & 0.72          & 2.75          & 2.05          & 0.25 \\
    w/o AAT-VAE     & 0.70          & 0.77          & 2.62          & 2.01          & 0.16 \\
    \bottomrule[1.5pt]
\end{NiceTabular}
}
\vspace{-4mm}
\end{wraptable}
\vspace{-1mm}
\subsection{Ablation Study}
\vspace{-3mm}
We conduct an ablation study to quantify the contribution of our core design components (Table~\ref{tab:ablation_study}). We evaluate five variants (100ns): (i) w/o Residual Vector: Predicting absolute latent coordinates rather than relative displacements anchored to the initial conformation; (ii) w/o AAT Prior: Replacing the target $\mbf{x}_{t,i}$ with the Gaussian distribution $\mathcal{N}(\mbf{0}, \mbf{I})$ to demonstrate the importance of the proposed prior.
  (iii) w/o Latent Flow: Replacing the residual flow with standard VAE sampling ($\mathbf{z} \sim \mathcal{N}(\mu, \sigma \mbf{I})$) to generate trajectories directly; (iv) w/o time dependent AAT Prior: Replacing the time-dependent target $\mbf{x}_{t,i}$ with the static initial frame $\rvx_{0,i}$ in Eq.~\ref{eq:HPVAE_loss} to assess the benefit of the dynamic structural prior; (v) w/o VAE: Replacing Variational Autoencoder with Deterministic Autoencoder;
and (vi) Cartesian Space Flow (w/o AAT-VAE): Applying flow matching directly in Cartesian space.
Removing any of these components leads to a significant performance drop, demonstrating that our architecture is tailored specifically for MD simulation rather than being a naive combination of modules.

\vspace{-2mm}
\section{Conclusion} 
\vspace{-3mm}
We introduced \ourM, a generative framework integrating an Atom-Anchored Thermal VAE and Residual Latent Flow for all-atom GPCR-ligand complex simulation. 
Crucially, we demonstrate that performing sampling in a regularized all-atom latent space, rather than in raw Cartesian coordinates, yields superior structural fidelity and ensemble consistency in dynamics simulation. Consequently, the proposed method simulates dynamics with unprecedented efficiency, compressing hundreds of nanoseconds of trajectories into seconds while achieving SOTA fidelity
in capturing ensemble distributions, structural flexibility, and critical interactions.
However, because the generated trajectories are anchored to the initial state, the model's performance may be sensitive to the physical plausibility of this starting conformation. To mitigate this, future pipelines could employ a brief energy relaxation step to refine the local structure prior to inference. Looking ahead, extending this framework to general GPCR monomer systems presents a highly promising direction for broader applications.



\bibliography{ref.bib}

\begin{thebibliography}{54}
\providecommand{\natexlab}[1]{#1}
\providecommand{\url}[1]{\texttt{#1}}
\expandafter\ifx\csname urlstyle\endcsname\relax
  \providecommand{\doi}[1]{doi: #1}\else
  \providecommand{\doi}{doi: \begingroup \urlstyle{rm}\Url}\fi

\bibitem[Abramson et~al.(2024)Abramson, Adler, Dunger, Evans, Green, Pritzel, Ronneberger, Willmore, Ballard, Bambrick, et~al.]{abramson2024accurate}
Josh Abramson, Jonas Adler, Jack Dunger, Richard Evans, Tim Green, Alexander Pritzel, Olaf Ronneberger, Lindsay Willmore, Andrew~J Ballard, Joshua Bambrick, et~al.
\newblock Accurate structure prediction of biomolecular interactions with alphafold 3.
\newblock \emph{Nature}, 630\penalty0 (8016):\penalty0 493--500, 2024.

\bibitem[Aranda-Garc{\'\i}a et~al.(2025)Aranda-Garc{\'\i}a, Stepniewski, Torrens-Fontanals, Garc{\'\i}a-Recio, Lopez-Balastegui, Medel-Lacruz, Morales-Pastor, Peralta-Garc{\'\i}a, Dieguez-Eceolaza, Sotillo-Nu{\~n}ez, et~al.]{aranda2025large}
David Aranda-Garc{\'\i}a, Tomasz~Maciej Stepniewski, Mariona Torrens-Fontanals, Adrian Garc{\'\i}a-Recio, Marta Lopez-Balastegui, Brian Medel-Lacruz, Adri{\'a}n Morales-Pastor, Alejandro Peralta-Garc{\'\i}a, Miguel Dieguez-Eceolaza, David Sotillo-Nu{\~n}ez, et~al.
\newblock Large scale investigation of gpcr molecular dynamics data uncovers allosteric sites and lateral gateways.
\newblock \emph{Nature communications}, 16\penalty0 (1):\penalty0 2020, 2025.

\bibitem[Ciancetta et~al.(2015)Ciancetta, Sabbadin, Federico, Spalluto, and Moro]{ciancetta2015advances}
Antonella Ciancetta, Davide Sabbadin, Stephanie Federico, Giampiero Spalluto, and Stefano Moro.
\newblock Advances in computational techniques to study gpcr--ligand recognition.
\newblock \emph{Trends in pharmacological sciences}, 36\penalty0 (12):\penalty0 878--890, 2015.

\bibitem[Conflitti et~al.(2025)Conflitti, Lyman, Sansom, Hildebrand, Guti{\'e}rrez-de Ter{\'a}n, Carloni, Ansell, Yuan, Barth, Robinson, et~al.]{conflitti2025functional}
Paolo Conflitti, Edward Lyman, Mark~SP Sansom, Peter~W Hildebrand, Hugo Guti{\'e}rrez-de Ter{\'a}n, Paolo Carloni, T~Bertie Ansell, Shuguang Yuan, Patrick Barth, Anne~S Robinson, et~al.
\newblock Functional dynamics of g protein-coupled receptors reveal new routes for drug discovery.
\newblock \emph{Nature Reviews Drug Discovery}, 24\penalty0 (4):\penalty0 251--275, 2025.

\bibitem[Dao \& Rahman(2025)Dao and Rahman]{dao2025deep}
Tuan~Minh Dao and Taseef Rahman.
\newblock Deep generative modeling of protein conformations: A comprehensive review.
\newblock \emph{BioChem}, 5\penalty0 (3):\penalty0 32, 2025.

\bibitem[Dong et~al.(2018)Dong, Liao, Meng, and Somero]{dong2018structural}
Yun-wei Dong, Ming-ling Liao, Xian-liang Meng, and George~N Somero.
\newblock Structural flexibility and protein adaptation to temperature: Molecular dynamics analysis of malate dehydrogenases of marine molluscs.
\newblock \emph{Proceedings of the National Academy of Sciences}, 115\penalty0 (6):\penalty0 1274--1279, 2018.

\bibitem[Feng et~al.(2025)Feng, Zhang, Zhang, Liu, and Li]{feng2025biomd}
Bin Feng, Jiying Zhang, Xinni Zhang, Zijing Liu, and Yu~Li.
\newblock Bio{MD}: All-atom generative model for biomolecular dynamics simulation.
\newblock In \emph{The Fourteenth International Conference on Learning Representations}, 2025.
\newblock URL \url{https://openreview.net/forum?id=LQDeJk6NOr}.

\bibitem[Frenkel \& Smit(2023)Frenkel and Smit]{frenkel2023understanding}
Daan Frenkel and Berend Smit.
\newblock \emph{Understanding molecular simulation: from algorithms to applications}.
\newblock Elsevier, 2023.

\bibitem[Geffner et~al.(2025)Geffner, Didi, Cao, Reidenbach, Zhang, Dallago, Kucukbenli, Kreis, and Vahdat]{geffner2025proteina}
Tomas Geffner, Kieran Didi, Zhonglin Cao, Danny Reidenbach, Zuobai Zhang, Christian Dallago, Emine Kucukbenli, Karsten Kreis, and Arash Vahdat.
\newblock La-proteina: Atomistic protein generation via partially latent flow matching.
\newblock \emph{arXiv preprint arXiv:2507.09466}, 2025.

\bibitem[Han et~al.(2023)Han, Zhang, Nazarova, Bernhard, Krumm, Zhao, Lam, Rangari, Majumdar, Nichols, et~al.]{han2023ligand}
Jianming Han, Jingying Zhang, Antonina~L Nazarova, Sarah~M Bernhard, Brian~E Krumm, Lei Zhao, Jordy~Homing Lam, Vipin~A Rangari, Susruta Majumdar, David~E Nichols, et~al.
\newblock Ligand and g-protein selectivity in the $\kappa$-opioid receptor.
\newblock \emph{Nature}, 617\penalty0 (7960):\penalty0 417--425, 2023.

\bibitem[Han et~al.(2024)Han, Xu, Lou, Ye, and Ermon]{han2024geometric}
Jiaqi Han, Minkai Xu, Aaron Lou, Haotian Ye, and Stefano Ermon.
\newblock Geometric trajectory diffusion models.
\newblock In \emph{The Thirty-eighth Annual Conference on Neural Information Processing Systems}, 2024.
\newblock URL \url{https://openreview.net/forum?id=OYmms5Mv9H}.

\bibitem[Hauser et~al.(2017)Hauser, Attwood, Rask-Andersen, Schi{\"o}th, and Gloriam]{hauser2017trends}
Alexander~S Hauser, Misty~M Attwood, Mathias Rask-Andersen, Helgi~B Schi{\"o}th, and David~E Gloriam.
\newblock Trends in gpcr drug discovery: new agents, targets and indications.
\newblock \emph{Nature reviews Drug discovery}, 16\penalty0 (12):\penalty0 829--842, 2017.

\bibitem[Hertrich et~al.(2025)Hertrich, Chambolle, and Delon]{hertrich2025relation}
Johannes Hertrich, Antonin Chambolle, and Julie Delon.
\newblock On the relation between rectified flows and optimal transport.
\newblock \emph{arXiv preprint arXiv:2505.19712}, 2025.

\bibitem[Hilger et~al.(2018)Hilger, Masureel, and Kobilka]{hilger2018structure}
Daniel Hilger, Matthieu Masureel, and Brian~K Kobilka.
\newblock Structure and dynamics of gpcr signaling complexes.
\newblock \emph{Nature structural \& molecular biology}, 25\penalty0 (1):\penalty0 4--12, 2018.

\bibitem[Hinsen(2005)]{hinsen2005normal}
Konrad Hinsen.
\newblock Normal mode theory and harmonic potential approximations.
\newblock In \emph{Normal Mode Analysis}, pp.\  25--40. Chapman and Hall/CRC, 2005.

\bibitem[Hoyt et~al.(2010)Hoyt, Trautt, and Upmanyu]{hoyt2010fluctuations}
JJ~Hoyt, ZT~Trautt, and M~Upmanyu.
\newblock Fluctuations in molecular dynamics simulations.
\newblock \emph{Mathematics and Computers in Simulation}, 80\penalty0 (7):\penalty0 1382--1392, 2010.

\bibitem[Jing et~al.(2024{\natexlab{a}})Jing, Berger, and Jaakkola]{jingAlphaFoldMeetsFlow2024}
Bowen Jing, Bonnie Berger, and Tommi Jaakkola.
\newblock Alphafold meets flow matching for generating protein ensembles.
\newblock In \emph{Proceedings of the 41st International Conference on Machine Learning}, pp.\  22277--22303, 2024{\natexlab{a}}.

\bibitem[Jing et~al.(2024{\natexlab{b}})Jing, St{\"a}rk, Jaakkola, and Berger]{jing2024generative}
Bowen Jing, Hannes St{\"a}rk, Tommi Jaakkola, and Bonnie Berger.
\newblock Generative modeling of molecular dynamics trajectories.
\newblock \emph{Advances in Neural Information Processing Systems}, 37:\penalty0 40534--40564, 2024{\natexlab{b}}.

\bibitem[Kapla et~al.(2021)Kapla, Rodr{\'\i}guez-Espigares, Ballante, Selent, and Carlsson]{kapla2021can}
Jon Kapla, Ismael Rodr{\'\i}guez-Espigares, Flavio Ballante, Jana Selent, and Jens Carlsson.
\newblock Can molecular dynamics simulations improve the structural accuracy and virtual screening performance of gpcr models?
\newblock \emph{PLoS Computational Biology}, 17\penalty0 (5):\penalty0 e1008936, 2021.

\bibitem[Karplus \& McCammon(2002)Karplus and McCammon]{karplus2002molecular}
Martin Karplus and J~Andrew McCammon.
\newblock Molecular dynamics simulations of biomolecules.
\newblock \emph{Nature structural biology}, 9\penalty0 (9):\penalty0 646--652, 2002.

\bibitem[Katritch et~al.(2013)Katritch, Cherezov, and Stevens]{katritch2013structure}
Vsevolod Katritch, Vadim Cherezov, and Raymond~C Stevens.
\newblock Structure-function of the g protein--coupled receptor superfamily.
\newblock \emph{Annual review of pharmacology and toxicology}, 53:\penalty0 531--556, 2013.

\bibitem[Kingma et~al.(2019)Kingma, Welling, et~al.]{kingma2019introduction}
Diederik~P Kingma, Max Welling, et~al.
\newblock An introduction to variational autoencoders.
\newblock \emph{Foundations and Trends{\textregistered} in Machine Learning}, 12\penalty0 (4):\penalty0 307--392, 2019.

\bibitem[Kong et~al.(2024)Kong, Jia, Huang, and Liu]{kong2024full}
Xiangzhe Kong, Yinjun Jia, Wenbing Huang, and Yang Liu.
\newblock Full-atom peptide design with geometric latent diffusion.
\newblock \emph{Advances in Neural Information Processing Systems}, 37:\penalty0 74808--74839, 2024.

\bibitem[Kong et~al.(2025)Kong, Zhang, Zhang, Jiao, Ma, Huang, Liu, and Liu]{kong2025unimomo}
Xiangzhe Kong, Zishen Zhang, Ziting Zhang, Rui Jiao, Jianzhu Ma, Wenbing Huang, Kai Liu, and Yang Liu.
\newblock Unimomo: Unified generative modeling of 3d molecules for de novo binder design.
\newblock In \emph{Forty-second International Conference on Machine Learning}, 2025.
\newblock URL \url{https://openreview.net/forum?id=KUN7A7Okb6}.

\bibitem[Latorraca et~al.(2017)Latorraca, Venkatakrishnan, and Dror]{latorraca2017gpcr}
Naomi~R Latorraca, AJ~Venkatakrishnan, and Ron~O Dror.
\newblock Gpcr dynamics: structures in motion.
\newblock \emph{Chemical reviews}, 117\penalty0 (1):\penalty0 139--155, 2017.

\bibitem[Lewis et~al.(2025)Lewis, Hempel, Jim{\'e}nez-Luna, Gastegger, Xie, Foong, Satorras, Abdin, Veeling, Zaporozhets, et~al.]{lewis2025scalable}
Sarah Lewis, Tim Hempel, Jos{\'e} Jim{\'e}nez-Luna, Michael Gastegger, Yu~Xie, Andrew~YK Foong, Victor~Garc{\'\i}a Satorras, Osama Abdin, Bastiaan~S Veeling, Iryna Zaporozhets, et~al.
\newblock Scalable emulation of protein equilibrium ensembles with generative deep learning.
\newblock \emph{Science}, pp.\  eadv9817, 2025.

\bibitem[Li et~al.(2025)Li, Zhang, Feng, Zeng, Chen, Pan, Li, Liu, and Yang]{li2025enhanced}
Maodong Li, Jiying Zhang, Bin Feng, Wenqi Zeng, Dechin Chen, Zhijun Pan, Yu~Li, Zijing Liu, and Yi~Isaac Yang.
\newblock Enhanced sampling, public dataset and generative model for drug-protein dissociation dynamics.
\newblock \emph{arXiv preprint arXiv:2504.18367}, 2025.

\bibitem[Lindorff-Larsen et~al.(2011)Lindorff-Larsen, Piana, Dror, and Shaw]{lindorff2011fast}
Kresten Lindorff-Larsen, Stefano Piana, Ron~O Dror, and David~E Shaw.
\newblock How fast-folding proteins fold.
\newblock \emph{Science}, 334\penalty0 (6055):\penalty0 517--520, 2011.

\bibitem[Lipman et~al.(2023)Lipman, Chen, Ben-Hamu, Nickel, and Le]{lipman2023flow}
Yaron Lipman, Ricky T.~Q. Chen, Heli Ben-Hamu, Maximilian Nickel, and Matthew Le.
\newblock Flow matching for generative modeling.
\newblock In \emph{ICLR}, 2023.

\bibitem[Liu et~al.(2023{\natexlab{a}})Liu, Du, Ma, Guo, and Tang]{liu2023group}
Shengchao Liu, Weitao Du, Zhi-Ming Ma, Hongyu Guo, and Jian Tang.
\newblock A group symmetric stochastic differential equation model for molecule multi-modal pretraining.
\newblock In \emph{International Conference on Machine Learning}, pp.\  21497--21526. PMLR, 2023{\natexlab{a}}.

\bibitem[Liu et~al.(2023{\natexlab{b}})Liu, Gong, and Liu]{liu2023flow}
Xingchao Liu, Chengyue Gong, and Qiang Liu.
\newblock Flow straight and fast: Learning to generate and transfer data with rectified flow.
\newblock In \emph{The Eleventh International Conference on Learning Representations (ICLR)}, 2023{\natexlab{b}}.

\bibitem[L{\'o}pez-Correa et~al.(2023)L{\'o}pez-Correa, K{\"o}nig, and Vellido]{lopez2023gpcr}
Juan~Manuel L{\'o}pez-Correa, Caroline K{\"o}nig, and Alfredo Vellido.
\newblock Gpcr molecular dynamics forecasting using recurrent neural networks.
\newblock \emph{Scientific reports}, 13\penalty0 (1):\penalty0 20995, 2023.

\bibitem[Lorente et~al.(2025)Lorente, Sokolov, Ferguson, Schi{\"o}th, Hauser, and Gloriam]{lorente2025gpcr}
Javier~S{\'a}nchez Lorente, Aleksandr~V Sokolov, Gavin Ferguson, Helgi~B Schi{\"o}th, Alexander~S Hauser, and David~E Gloriam.
\newblock Gpcr drug discovery: new agents, targets and indications.
\newblock \emph{Nature Reviews Drug Discovery}, pp.\  1--22, 2025.

\bibitem[Lu et~al.(2024)Lu, Zhong, Zhang, and Tang]{str2str}
Jiarui Lu, Bozitao Zhong, Zuobai Zhang, and Jian Tang.
\newblock Str2str: A score-based framework for zero-shot protein conformation sampling.
\newblock In \emph{The Twelfth International Conference on Learning Representations}, 2024.

\bibitem[Lu et~al.(2025)Lu, Chen, Lu, Lozano, Chenthamarakshan, Das, and Tang]{lu2025aligning}
Jiarui Lu, Xiaoyin Chen, Stephen~Zhewen Lu, Aurelie Lozano, Vijil Chenthamarakshan, Payel Das, and Jian Tang.
\newblock Aligning protein conformation ensemble generation with physical feedback.
\newblock In \emph{Forty-second International Conference on Machine Learning}, 2025.
\newblock URL \url{https://openreview.net/forum?id=Asr955jcuZ}.

\bibitem[Mansoor et~al.(2024)Mansoor, Baek, Park, Lee, and Baker]{mansoor2024protein}
Sanaa Mansoor, Minkyung Baek, Hahnbeom Park, Gyu~Rie Lee, and David Baker.
\newblock Protein ensemble generation through variational autoencoder latent space sampling.
\newblock \emph{Journal of Chemical Theory and Computation}, 20\penalty0 (7):\penalty0 2689--2695, 2024.

\bibitem[Newport et~al.(2019)Newport, Sansom, and Stansfeld]{newport2019memprotmd}
Thomas~D Newport, Mark S~P Sansom, and Phillip~J Stansfeld.
\newblock The memprotmd database: a resource for membrane-embedded protein structures and their lipid interactions.
\newblock \emph{Nucleic acids research}, 47\penalty0 (D1):\penalty0 D390--D397, 2019.

\bibitem[Peebles \& Xie(2023)Peebles and Xie]{peebles2023scalable}
William Peebles and Saining Xie.
\newblock Scalable diffusion models with transformers.
\newblock In \emph{Proceedings of the IEEE/CVF international conference on computer vision}, pp.\  4195--4205, 2023.

\bibitem[Rodr{\'\i}guez-Espigares et~al.(2020)Rodr{\'\i}guez-Espigares, Torrens-Fontanals, Tiemann, Aranda-Garc{\'\i}a, Ram{\'\i}rez-Anguita, Stepniewski, Worp, Varela-Rial, Morales-Pastor, Medel-Lacruz, et~al.]{rodriguez2020gpcrmd}
Ismael Rodr{\'\i}guez-Espigares, Mariona Torrens-Fontanals, Johanna~KS Tiemann, David Aranda-Garc{\'\i}a, Juan~Manuel Ram{\'\i}rez-Anguita, Tomasz~Maciej Stepniewski, Nathalie Worp, Alejandro Varela-Rial, Adri{\'a}n Morales-Pastor, Brian Medel-Lacruz, et~al.
\newblock Gpcrmd uncovers the dynamics of the 3d-gpcrome.
\newblock \emph{Nature Methods}, 17\penalty0 (8):\penalty0 777--787, 2020.

\bibitem[Roessner et~al.(2025)Roessner, Floquet, and Louet]{roessner2025unveiling}
Rita~A Roessner, Nicolas Floquet, and Maxime Louet.
\newblock Unveiling g-protein-coupled receptor conformational dynamics via metadynamics simulations and markov state models.
\newblock \emph{Journal of Chemical Information and Modeling}, 65\penalty0 (9):\penalty0 4630--4642, 2025.

\bibitem[Samaddar et~al.(2025)Samaddar, Sun, Nilsson, and Madireddy]{samaddar2025efficient}
Anirban Samaddar, Yixuan Sun, Viktor Nilsson, and Sandeep Madireddy.
\newblock Efficient flow matching using latent variables.
\newblock \emph{arXiv preprint arXiv:2505.04486}, 2025.

\bibitem[Sengar et~al.(2025{\natexlab{a}})Sengar, Hariri, Probst, Barth, and Vandergheynst]{sengar2025generative}
Aditya Sengar, Ali Hariri, Daniel Probst, Patrick Barth, and Pierre Vandergheynst.
\newblock Generative modeling of full-atom protein conformations using latent diffusion on graph embeddings.
\newblock \emph{arXiv preprint arXiv:2506.17064}, 2025{\natexlab{a}}.

\bibitem[Sengar et~al.(2025{\natexlab{b}})Sengar, Zhang, Vandergheynst, and Barth]{sengar2025beyond}
Aditya Sengar, Jiying Zhang, Pierre Vandergheynst, and Patrick Barth.
\newblock Beyond ensembles: Simulating all-atom protein dynamics in a learned latent space.
\newblock In \emph{The Fourteenth International Conference on Learning Representations}, 2025{\natexlab{b}}.
\newblock URL \url{https://openreview.net/forum?id=AwowReRWXI}.

\bibitem[Shen et~al.(2025)Shen, Wang, Yuan, Wang, Yang, and Gu]{shen2025simultaneous}
Yuning Shen, Lihao Wang, Huizhuo Yuan, Yan Wang, Bangji Yang, and Quanquan Gu.
\newblock Simultaneous modeling of protein conformation and dynamics via autoregression.
\newblock \emph{arXiv preprint arXiv:2505.17478}, 2025.

\bibitem[Siebenmorgen et~al.(2024)Siebenmorgen, Menezes, Benassou, Merdivan, Didi, Mour{\~a}o, Kitel, Li{\`o}, Kesselheim, Piraud, et~al.]{siebenmorgen2024misato}
Till Siebenmorgen, Filipe Menezes, Sabrina Benassou, Erinc Merdivan, Kieran Didi, Andr{\'e} Santos~Dias Mour{\~a}o, Rados{\l}aw Kitel, Pietro Li{\`o}, Stefan Kesselheim, Marie Piraud, et~al.
\newblock Misato: machine learning dataset of protein--ligand complexes for structure-based drug discovery.
\newblock \emph{Nature Computational Science}, pp.\  1--12, 2024.

\bibitem[Stark et~al.(2023)Stark, Jing, Barzilay, and Jaakkola]{stark2023harmonic}
Hannes Stark, Bowen Jing, Regina Barzilay, and Tommi Jaakkola.
\newblock Harmonic prior self-conditioned flow matching for multi-ligand docking and binding site design.
\newblock In \emph{NeurIPS 2023 AI for Science Workshop}, 2023.
\newblock URL \url{https://openreview.net/forum?id=3WF88uMjGz}.

\bibitem[Torrens-Fontanals et~al.(2020)Torrens-Fontanals, Stepniewski, Aranda-Garc{\'\i}a, Morales-Pastor, Medel-Lacruz, and Selent]{torrens2020molecular}
Mariona Torrens-Fontanals, Tomasz~Maciej Stepniewski, David Aranda-Garc{\'\i}a, Adri{\'a}n Morales-Pastor, Brian Medel-Lacruz, and Jana Selent.
\newblock How do molecular dynamics data complement static structural data of gpcrs.
\newblock \emph{International journal of molecular sciences}, 21\penalty0 (16):\penalty0 5933, 2020.

\bibitem[Verlet(1967)]{verlet1967}
Loup Verlet.
\newblock Computer "experiments" on classical fluids. i. thermodynamical properties of lennard-jones molecules.
\newblock \emph{Physical Review}, 159\penalty0 (1):\penalty0 98--103, 1967.

\bibitem[Wang et~al.(2026)Wang, Xu, Li, Yu, Tan, Li, Huang, Chen, Wu, Wu, et~al.]{wang2026learning}
Yusong Wang, Youjun Xu, Wentao Li, Haoyu Yu, Wenjuan Tan, Shaoning Li, Qiaojing Huang, Nanjun Chen, Xuan Wu, Qilong Wu, et~al.
\newblock Learning the all-atom equilibrium distribution of biomolecular interactions at scale.
\newblock \emph{bioRxiv}, pp.\  2026--03, 2026.

\bibitem[Wu \& Li(2023)Wu and Li]{wu2023diffmd}
Fang Wu and Stan~Z Li.
\newblock Diffmd: A geometric diffusion model for molecular dynamics simulations.
\newblock In \emph{Proceedings of the AAAI Conference on Artificial Intelligence}, volume~37, pp.\  5321--5329, 2023.

\bibitem[Xu et~al.(2023)Xu, Powers, Dror, Ermon, and Leskovec]{xu2023geometric}
Minkai Xu, Alexander~S Powers, Ron~O Dror, Stefano Ermon, and Jure Leskovec.
\newblock Geometric latent diffusion models for 3d molecule generation.
\newblock In \emph{International Conference on Machine Learning}, pp.\  38592--38610. PMLR, 2023.

\bibitem[y~Schnitzler et~al.(2008)y~Schnitzler, Storch, Meibers, Nurwakagari, Breit, Essin, Gollasch, and Gudermann]{y2008gq}
Michael~Mederos y~Schnitzler, Ursula Storch, Simone Meibers, Pascal Nurwakagari, Andreas Breit, Kirill Essin, Maik Gollasch, and Thomas Gudermann.
\newblock Gq-coupled receptors as mechanosensors mediating myogenic vasoconstriction.
\newblock \emph{The EMBO journal}, 27\penalty0 (23):\penalty0 3092, 2008.

\bibitem[Zhang et~al.(2024)Zhang, Chen, Lu, Lan, Chen, and Lu]{zhang2024g}
Mingyang Zhang, Ting Chen, Xun Lu, Xiaobing Lan, Ziqiang Chen, and Shaoyong Lu.
\newblock G protein-coupled receptors (gpcrs): advances in structures, mechanisms and drug discovery.
\newblock \emph{Signal transduction and targeted therapy}, 9\penalty0 (1):\penalty0 88, 2024.

\bibitem[Zhou et~al.(2024)Zhou, Lopez, and Giovambattista]{zhou2024harmonic}
Yang Zhou, Gustavo~E Lopez, and Nicolas Giovambattista.
\newblock The harmonic and gaussian approximations in the potential energy landscape formalism for quantum liquids.
\newblock \emph{Journal of chemical theory and computation}, 20\penalty0 (5):\penalty0 1847--1861, 2024.

\end{thebibliography}
\bibliographystyle{iclr2027_conference}

\newpage

\onecolumn
\appendix
\begin{center}
\Large
\textbf{Technical Appendices and Supplementary Material}
 \\[20pt]
\end{center}
\etocdepthtag.toc{mtappendix}
\etocsettagdepth{mtchapter}{none}
\etocsettagdepth{mtappendix}{subsection}
\tableofcontents

\begin{figure*}[ht]
    \centering
    \includegraphics[width=0.99950\linewidth]{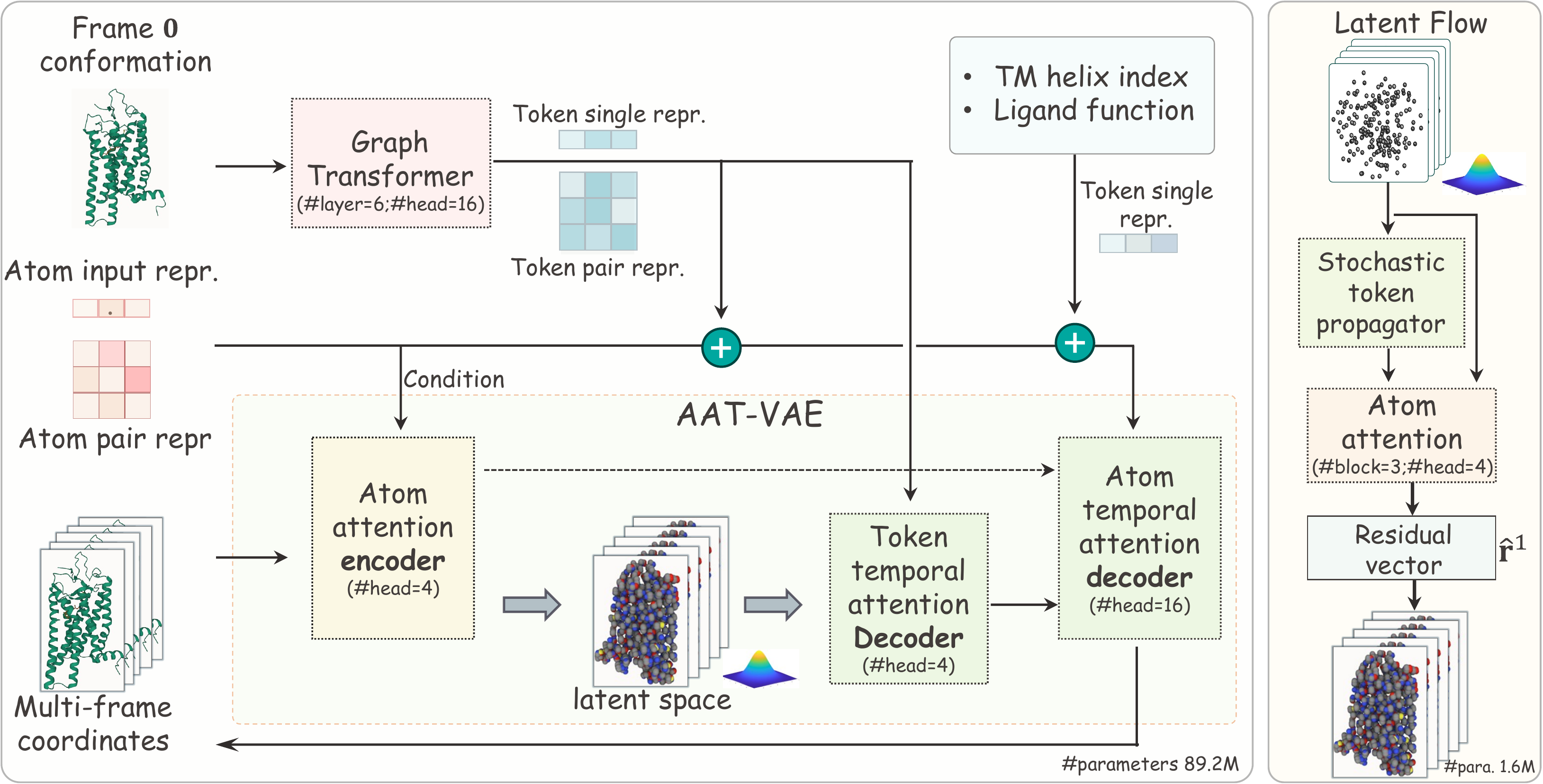}
    \caption{Schematic diagram of the model architecture. \textbf{Left}) The whole framework of \ourM. The atomic structure information will be encoded in the latent space, where the latent coordinate can be decoded to the original Cartesian space. The frame $0$ structure information, sequence/atom type, as well as GPCR Transmembrane helix indices (TM index) and ligand function information are the conditional information involved in both the encoder and decoders. The TM index and ligand function information will be concatenated to the token single representations. 
    The dotted arrow in VAE between the encoder and the decoder indicates information without structure. \textbf{Right}). The residual latent flow will input the Gaussian noise and output the relative latent coordinates. To bridge the information gap for future time steps, we employ a Stochastic Token Propagator. This module actively transfers the structural context from the initial frame tokens ($\mathbf{a}_0$) into the temporal domain by mixing them with time-scaled diffusion noise. The details of the velocity network can be seen in ~\autoref{alg:latent_velocity_network}.
    }
    \label{fig:architecture_detail}
\end{figure*}

\section{Data Pipeline and Featurization}

To specialize our model for GPCR-ligand interactions, we introduce domain-specific features alongside standard protein featurization.

\paragraph{GPCR Transmembrane (TM) Topology.} To explicitly encode the receptor's topological structure, we assign a Transmembrane Helix Index to each residue. 

\begin{itemize}
\item \textbf{Categorization:} Residues belonging to the transmembrane helices are labeled as \texttt{TM1} through \texttt{TM7}. Residues in loop regions or termini (non-TM) and ligand atoms are assigned distinct Non-TM and Ligand tokens, respectively. 
\item \textbf{Encoding:} These discrete categorical indices are projected into continuous feature vectors via a learnable lookup table (implemented as an \texttt{Embedding} layer), which is optimized jointly with the model. \end{itemize}

\paragraph{Ligand Pharmacological Classification.} We incorporate the functional role of the ligand as a global conditioning signal. To handle diverse annotation terminology, we map raw ligand descriptions into four primary pharmacological classes: Agonist, Antagonist, Inverse Agonist, and Unknown. The detailed mapping strategy is as follows: \begin{itemize} 
\item \textbf{Agonist:} Includes Agonist, Partial Agonist, Allosteric Agonist, and PAM (Positive Allosteric Modulator).

\item \textbf{Antagonist:} Includes Antagonist, NAM (Negative Allosteric Modulator), and Allosteric Antagonist.

\item \textbf{Inverse Agonist:} Includes Inverse Agonist. 

\item \textbf{Unknown:} Used for ligands with unclassified or ambiguous functions. 
\end{itemize} 

Similar to the structural features, these functional categories are encoded via learnable embeddings and concatenated to the global token representation.

\paragraph{General Protein Features.} Beyond domain-specific attributes, we adopt the standard featurization pipeline from AlphaFold3~\cite{abramson2024accurate}. This includes amino acid identity, relative token indices, and chain identifiers (asym\_id), ensuring the model captures fundamental protein geometry and sequence context.

\section{More Experimental Details}

\subsection{Hyper-parameters.}
The main hyper-parameters of \ourM are shown in Table \ref{tab:hparam_training},\ref{tab:flow_matchinghyperparameter},\ref{tab:VAE_hyperparameter}.

In particular, the parameter $\sigma$ in the AAT Prior $p(\rvz_i)$ in Eq. \ref{eq:vae_ac_elbo} physically controls the width of the local harmonic energy well. In practice, its optimal value scales with the dataset's conformational diversity: smaller values are suited for highly stable unbiased MD, while larger values should be used for enhanced sampling or highly flexible systems. For our benchmark, we empirically found $\sigma=16$ strikes an optimal balance.

\begin{table*}[hbt!]

    \centering
    \caption{Training hyperparameters}
    \setlength{\tabcolsep}{16mm
    }
    \begin{tabular}{ll}
    \toprule[1.2pt]
    
    \textbf{Hyperparameters} & \textbf{Values}\\ 
     \midrule[0.5pt]
    Batch Size  &  2 \\
    Frames Num & 50 \\ 
    EMA decay & 0.999 \\
    Temporal stride & 10 \\
    Learning Rate   & 1$\times 10^{-4}$  \\
    Optimizer  &  Adam (weight decay = 0.)\\
    \bottomrule[1.2pt]
    \end{tabular}
     \label{tab:hparam_training}
\end{table*}

\begin{table*}[hbt!]
     
    \centering
    \caption{Flow matching hyperparameters}
    \setlength{\tabcolsep}{16mm
    }
    \begin{tabular}{lc}
    \toprule[1.2pt]
    
    \textbf{Hyperparameters} & \textbf{Values}\\ 
     \midrule[0.5pt]
    Sampling step in Flow &  10 \\
    Velocity network blocks Num & 3 \\ 
    Velocity network head Num & 4 \\
    \bottomrule[1.2pt]
    \end{tabular}
\label{tab:flow_matchinghyperparameter}
\end{table*}

\begin{table*}[hbt!]
    \centering
    \caption{AAT-VAE hyperparameters}
    \setlength{\tabcolsep}{16mm
    }
    \begin{tabular}{lc}
    \toprule[1.2pt]
    
    \textbf{Hyperparameters} & \textbf{Values}\\ 
     \midrule[0.5pt]
    Dim of TM index embeddings & 8 \\
    Dim of ligand function embeddings  &  8 \\
  Atom attn encoder blocks Num & 3 \\ 
  Atom attn encoder head Num & 4 \\ 
    Token temporal attn decoder blocks Num & 3 \\ 
    Token temporal attn decoder head Num & 4 \\
   Atom temporal attn decoder blocks Num & 6 \\  
   Atom temporal attn decoder head Num & 16 \\ 
   $\sigma_p^2$ in  AAT Prior $p(\rvz_i)$ in Eq. \ref{eq:vae_ac_elbo} & 16 \\
        Latent dim in VAE & 3 \\
    \bottomrule[1.2pt]
    \end{tabular}
    \label{tab:VAE_hyperparameter}
\end{table*}

\subsection{Data Curation}
The training, validation, and test datasets are summarized in Table~\ref{tab:data_summary}. All training and validation samples, as well as a subset of 19 test samples, were collected from the public GPCRMD dataset contributed by the \textit{GPCRMD community}. For the remaining 96 test samples, we adopted the same MD configurations (e.g., temperature and pressure) as GPCRMD~\cite{rodriguez2020gpcrmd} and generated a 500 ns (2,500 frames) trajectory for each system (protocol details refers to sec. \ref{appsec:simulation_protocal}). Generating these test samples required approximately 5,000 GPU hours in total, highlighting that the production of a single traditional 500 ns all-atom MD trajectory takes over 50 hours. Notably, these 96 newly generated test samples share $<30\%$ sequence similarity and feature distinct ligand scaffolds compared to the training set.

\begin{table*}[hbt!]
    \centering
    \caption{Data summary across Train, Validation, and Test sets. The test set contains the samples with < 50\% sequence similarity compared to the training set. Also,  over 95\% of the test samples feature ligand scaffolds distinct from those in the training set.}
    \setlength{\tabcolsep}{6mm}    
    \begin{tabular}{l |r r r}
    \toprule[1.2pt]
    \textbf{Statistical variable} & \textbf{Train} & \textbf{Val} & \textbf{Test} \\ 
    \midrule[0.5pt]
    \# Traj & 531 & 57 & 115 \\
    \# Simulation time for each traj & 500 ns & 500 ns & 500 ns \\
    \# Frames Num & 1,327,500 & 142,500 & 287,500 \\ 
    \# PDB id & 177 & 19 & 115 \\
    Receptor sequence length range & 276 $\sim$ 501 & 259 $\sim$ 327 & 253 $\sim$ 372 \\
    Averaged receptor Sequence length & 317.9 & 304.4 & 293.87 \\
    \bottomrule[1.2pt]
    \end{tabular}
    \label{tab:data_summary}
\end{table*}

\subsection{Baselines.}

Since GPCRLMD is specifically designed for GPCR--ligand complex dynamics, we compare it against state-of-the-art methods from both protein--ligand and protein-only molecular dynamics modeling. For protein--ligand systems, we select \textbf{AnewSampling}~\cite{wang2026learning}, a recent state-of-the-art protein--ligand ensemble generation method, and \textbf{BioMD}~\cite{feng2025biomd}, a generative model for all-atom protein--ligand trajectory generation.

We further include several representative protein-only dynamics models, including \textbf{BioEmu}~\cite{lewis2025scalable}, \textbf{Str2Str}~\cite{str2str}, \textbf{MDGen}~\cite{jing2024generative}, and \textbf{AlphaFlow}~\cite{jingAlphaFoldMeetsFlow2024}. These methods are designed for protein monomer conformational ensemble or dynamics generation and do not explicitly model bound ligands. For these protein-only baselines, we use the publicly released checkpoints provided in their official GitHub repositories and perform inference directly on the receptor structures in our test set, with the ligand excluded.

For \textbf{BioMD}, we retrain the model from scratch on our GPCR--ligand training set and evaluate it on the same test set as \ourM. For \textbf{AnewSampling}, we perform evaluation using its official API. For all protein-only baselines, including BioEmu, Str2Str, MDGen, and AlphaFlow, we use their released pretrained checkpoints without retraining.

The experimental results are shown in Table ~\ref{tab:gpc_emsemble_bioemu}.

We omit comparisons to recent GPCR-related generative models~\cite{sengar2025generative, sengar2025beyond} as their architectures are typically restricted to single-system modeling. These methods lack the generalizability to handle unseen complexes without retraining.

\subsection{Training Details.}
We train \ourM in two separate stages. In the first stage, we optimize only the VAE (encoder and decoder, etc, except for the latent flow velocity network) using the KL-divergence loss together with the auxiliary losses. In the second stage, we freeze the VAE parameters and optimize only the latent velocity network for latent flow matching.

The total number of parameters in \ourM is 90.8M, of which the latent flow network accounts for 1.6M. We use the Adam optimizer to train the model on 16 H100 GPUs. Training the VAE takes approximately 6 days for 80,000 steps. For latent flow matching, training takes approximately 2 days for 50,000 steps.

For the baseline BioMD, the total number of parameters is 89.2M, and the velocity network required for sampling occupies 78M parameters, meaning that it requires more computation time to generate the final trajectory compared with \ourM.

\paragraph{Long-Trajectory Generation Paradigm.}

\begin{figure*}[ht]
    \centering
    \includegraphics[width=0.7\linewidth]{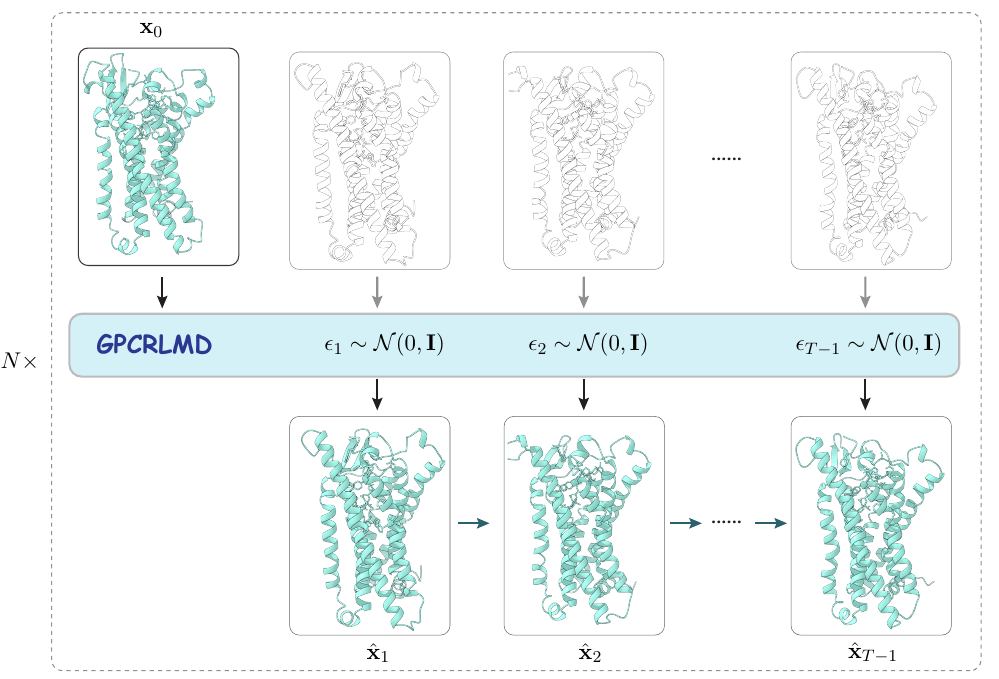}
    \caption{Schematic diagram of the trajectory generative process. The proposed model, \ourM, utilizes an initial frame to generate the subsequent $T-1$ frames in parallel. This process is repeated $N$ times to obtain $N \times (T - 1)$ samples, where the final frame of the previous iteration serves as the initial frame for the next.}
    \label{fig:generative_process}
\end{figure*}

As illustrated in Figure \ref{fig:generative_process}, once the initial batch of frames is generated from the starting frame $\x_0$, we can leverage \ourM to extend the trajectory. In this process, the final frame of the current iteration serves as the initial frame for the subsequent iteration, allowing for the continuous generation of long-trajectory simulations.

\subsection{Simulation Protocol}~\label{appsec:simulation_protocal}

    The all-atom molecular dynamics (MD) simulations for the 96 GPCR-ligand test samples are conducted using a dedicated Python-based \href{https://github.com/GPCRmd/simulation_pipeline/}{simulation pipeline}. The systems are parameterized with the \textbf{CHARMM36} force field (July 2020 version). Following initial solvation and neutralization, the relaxation of each system began with an energy minimization of 5,000 steps.

Following minimization, the systems undergo a 40~ns equilibration phase in the $NPT$ (constant pressure) ensemble at a physiological temperature of 310~K and a pressure of 1~atm. The temperature is controlled using a \textbf{Langevin thermostat} with a damping coefficient of 1~$\text{ps}^{-1}$, while the pressure is maintained using a \textbf{Monte Carlo barostat}. The equilibration utilizes a 2~fs time step, with trajectory frames save every 0.05~ns (25,000 integration steps). 

During this phase, flat-bottom harmonic position restraints with a width of 2~\AA\ were applied along the three axes to the protein, co-solved ligand molecules, HOMOLWAT, and co-solved waters. The restraint strength followed a specific schedule:
\begin{itemize}
    \item \textbf{0--20 ns:} Constant at 1~$\text{kcal/mol/\AA}^2$.
    \item \textbf{20--30 ns:} Linearly tapered from 1 to 0~$\text{kcal/mol/\AA}^2$.
    \item \textbf{30--40 ns:} Completely removed (0~$\text{kcal/mol/\AA}^2$) to allow for final system relaxation.
\end{itemize}

Upon successful equilibration, production simulations were conducted in the $NVT$ (constant volume) ensemble for 500~ns per sample. For the production runs, the integration time step was increased to 4~fs, and the Langevin thermostat damping coefficient was reduced to 0.1~$\text{ps}^{-1}$. Production trajectories were saved every 0.2~ns (50,000 integration steps).

Finally, the resulting trajectories were wrapped and aligned around the receptor using the \texttt{MDAnalysis} module (version 2.1.0) to ensure data consistency for downstream structural analyses.

\section{Theoretical Motivation of AAT-VAE}

The atom-anchored KL regularization induces a natural reference embedding from the Cartesian conformational space into a space of probability distributions. Specifically, for a conformation
$x\in\mathbb{R}^{3N}$, we define
\[
F:x\mapsto P_x,
\qquad
P_x=\mathcal{N}(x,\sigma_p^2 I),
\]
where $\sigma_p^2$ is fixed across conformations. The learned encoder produces a posterior distribution
\[
Q_x=q_\theta(z\mid x),
\]
which is regularized toward the geometry-anchored reference distribution $P_x$ through the KL term.

A key property of this construction is that the reference map $F$ is exactly isometric under the $2$-Wasserstein metric. For any two conformations $x$ and $y$ represented in the same Cartesian frame,
\[
W_2^2(P_x,P_y)
=
\|x-y\|_2^2,
\]
because $P_x$ and $P_y$ share the same isotropic covariance $\sigma_p^2 I$. Therefore,
\[
W_2(P_x,P_y)=\|x-y\|_2,
\]
meaning that pairwise Euclidean distances in Cartesian space are preserved exactly in the corresponding Gaussian distribution space.

The learned posteriors $Q_x$ are not guaranteed to satisfy this isometry exactly. However, the KL regularization encourages them to remain close to their corresponding reference distributions. In particular, since $P_x$ is Gaussian, the transport inequality gives
\[
W_2^2(Q_x,P_x)
\leq
2\sigma_p^2
D_{\mathrm{KL}}(Q_x\|P_x).
\]
Thus, minimizing the KL divergence also constrains the Wasserstein displacement of the learned posterior from its geometry-anchored reference.

Combining this bound with the triangle inequality yields
\[
\left|
W_2(Q_x,Q_y)-\|x-y\|_2
\right|
\leq
W_2(Q_x,P_x)+W_2(Q_y,P_y),
\]
and consequently,
\[
\left|
W_2(Q_x,Q_y)-\|x-y\|_2
\right|
\leq
\sqrt{2\sigma_p^2D_{\mathrm{KL}}(Q_x\|P_x)}
+
\sqrt{2\sigma_p^2D_{\mathrm{KL}}(Q_y\|P_y)}.
\]

Therefore, when the KL regularization is sufficiently small, pairwise distances between learned latent distributions remain close to the corresponding Cartesian distances. This provides a geometric interpretation of the AAT prior: rather than collapsing conformations toward a common isotropic Gaussian, it establishes a geometry-preserving probabilistic reference space and encourages the learned latent representation to remain close to this structure.

\section{Additional Experimental Results}\label{app:additional_exp}

\subsection{Experimental Results on MISATO}~\label{appsec:misato}

To definitively prove that our architecture \ourM is inherently generalizable to universal protein-ligand systems (by simply omitting these domain-specific token embeddings), we train \ourM from scratch on MISATO~\cite{siebenmorgen2024misato}, a massive universal machine learning dataset containing 16,972 simulated protein-ligand complexes (10 ns each).

\textbf{Setting}. Following the BioMD protocol, we trained our model from scratch on the MISATO~\cite{siebenmorgen2024misato} training set and evaluated it on 1,031 test targets (protein length $\leq$ 800, ligand size $\leq$ 100 heavy atoms).

\textbf{Baselines}. (1) BioMD: A concurrent SOTA framework designed for general protein-ligand dynamics. (2) DenoisingLD, GNNMD, NeuralMD-ODE, NeuralMD-SDE, VerletMD: SOTA models predicting binding dynamics, proposed and implemented by ~\citet{liu2023group}.

\textbf{Results}. As shown in Table~\ref{tab:misato_results}, \ourM significantly outperforms baseline methods on 1,031 test samples across main RMSF metrics. Remarkably, despite being primarily designed for GPCR systems, it surpasses even the concurrent SOTA general protein-ligand MD model, BioMD, in terms of ligand RMSF metrics. These strong empirical results validate that our core physics-informed architecture inherently generalizes well beyond the GPCR domain.

\begin{table*}[ht]\centering
\caption{Results on the MISATO test set (1031 samples). Mean $\pm$ std values are reported. Values closer to those of the reference (in parentheses) indicate better performance.}
\vspace{-2mm}
\label{tab:misato_results}
\small
\setlength{\tabcolsep}{12pt} 
\scalebox{0.908}{
\begin{NiceTabular}{l|cc|cc}
    \CodeBefore
        \rowcolor{blue!5}{9}
    \Body
    \toprule[1.5pt]
    \Block[l]{2-1}{Method} & \Block{1-2}{RMSF Correlation $\uparrow$} & & \Block{1-2}{RMSF Value (closer to Ref.)} & \\
    \cmidrule(lr){2-3} \cmidrule(lr){4-5}
    & Ligand & Protein & Ligand (3.375) & Protein (3.301)* \\
    \midrule
    DenoisingLD~\cite{liu2023group}     & $-0.0290$ & --- & $> 10^{12}$ & --- \\
    GNNMD~\cite{liu2023group}           & $-0.0103$ & --- & 0.2165 & --- \\
    NeuralMD-ODE~\cite{liu2023group}    & 0.3405 & --- & 0.3220 & --- \\
NeuralMD-SDE~\cite{liu2023group}    & 0.3405 & --- & 0.3220 & --- \\
    VerletMD~\cite{liu2023group}        & 0.3356 & --- & 0.3226 & --- \\
    BioMD~\cite{feng2025biomd}               & 0.4861 & 0.5945 & 0.5369 & 0.5177 \\
    \midrule
    \textbf{GPCRLMD (ours)} & \textbf{0.6268 $\pm$ 0.1381} & 0.5831 $\pm$ 0.4228 & \textbf{3.2194 $\pm$ 2.2493} & \textbf{1.2990 $\pm$ 0.35} \\
    \bottomrule[1.5pt]
\end{NiceTabular}
}
\par 
    \parbox{1.0\linewidth}{
        \scriptsize 
        {Note:} \textsuperscript{*} RMSF values for reference trajectories are given in parentheses. Baseline results are copied from the BioMD~\cite{feng2025biomd} paper.
    }
\end{table*}

\subsection{Comparison with BioEmu}\label{appsec:bioemu}

To rigorously assess the quality of the generated receptor ensembles, we benchmark \ourM against BioEmu, a state-of-the-art protein-only generative model. As shown in Table~\ref{tab:gpc_emsemble_bioemu}, \ourM demonstrates superior performance across most kinetic and computational metrics.

\textbf{Capturing Conformational Flexibility.} \ourM exhibits a consistently stronger correlation with ground-truth flexibility profiles. In the \textit{Predicting Flexibility} task, \ourM achieves a high Pairwise RMSD correlation of $\mathbf{0.68}$, outperforming BioEmu ($0.58$). Similarly, our method surpasses BioEmu in reproducing both global ($\mathbf{0.77}$ vs.\ $0.70$) and per-target ($\mathbf{0.82}$ vs.\ $0.76$) RMSF profiles. This indicates that \ourM more accurately models the magnitude and localization of atomic fluctuations during GPCR activation.

\textbf{Distributional Fidelity and Observables.} In terms of distributional accuracy, \ourM achieves the lowest Root Mean $\mathcal{W}_2$ distance ($\mathcal{W}_2 = \mathbf{2.36}$) compared to BioEmu ($3.22$), suggesting a more precise recovery of the equilibrium Boltzmann distribution. This advantage extends to specific biophysical observables; our model excels in recovering weak contacts ($J=\mathbf{0.52}$) and exposed residue patterns ($J=\mathbf{0.54}$), outperforming BioEmu ($0.45$ and $0.47$, respectively). While BioEmu shows a slight edge in capturing high-similarity principal components (\% PC-sim $> 0.5$), \ourM overall better preserves the subtle internal packing and surface geometry required for realistic protein dynamics.

\textbf{Computational Efficiency.} Beyond accuracy, \ourM offers a transformative advantage in efficiency. While BioEmu requires substantial computational overhead ($\approx 1800$ GPU seconds per sample) due to its denoising sampling process, \ourM generates equivalent ensembles in merely $\mathbf{10.0}$ seconds. This represents a $\mathbf{180\times}$ speedup, making \ourM uniquely scalable for high-throughput virtual screening campaigns where rapid ensemble generation is critical.

\textbf{Importance of Ligand-Aware Modeling.} Ultimately, this comparison rigorously validates the necessity of explicit ligand modeling. The substantial performance gap between \ourM and BioEmu across all receptor-specific metrics provides a critical biophysical insight: general protein models that ignore the ligand fail to capture the specific conformational rearrangements, such as induced fit, driven by binding events. By explicitly modeling the entire complex, \ourM successfully recovers the functional ligand-bound receptor ensemble---a task where ``ligand-blind'' baselines fundamentally struggle.

\begin{table*}[ht]
\centering
\caption{
Statistical metrics on the MD ensemble benchmark using the GPCRMD test set
(sequence similarity $<$ 50\%).
We compare Str2str~\cite{str2str}, AlphaFlow\cite{jingAlphaFoldMeetsFlow2024}, MDgen~\cite{jing2024generative}, AnewSampling, BioEmu, BioMD, and \ourM
on receptor conformational ensemble tasks, reporting the median metric across
all test targets.
Each method generates 250 conformations per sample.
To ensure a fair comparison with protein-only models, metrics for all methods
are calculated solely on the receptor, excluding ligand coordinates.
Runtime is reported as GPU seconds required per sample, averaged across all
targets. The best results are highlighted in \textbf{bold}.
}
\label{tab:gpc_emsemble_bioemu}

\small
\setlength{\tabcolsep}{6pt}

\scalebox{0.8}{
\begin{NiceTabular}{cl|ccccccc}
    \CodeBefore
        \columncolor{blue!5}{9}
    \Body

    \toprule[1.5pt]

    \Block[]{2-2}{Metrics / Methods}
    & & \multicolumn{7}{c}{250 frames} \\

    \cmidrule(lr){3-9}

    & & Str2str & AlphaFlow~& MDGen & BioMD & BioEmu & AnewSampling & \ourM \\

    \midrule

    \Block[]{3-1}{Predicting\\flexibility}

    & Pairwise RMSD $r$ $\uparrow$
    & 0.57 & 0.64 & 0.01 & 0.49 & 0.58 & \textbf{0.68} & \textbf{0.68} \\

    & Global RMSF $r$ $\uparrow$
    & 0.59 & 0.75 & 0.59 & 0.72 & 0.70 & 0.76 & \textbf{0.77} \\

    & Per-target RMSF $r$ $\uparrow$
    & 0.62 & 0.79 & 0.70 & 0.78 & 0.76 & 0.78 & \textbf{0.82} \\

    \midrule

    \Block[]{6-1}{Distributional\\accuracy}

    & Root mean $\mathcal{W}_2$-dist. $\downarrow$
    & 3.72 & 3.30 & 2.75 & 2.71 & 3.22 & 3.71 & \textbf{2.36} \\

    & $\hookrightarrow$ Trans. contrib. $\downarrow$
    & 3.54 & 3.05 & 2.58 & 2.40 & 3.04 & 3.61 & \textbf{2.18} \\

    & $\hookrightarrow$ Var. contrib. $\downarrow$
    & 1.26 & 1.12 & 1.00 & 1.39 & 1.00 & \textbf{0.86} & \textbf{0.86} \\

    & MD PCA $\mathcal{W}_2$-dist. $\downarrow$
    & 1.22 & 1.21 & 1.24 & 1.29 & 1.18 & 1.35 & \textbf{1.16} \\

    & Joint PCA $\mathcal{W}_2$-dist. $\downarrow$
    & 3.59 & 2.80 & 2.29 & 2.15 & 2.65 & 3.23 & \textbf{1.91} \\

    & \% PC-sim $>0.5$ $\uparrow$
    & 0.87 & 12.17 & 0.00 & 8.70 & \textbf{17.39} & 13.16 & 15.65 \\

    \midrule

    \Block[]{4-1}{Ensemble\\observables}

    & Weak contacts $J$ $\uparrow$
    & 0.44 & \textbf{0.54} & 0.48 & 0.11 & 0.45 & 0.49 & 0.52 \\

    & Transient contacts $J$ $\uparrow$
    & 0.15 & 0.29 & 0.23 & 0.26 & \textbf{0.30} & 0.29 & \textbf{0.30} \\

    & Exposed residue $J$ $\uparrow$
    & NaN$^{*}$ & \textbf{0.56} & 0.50 & 0.29 & 0.47 & 0.49 & 0.54 \\

    & Exposed MI matrix $\rho$ $\uparrow$
    & NaN$^{*}$ & 0.31 & \textbf{0.34} & 0.17 & 0.27 & 0.31 & 0.27 \\

    \midrule

    \Block[]{1-1}{Runtime}

    & GPU sec. per sample
    & 3600 & 4800 & 60 & 50 & 1800 & -- & \textbf{10.0} \\

    \bottomrule[1.5pt]

\end{NiceTabular}
}

\vspace{2pt}
\begin{minipage}{0.98\textwidth}
\footnotesize
$^{*}$Exposed residue and Exposed MI matrix metrics for str2str are reported as
NaN because str2str generates backbone-only coordinates, making accurate
side-chain-dependent solvent-accessibility calculations infeasible.
\end{minipage}

\end{table*}

\subsection{RMSF Comparison.}
\begin{figure*}
    \centering
    \includegraphics[width=
    1.0\linewidth]{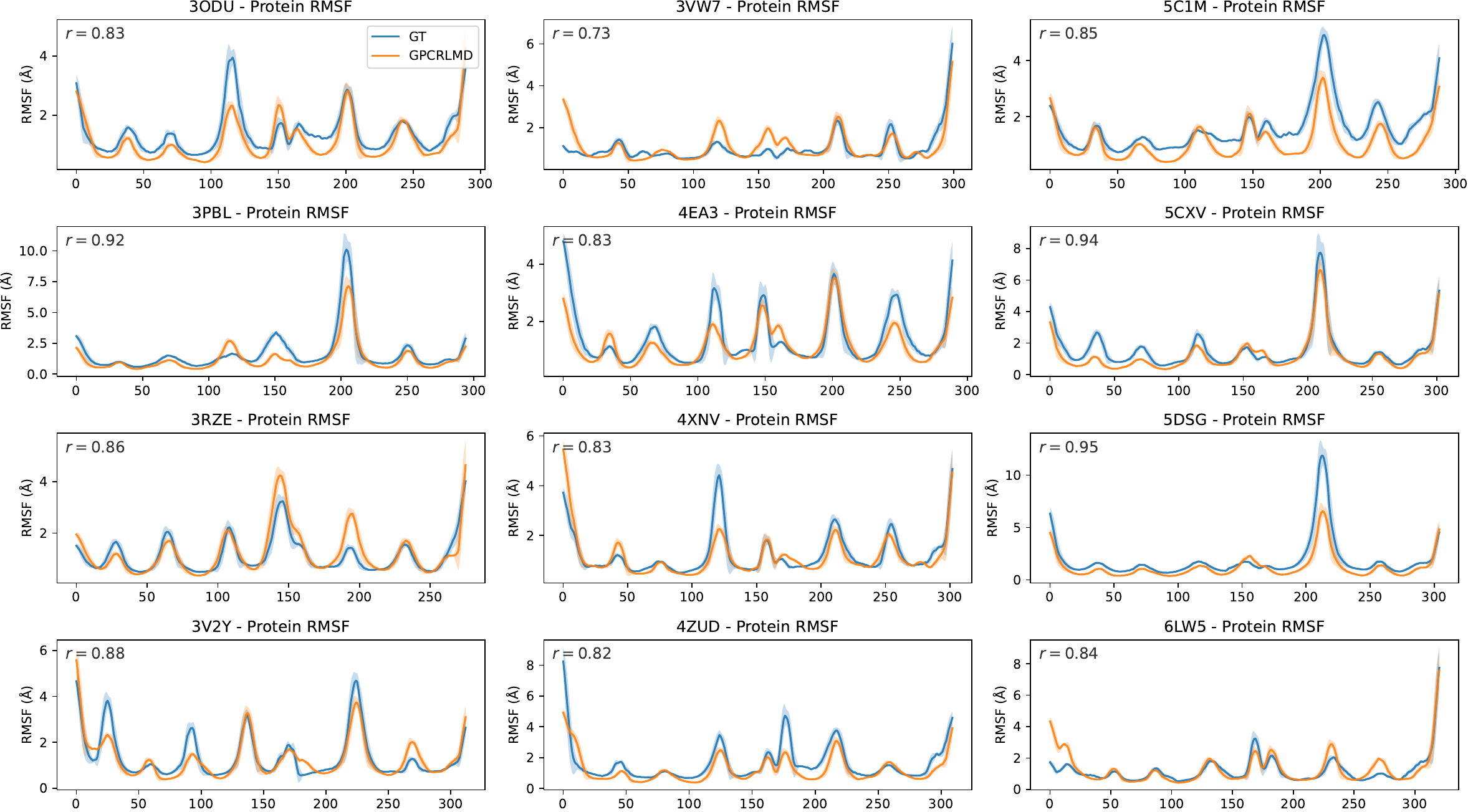}
     \caption{RMSF visualization of GPCR receptor on the test set. 
     This results show that our method achieves high RMSF similarity with the original MD, especially for the receptor. Notably, our method can maintain the low flexibility in the seven helix region while reproducing high diversity on the loop area. 
}
    \label{fig:rmsf_protein}
\end{figure*}

\begin{figure*}
    \centering
    \includegraphics[width=1.0\linewidth]{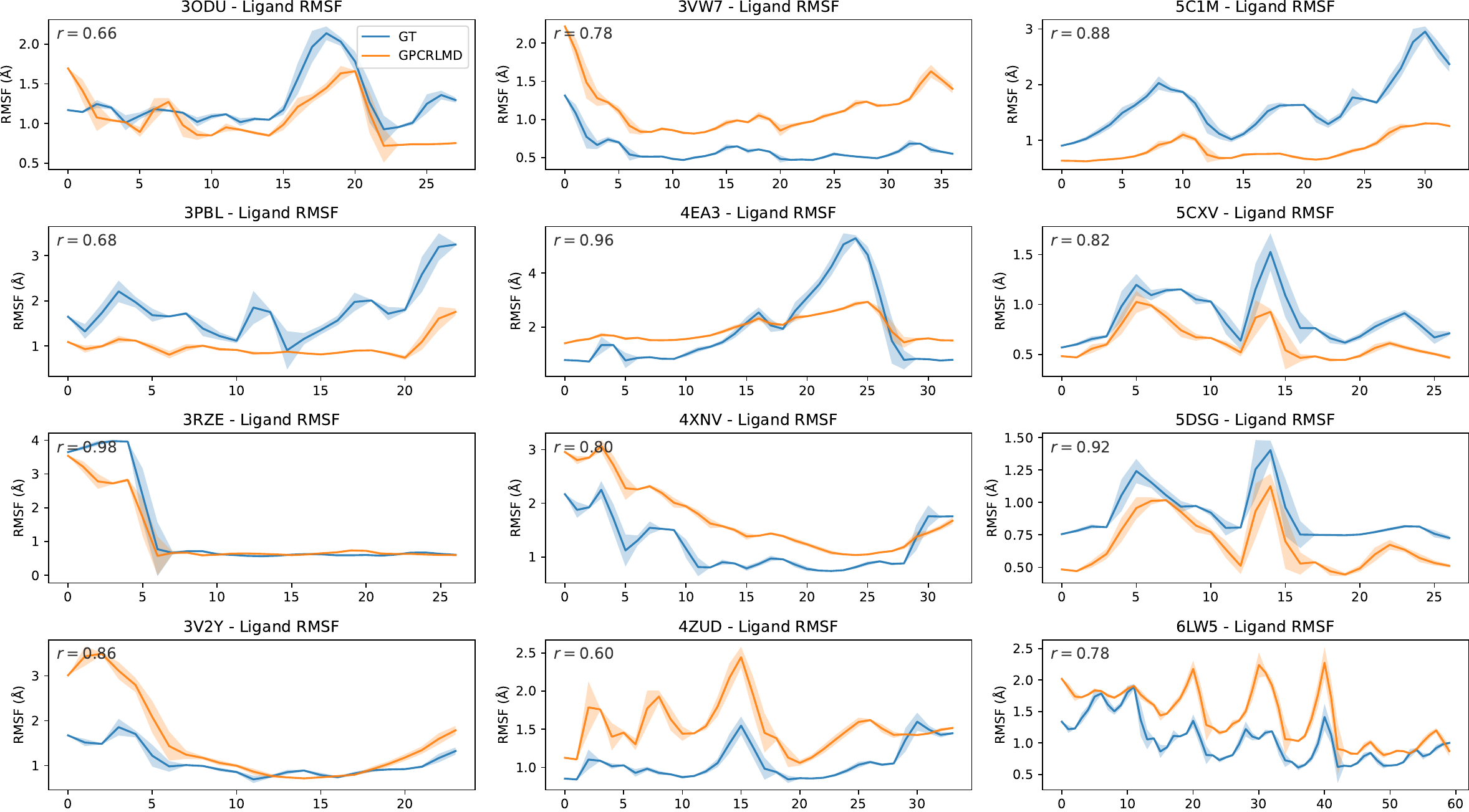}
    \caption{RMSF visualization of ligand on the test set. 
}
    \label{fig:rmsf_ligand}
\end{figure*}

\begin{figure*}
    \centering
    \vspace{-5mm}
    \includegraphics[width=0.75\linewidth]{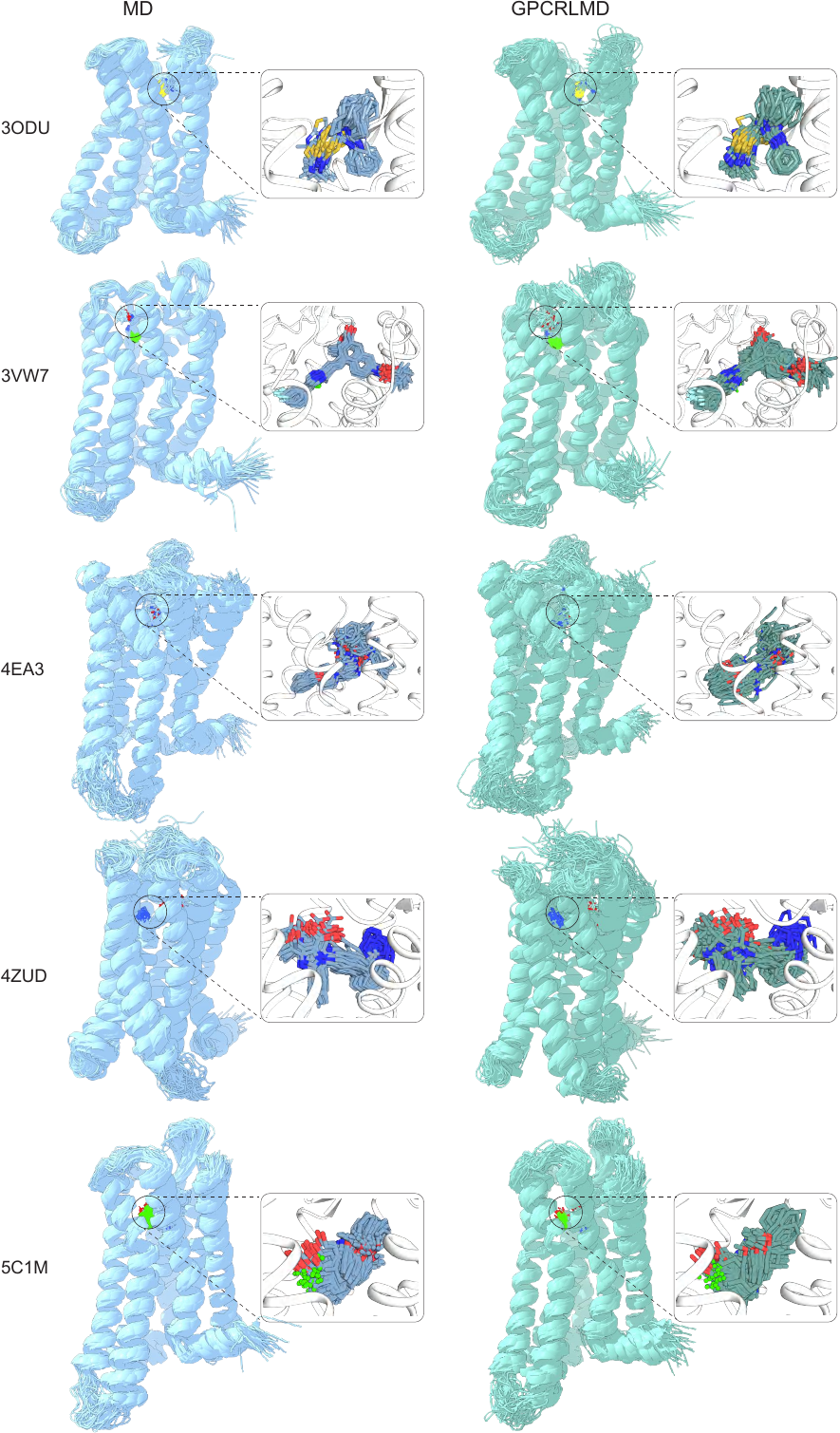}
    \caption{Ensemble visualization. We compare \ourM with the MD reference (100ns).
}
    \label{fig:gpcr_ligand_ensemble}
\end{figure*}

\autoref{fig:rmsf_protein} and \autoref{fig:rmsf_ligand} demonstrate that our method achieves high RMSF similarity with the reference MD, particularly for the receptor. Notably, \ourM accurately captures the distinct dynamical profiles of the GPCR structure: it maintains the requisite structural stability (low flexibility) of the seven transmembrane helices while successfully reproducing the high conformational diversity of the loop regions. Qualitative visualizations of randomly selected cases from the test set are provided in \autoref{fig:gpcr_ligand_ensemble}.

\subsection{Ablation Study}
We conduct an additional ablation study to verify the effectiveness of our designed components for long trajectory generation (500 ns), as shown in \autoref{tab:ablation_study500ns}.

\begin{table*}[ht]\centering
\caption{Ablation study on the test set (115 samples) at \textbf{500 ns}, with protein sequence similarity $ < 50\%$.}
\vspace{-2mm}
\label{tab:ablation_study500ns}
\small
\setlength{\tabcolsep}{6pt} 
\scalebox{0.8}{
\begin{NiceTabular}{cl|cccc}
    \CodeBefore
        \columncolor{blue!5}{3-6}
    \Body
    \toprule[1.5pt]
    \Block[]{2-2}{Metrics / Methods} & & \Block[]{1-4}{500 $ns$} \\
    \cmidrule(lr){3-6}
    & & \ourM  & w/o Time Dep. AAT & w/o AAT Prior (Std. Gauss) & w/o Residual Vector \\
    \midrule
    \Block[]{3-1}{Predicting\\flexibility}
    & Pairwise RMSD $r$ $\uparrow$     & 0.64 & 0.60 & \textbf{0.77} & 0.39 \\
    & Global RMSF $r$ $\uparrow$       & \textbf{0.75} & 0.71 & 0.71 & 0.61 \\
    & Per-target RMSF $r$  $\uparrow$  & \textbf{0.79} & 0.76 & 0.70 & 0.71 \\
    \midrule
    \Block[]{6-1}{Distributional\\accuracy}
    & Root mean $\mathcal{W}_2$-dist. $\downarrow$ & \textbf{2.92} & 3.14 & 6.18 & 6.40 \\
    & $\hookrightarrow$ Trans. contrib. $\downarrow$ & \textbf{2.68} & 2.82 & 5.95 & 6.19 \\
    & $\hookrightarrow$ Var. contrib. $\downarrow$   & \textbf{1.10} & 1.33 & 1.69 & 1.55 \\
    & MD PCA $\mathcal{W}_2$-dist. $\downarrow$    & \textbf{1.10} & {1.10} & 1.40 & 1.36 \\
    & Joint PCA $\mathcal{W}_2$-dist. $\downarrow$ & \textbf{2.41} & 2.52 & 5.87 & 6.10 \\
    & \% PC-sim $>0.5$ $\uparrow$        & \textbf{1.74} & {1.74} & 0.00 & 0.87 \\
    \midrule
    \Block[]{4-1}{Ensemble\\observables}
    & Weak contacts $J$ $\uparrow$       & \textbf{0.48} & {0.48} & 0.36 & 0.08 \\
    & Transient contacts $J$ $\uparrow$ & \textbf{0.22} & 0.19 & 0.03 & 0.07 \\
    & Exposed residue $J$ $\uparrow$    & \textbf{0.53} & 0.51 & 0.37 & 0.14 \\
    & Exposed MI matrix $\rho$ $\uparrow$ & \textbf{0.24} & 0.15 & 0.13 & 0.04 \\
\bottomrule[1.5pt]
\end{NiceTabular}
}
\end{table*}

\begin{table*}[ht]\centering
\caption{Ablation study on the test set (115 samples) at \textbf{500 ns}, with protein sequence similarity $ < 50\%$.}
\vspace{-2mm}
\label{tab:ablation_study500ns}
\small
\setlength{\tabcolsep}{6pt} 
\scalebox{0.8}{
\begin{NiceTabular}{cl|cccc}
    \CodeBefore
        \columncolor{blue!5}{3-6}
    \Body
    \toprule[1.5pt]
    \Block[]{2-2}{Metrics / Methods} & & \Block[]{1-4}{500 $ns$} \\
    \cmidrule(lr){3-6}
    & & \ourM  & w/o Time Dep. AAT & w/o AAT Prior (Std. Gauss) & w/o Residual Vector \\
    \midrule
    \Block[]{3-1}{Predicting\\flexibility}
    & Pairwise RMSD $r$ $\uparrow$     & 0.64 & 0.60 & \textbf{0.77} & 0.39 \\
    & Global RMSF $r$ $\uparrow$       & \textbf{0.75} & 0.71 & 0.71 & 0.61 \\
    & Per-target RMSF $r$  $\uparrow$  & \textbf{0.79} & 0.76 & 0.70 & 0.71 \\
    \midrule
    \Block[]{6-1}{Distributional\\accuracy}
    & Root mean $\mathcal{W}_2$-dist. $\downarrow$ & \textbf{2.92} & 3.14 & 6.18 & 6.40 \\
    & $\hookrightarrow$ Trans. contrib. $\downarrow$ & \textbf{2.68} & 2.82 & 5.95 & 6.19 \\
    & $\hookrightarrow$ Var. contrib. $\downarrow$   & \textbf{1.10} & 1.33 & 1.69 & 1.55 \\
    & MD PCA $\mathcal{W}_2$-dist. $\downarrow$    & \textbf{1.10} & {1.10} & 1.40 & 1.36 \\
    & Joint PCA $\mathcal{W}_2$-dist. $\downarrow$ & \textbf{2.41} & 2.52 & 5.87 & 6.10 \\
    & \% PC-sim $>0.5$ $\uparrow$        & \textbf{1.74} & {1.74} & 0.00 & 0.87 \\
    \midrule
    \Block[]{4-1}{Ensemble\\observables}
    & Weak contacts $J$ $\uparrow$       & \textbf{0.48} & {0.48} & 0.36 & 0.08 \\
    & Transient contacts $J$ $\uparrow$ & \textbf{0.22} & 0.19 & 0.03 & 0.07 \\
    & Exposed residue $J$ $\uparrow$    & \textbf{0.53} & 0.51 & 0.37 & 0.14 \\
    & Exposed MI matrix $\rho$ $\uparrow$ & \textbf{0.24} & 0.15 & 0.13 & 0.04 \\
\bottomrule[1.5pt]
\end{NiceTabular}
}
\end{table*}

\paragraph{Deterministic autoencoder ablation.}

To further distinguish the contribution of the variational formulation
from generic latent-space modeling, we trained the same residual latent
flow on a deterministic autoencoder latent space. Compared with a
deterministic autoencoder, the variational formulation introduces
stochastic latent representations, which encourage the decoder to
reconstruct structures from a neighborhood around each latent state and
provide a robust latent space for subsequent flow matching.
Interestingly, the deterministic autoencoder achieves a lower
reconstruction MSE than AAT-VAE ($0.002$ vs.\ $0.015$), indicating that
the advantage of AAT-VAE does not come from reconstruction accuracy
alone. However, the generated trajectory results in
Table~\ref{tab:deterministic_ae} show that AAT-VAE achieves better
dynamics modeling performance, supporting the benefit of the stochastic
geometry-anchored latent representation.

\begin{table*}[ht]
\centering

\caption{Ablation study comparing the deterministic autoencoder with AAT-VAE at trajectory lengths of \textbf{100 ns} and \textbf{500 ns}.}

\vspace{-2mm}

\label{tab:deterministic_ae}

\small

\setlength{\tabcolsep}{6pt}

\scalebox{0.95}{
\begin{NiceTabular}{cl|cc|cc}
\CodeBefore
    \columncolor{blue!5}{3-6}
\Body
    \toprule[1.5pt]

    \Block[]{2-2}{Metrics / Methods}
    &
    & \Block[]{1-2}{100 ns}
    &
    & \Block[]{1-2}{500 ns}
    & \\

    \cmidrule(lr){3-4}
    \cmidrule(lr){5-6}

    &
    & AAT-VAE
    & Deterministic AE
    & AAT-VAE
    & Deterministic AE \\

    \midrule

    \Block[]{3-1}{Predicting\\flexibility}

    & Pairwise RMSD $r$ $\uparrow$
    & \textbf{0.67}
    & 0.45
    & \textbf{0.64}
    & 0.48 \\

    & Global RMSF $r$ $\uparrow$
    & \textbf{0.76}
    & 0.70
    & \textbf{0.75}
    & 0.67 \\

    & Per-target RMSF $r$ $\uparrow$
    & \textbf{0.81}
    & 0.77
    & \textbf{0.79}
    & 0.72 \\

    \midrule

    \Block[]{4-1}{Distributional\\accuracy}

    & Root mean $\mathcal{W}_2$-dist. $\downarrow$
    & \textbf{2.40}
    & 2.62
    & \textbf{2.92}
    & 3.78 \\

    & $\hookrightarrow$ Trans. contrib. $\downarrow$
    & \textbf{2.24}
    & 2.32
    & \textbf{2.68}
    & 3.36 \\

    & $\hookrightarrow$ Var. contrib. $\downarrow$
    & \textbf{0.87}
    & 1.21
    & \textbf{1.10}
    & 1.74 \\

    & Joint PCA $\mathcal{W}_2$-dist. $\downarrow$
    & \textbf{1.92}
    & 2.01
    & \textbf{2.41}
    & 3.14 \\

    \midrule

    \Block[]{4-1}{Ensemble\\observables}

    & Weak contacts $J$ $\uparrow$
    & \textbf{0.52}
    & 0.34
    & \textbf{0.48}
    & 0.24 \\

    & Transient contacts $J$ $\uparrow$
    & \textbf{0.30}
    & 0.16
    & \textbf{0.22}
    & 0.10 \\

    & Exposed residue $J$ $\uparrow$
    & \textbf{0.54}
    & \textbf{0.54}
    & \textbf{0.53}
    & 0.41 \\

    & Exposed MI matrix $\rho$ $\uparrow$
    & \textbf{0.27}
    & 0.06
    & \textbf{0.24}
    & 0.02 \\

    \bottomrule[1.5pt]
\end{NiceTabular}
}

\end{table*}

\paragraph{Ablation Study for AAT-VAE Module.}
To isolate the contribution of
AAT-VAE, we trained a parameter-matched Cartesian residual-flow baseline
from scratch on the same GPCR benchmark. This baseline uses the same
residual prediction target, conditioning information, training protocol,
and comparable network capacity, but removes AAT-VAE and performs flow
matching directly in Cartesian space. The results are shown in
Table~\ref{tab:aatvae_ablation}. From the table, we observe that the AAT-VAE model improves most metrics, with particularly clear gains
in conformational flexibility, receptor--ligand contacts, and
exposed-residue observables. Notably, the two models achieve similar
coordinate MSE losses during training, suggesting that the improvement
does not simply arise from better pointwise coordinate fitting.
Coordinate MSE may remain low while failing to capture ensemble
variance, long-range correlations, and transient interactions.
AAT-VAE instead provides a geometry-regularized stochastic
representation together with a learned decoder projection, allowing
the flow model to focus on molecular fluctuations without
simultaneously reconstructing and maintaining all Cartesian geometric
constraints.

\begin{table*}[ht]
\centering

\caption{Ablation study on AAT-VAE at trajectory lengths of \textbf{100 ns} and \textbf{500 ns}. The Cartesian residual-flow baseline uses the same residual prediction target, conditioning information, training protocol, and comparable network capacity, but performs flow matching directly in Cartesian space without AAT-VAE.}

\vspace{-2mm}

\label{tab:aatvae_ablation}

\small

\setlength{\tabcolsep}{6pt}

\scalebox{0.95}{
\begin{NiceTabular}{cl|cc|cc}
\CodeBefore
    \columncolor{blue!5}{3-6}
\Body
    \toprule[1.5pt]

    \Block[]{2-2}{Metrics / Methods}
    &
    & \Block[]{1-2}{100 ns}
    &
    & \Block[]{1-2}{500 ns}
    & \\

    \cmidrule(lr){3-4}
    \cmidrule(lr){5-6}

    &
    & GPCRLMD
    & w/o AAT-VAE
    & GPCRLMD
    & w/o AAT-VAE \\

    \midrule

    \Block[]{3-1}{Predicting\\flexibility}

    & Pairwise RMSD $r$ $\uparrow$
    & \textbf{0.67}
    & 0.23
    & \textbf{0.64}
    & 0.58 \\

    & Global RMSF $r$ $\uparrow$
    & \textbf{0.76}
    & 0.63
    & \textbf{0.75}
    & 0.54 \\

    & Per-target RMSF $r$ $\uparrow$
    & \textbf{0.81}
    & 0.72
    & \textbf{0.79}
    & 0.58 \\

    \midrule

    \Block[]{4-1}{Distributional\\accuracy}

    & Root mean $\mathcal{W}_2$-dist. $\downarrow$
    & \textbf{2.40}
    & 2.75
    & \textbf{2.92}
    & 3.02 \\

    & $\hookrightarrow$ Trans. contrib. $\downarrow$
    & \textbf{2.24}
    & 2.36
    & \textbf{2.68}
    & 2.75 \\

    & $\hookrightarrow$ Var. contrib. $\downarrow$
    & \textbf{0.87}
    & 1.40
    & \textbf{1.10}
    & 1.21 \\

    & Joint PCA $\mathcal{W}_2$-dist. $\downarrow$
    & \textbf{1.92}
    & 2.05
    & 2.41
    & \textbf{2.40} \\

    \midrule

    \Block[]{4-1}{Ensemble\\observables}

    & Weak contacts $J$ $\uparrow$
    & \textbf{0.52}
    & 0.27
    & \textbf{0.48}
    & 0.27 \\

    & Transient contacts $J$ $\uparrow$
    & \textbf{0.30}
    & 0.25
    & \textbf{0.22}
    & 0.19 \\

    & Exposed residue $J$ $\uparrow$
    & \textbf{0.54}
    & 0.27
    & \textbf{0.53}
    & 0.31 \\

    & Exposed MI matrix $\rho$ $\uparrow$
    & \textbf{0.27}
    & 0.18
    & \textbf{0.24}
    & 0.23 \\

    \bottomrule[1.5pt]
\end{NiceTabular}
}

\end{table*}

\subsection{Sensitivity Study on the Sigma in AAT Prior}

We performed a sensitivity analysis by varying the prior variance $\sigma_p^2$ in the AAT prior:
\begin{equation}
    p(z_i \mid x_i)
    =
    \mathcal{N}\!\left(x_i,\sigma_p^2 I_3\right),
\end{equation}
with $\sigma_p^2 \in \{16, 8, 1\}\,\text{\AA}^2$.
The results are summarized in Table~\ref{tab:sigma_sensitivity}.

\begin{table*}[ht]
\centering
\caption{Sensitivity study on the prior variance $\sigma_p^2$ in the AAT prior. Results are reported for trajectory lengths of \textbf{100 ns} and \textbf{500 ns}. The default setting uses $\sigma_p^2=16\,\text{\AA}^2$.}

\vspace{-2mm}

\label{tab:sigma_sensitivity}

\small
\setlength{\tabcolsep}{6pt}
\scalebox{0.95}{
\begin{NiceTabular}{cl|ccc|ccc}
    \CodeBefore
        \columncolor{blue!5}{3-8}
    \Body

    \toprule[1.5pt]

    \Block[]{2-2}{Metrics / Settings}
    & 
    & \Block[]{1-3}{100 ns} 
    & 
    & 
    & \Block[]{1-3}{500 ns} 
    & 
    & \\

    \cmidrule(lr){3-5}
    \cmidrule(lr){6-8}

    &
    & $\sigma_p^2=16$
    & $\sigma_p^2=8$
    & $\sigma_p^2=1$
    & $\sigma_p^2=16$
    & $\sigma_p^2=8$
    & $\sigma_p^2=1$ \\

    \midrule

    \Block[]{3-1}{Predicting\\flexibility}

    & Pairwise RMSD $r$ $\uparrow$
    & \textbf{0.67}
    & 0.63
    & 0.10
    & \textbf{0.64}
    & 0.63
    & 0.26 \\

    & Global RMSF $r$ $\uparrow$
    & \textbf{0.76}
    & 0.74
    & 0.58
    & \textbf{0.75}
    & 0.72
    & 0.58 \\

    & Per-target RMSF $r$ $\uparrow$
    & \textbf{0.81}
    & 0.80
    & 0.67
    & \textbf{0.79}
    & 0.78
    & 0.64 \\

    \midrule

    \Block[]{4-1}{Distributional\\accuracy}

    & Root mean $\mathcal{W}_2$-dist. $\downarrow$
    & \textbf{2.40}
    & 2.48
    & 4.35
    & \textbf{2.92}
    & 3.51
    & 6.48 \\

    & $\hookrightarrow$ Trans. contrib. $\downarrow$
    & 2.24
    & \textbf{2.20}
    & 3.08
    & \textbf{2.68}
    & 3.09
    & 5.26 \\

    & $\hookrightarrow$ Var. contrib. $\downarrow$
    & \textbf{0.87}
    & 1.13
    & 3.12
    & \textbf{1.10}
    & 1.64
    & 3.75 \\

    & Joint PCA $\mathcal{W}_2$-dist. $\downarrow$
    & \textbf{1.92}
    & 1.93
    & 2.75
    & \textbf{2.41}
    & 2.89
    & 5.26 \\

    \midrule

    \Block[]{4-1}{Ensemble\\observables}

    & Weak contacts $J$ $\uparrow$
    & 0.52
    & \textbf{0.53}
    & 0.14
    & \textbf{0.48}
    & 0.46
    & 0.15 \\

    & Transient contacts $J$ $\uparrow$
    & \textbf{0.30}
    & 0.23
    & 0.07
    & \textbf{0.22}
    & 0.14
    & 0.04 \\

    & Exposed residue $J$ $\uparrow$
    & 0.54
    & \textbf{0.59}
    & 0.39
    & \textbf{0.53}
    & 0.51
    & 0.38 \\

    & Exposed MI matrix $\rho$ $\uparrow$
    & \textbf{0.27}
    & 0.09
    & 0.01
    & \textbf{0.24}
    & 0.07
    & 0.00 \\

    \bottomrule[1.5pt]

\end{NiceTabular}

}

\end{table*}

The results show that the default setting
$\sigma_p^2=16\,\text{\AA}^2$ achieves the best overall performance
across flexibility prediction, distributional accuracy, and ensemble
observables. A moderate reduction to
$\sigma_p^2=8\,\text{\AA}^2$ maintains comparable performance,
suggesting that \textbf{the method is robust within a reasonable range
of prior scales}.

\subsection{TICA Analysis}
We assess kinetic mode recovery using Time-lagged Independent Component Analysis (TICA). The ground-truth free energy landscape is established using the combined trajectories of MD Replicas 2 and 3. Because certain test targets lack multi-replica sampling, we specifically curated a subset of 17 systems (PDB: 4EA3, 4XNV, 5C1M, 3V2Y, 5CXV, 4XNW, 4ZUD, 3ODU, 3PBL, 3RZE, 5DSG, 4Z36, 4Z34, 4DJH, 3VW7, 6LW5, 5N2S) that possess three independent replicas; these datasets are also publicly available via the GPCRMD database. The TICA model is fitted on these high-resolution trajectories ($1.0\mu s$, 5000 frames, stride $\Delta t=0.2$ns, lag time $\tau=4$ns) to identify the slowest relaxation modes. This pre-calculated landscape serves as the background density in our visualizations.

\begin{figure*}
    \centering
    \includegraphics[width=0.9999\linewidth]{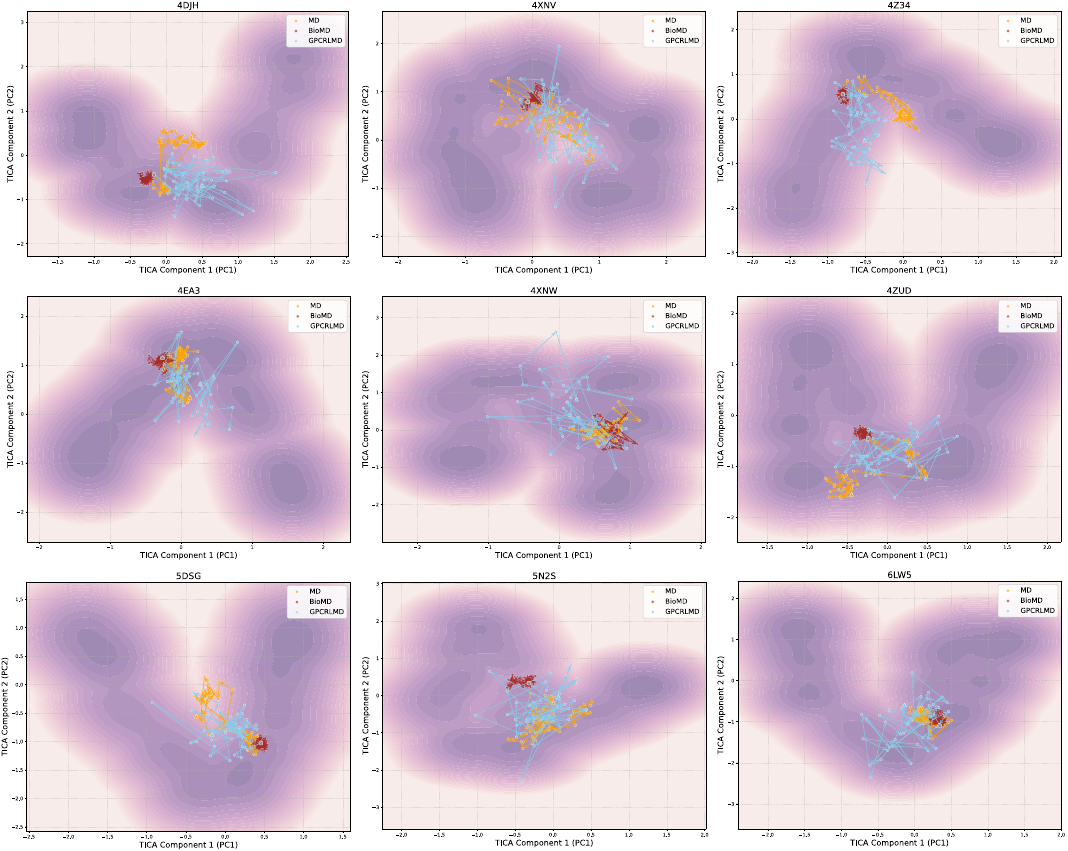}
    \caption{TICA Visualization of the 500ns trajectories. The TICA features only contain the backbone. Baseline MD is a 500ns trajectory, downsampling by 2ns to get 250 frames to fairly align the generative method BioMD and \ourM. The pink
background indicates the density of the ground-truth conformation distribution from the MD references (replica 2 and replica 3, total 1.0 $\mu s$).
\ourM shows high recovery of conformational states in most cases, sampling more diverse conformations. The quantization results (JSD) on all test samples are shown in the \autoref{tab:tica_js_divergence}. }
    \label{fig:tica_bb_500ns}
\end{figure*}

\begin{figure*}
    \centering
    \includegraphics[width=0.9999\linewidth]{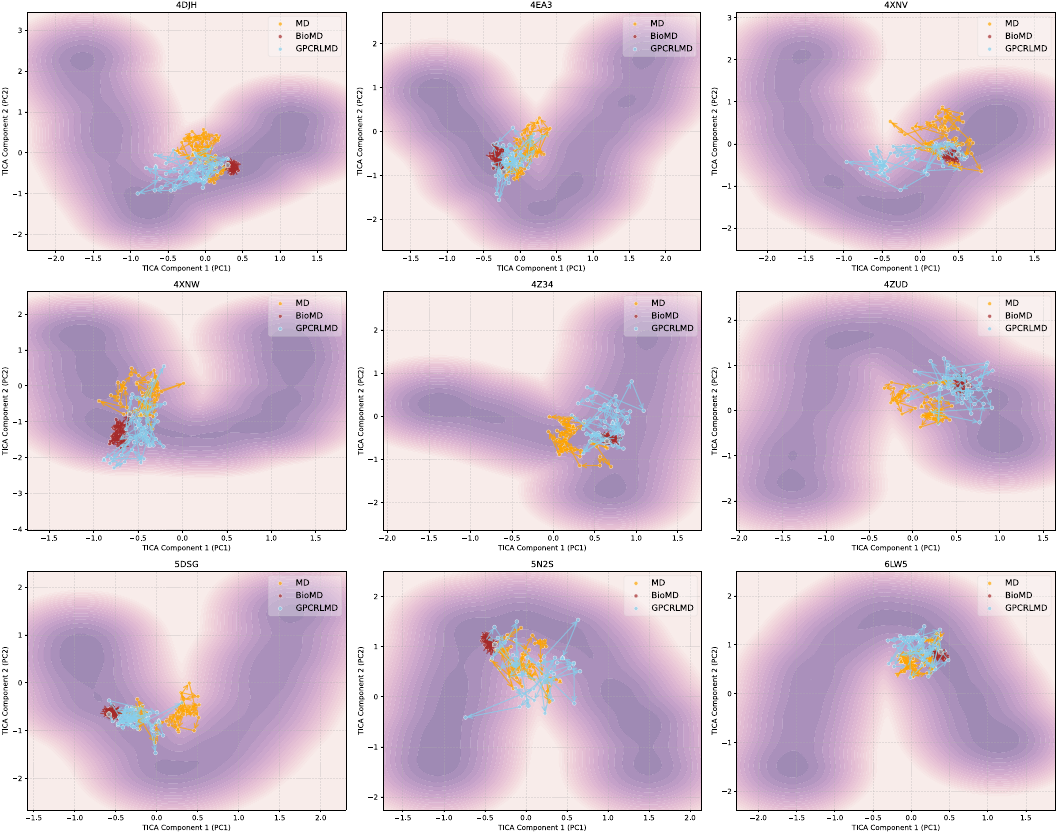}
    \caption{TICA Visualization of the 500ns trajectories. Baseline MD is a 500 ns trajectory, downsampled by 2 ns to obtain 250 frames to fairly align the generative methods BioMD and \ourM. The TICA features include the backbone and side chain. The pink background indicates the density of the ground-truth conformation distribution from the MD references (replica 2 and replica 3, total 1.0 $\mu s$).
\ourM shows high recovery of conformational states in most cases, sampling more diverse conformations. The quantization results (JSD) on all test samples are shown in the \autoref{tab:tica_js_divergence}.}
    \label{fig:tica_sidechain_500ns}
\end{figure*}

For evaluation, we project three distinct sets of trajectories onto this fixed TICA space: (1) the unseen Replica 1 (serving as the MD baseline), (2) BioMD, and (3) \ourM. All projected trajectories are normalized to a 500 ns duration (250 frames, 2 ns stride) for fair comparison (Figure~\ref{fig:tica_bb_500ns},\ref{fig:tica_sidechain_500ns}). The quantitative fidelity is measured using the Jensen-Shannon Divergence (JSD) between the TICA distribution of each generated trajectory and the ground-truth reference distribution (Replicas 2+3).

\textbf{Results}. As shown in \autoref{tab:tica_js_divergence}, \ourM consistently achieves the lowest Jensen-Shannon Divergence (JSD) scores compared to BioMD, outperforming even the held-out MD baseline (Replica 1) in three out of four metrics. This indicates that \ourM does not merely memorize a single trajectory path but effectively samples the broader, biologically plausible conformational ensemble defined by the extensive ground-truth simulations. By covering a more comprehensive region of the valid energy landscape, \ourM demonstrates superior generative diversity and thermodynamic fidelity.

These quantitative findings are corroborated by the TICA projections in Figures~\ref{fig:tica_bb_500ns} and \ref{fig:tica_sidechain_500ns}. Visual inspection reveals that \ourM achieves high conformational state recovery across most test cases. Crucially, the model demonstrates the capacity to sample more diverse conformations than the individual reference trajectory segments. By effectively populating broader low-energy basins, \ourM captures a more comprehensive view of the equilibrium thermodynamic distribution.

\begin{table}[hbt!]
\centering
\caption{Jensen-Shannon divergence (JSD) on TICA projections (500ns simulation). The table reports the divergence between the generated ensembles and the reference distribution on the first TICA component (TICA-0) and the joint first two components (TICA-0,1). Lower values indicate better performance. We report the mean and variance for all samples in the test set.}
\label{tab:tica_js_divergence}
\small
\setlength{\tabcolsep}{10pt}
\scalebox{0.88}{
\begin{NiceTabular}{l|cc|cc}
    \CodeBefore
    \rowcolor{blue!5}{5}
    \Body
    \toprule[1.5pt]
    \RowStyle{\bfseries}
    \Block{2-1}{Method} & \Block{1-2}{Backbone + Sidechain} & & \Block{1-2}{Backbone Only} & \\
    \RowStyle{\bfseries}
    & TICA-0 (JSD $ \downarrow$) & TICA-0,1 (JSD $\downarrow$) & TICA-0 (JSD $\downarrow$) & TICA-0,1 (JSD $\downarrow$) \\
    \midrule
    MD (500ns)  & 0.50 $\pm$ 0.06 & 0.66 $\pm$ 0.02 & 0.44 $\pm$ 0.10 & \textbf{0.62 $\pm$ 0.06} \\
    BioMD           & 0.56 $\pm$ 0.08 & \textbf{0.64 $\pm$ 0.04} & 0.51 $\pm$ 0.10 & 0.63 $\pm$ 0.04 \\
    GPCRLMD (Ours)  & \textbf{0.44 $\pm$ 0.05} & \textbf{0.64 $\pm$ 0.03} &\textbf{ 0.41 $\pm$ 0.10} & {0.63 $\pm$ 0.04} \\
    \bottomrule[1.5pt]
\end{NiceTabular}
}
\end{table}


\subsection{Salt Bridge Analysis}\label{appsec:salt_bridge}
\begin{figure*}
    \centering
    \includegraphics[width=0.995\linewidth]{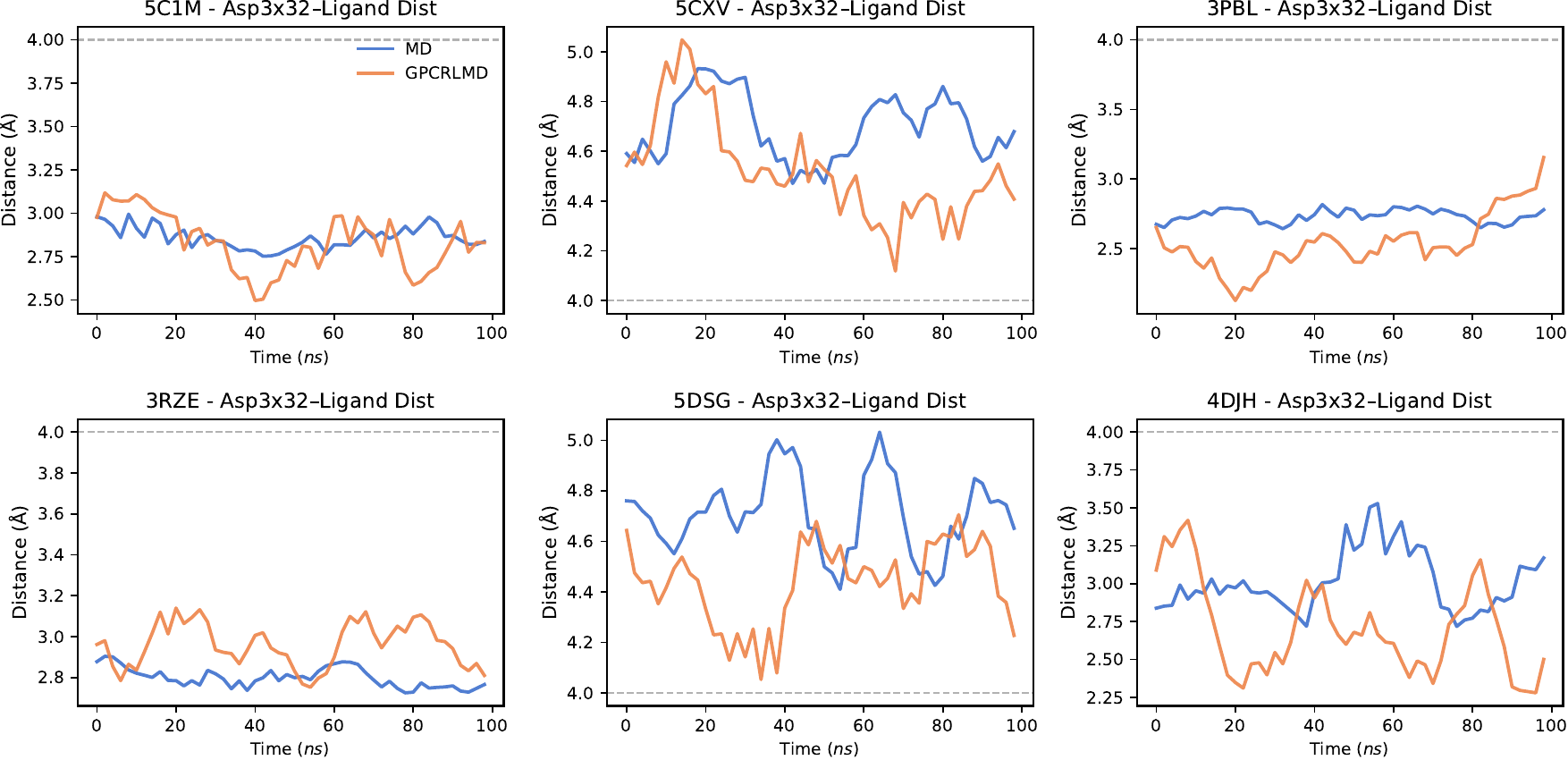}
    \caption{\textbf{Salt bridge stability analysis.} We plot the distance between the carboxylate group of residue Asp$^{3.32}$ and the closest nitrogen atom of the ligand over time. A distance $< 4.0$ \AA\ (dashed grey line) indicates the existence of a salt bridge. The results indicate that \ourM effectively maintains the salt bridge structure only if it is present in the initial conformation; otherwise, the distance fluctuates outside the interaction threshold, consistent with the starting structure.}
    \label{fig:asp332_salt_bridge}
\end{figure*}

Salt bridges are critical electrostatic interactions that anchor the ligand within the binding pocket. We specifically measure the distance between the ligand's cationic nitrogen and the conserved aspartate residue ($D^{3.32}$) on Transmembrane Helix 3~\cite{katritch2013structure,han2023ligand}. Physiologically, the side chain of Asp$^{3.32}$ contains a carboxylate group that is deprotonated (negatively charged), while the ligand's amine group is typically protonated (positively charged). This electrostatic attraction creates a stable salt bridge essential for receptor activation. The results in \autoref{fig:asp332_salt_bridge} demonstrate that conditional on the presence of this salt bridge in the initial conformation ($\mathbf{x}_0$), \ourM effectively preserves the interaction within the canonical range ($< 4$ \AA) throughout the trajectory. Conversely, if the interaction is absent initially (as seen in the 5CXV case), the model faithfully propagates the non-interacting state.

\subsection{Verifying Topology Preservation}
For each trajectory frame in the test set, we computed the pairwise Euclidean distance matrix for all atoms in both Cartesian space and latent spaces. We observe a strong positive Pearson correlation of ~0.80. The slight variance from 1.0 is physically expected, reflecting the localized thermal fluctuations intentionally introduced by the AAT Prior.

\subsection{Posterior Activity of AAT-VAE}\label{appsec:Posterior}
 To quantify posterior activity, we  conduct an active-unit analysis on our 115 test trajectories. 
 of the residual dimensions 

\begin{table}[t]
    \centering
    \caption{Active-unit statistics for AAT-VAE.}
    \label{tab:active_units}
    \begin{NiceTabular}{lcc}
        \toprule
        \textbf{Metric} 
        & \textbf{Threshold $\tau=0.01$} 
        & \textbf{Threshold $\tau=0.05$} \\
        \midrule
        Raw active-unit ratio
        & $0.922 \pm 0.037$
        & $0.435 \pm 0.072$ \\
        
        Residual active-unit ratio
        & $0.806 \pm 0.059$
        & $0.360 \pm 0.050$ \\
        \bottomrule
    \end{NiceTabular}
\end{table}

At the commonly used threshold $\tau=0.01$, $92.2\%$ of the raw latent dimensions and $80.6\%$ of the residual latent dimensions are active. Importantly, after removing the variation directly inherited from the coordinate-dependent prior mean, most latent dimensions remain active. Even under the more conservative threshold $\tau=0.05$, $36.0\%$ of the residual dimensions remain strongly active.

These results indicate that posterior activity is not solely explained by the varying coordinate anchor and that the learned posterior retains substantial frame-dependent information beyond the reference prior mean. We will add this analysis to the revised manuscript.

\paragraph{Metrics Definition}
For each atom-coordinate latent dimension $(i,d)$, we define the raw active-unit ratio as
$$R_{raw}=\frac{1}{3N}\sum_{i=1}^{N}\sum_{d=1}^{3}\mathbf{1}[\frac{Var_t(\mu_{t,i,d})}{\sigma_p}>\tau]$$ where $\mu_{t,i,d}$ is the encoder-predicted posterior mean and $\tau$ is the threshold. Because the prior mean is anchored at the input coordinate $x_{t,i,d}$, we additionally define a residual active-unit ratio: 
$$R_{res}=\frac{1}{3N}\sum_{i=1}^{N}\sum_{d=1}^{3}\mathbf{1}[\frac{Var_t(\mu_{t,i,d}-x_{t,i,d})}{\sigma_p}>\tau].$$ 
Here, $N$ is the number of atoms, $i$ indexes atoms, $d$ indexes the three Cartesian coordinates, $t$ indexes trajectory frames, $\mu_{t,i,d}$ is the encoder-predicted posterior mean, $x_{t,i,d}$ is the corresponding input coordinate and prior mean, $\sigma_p$ is the anchored-prior variance (16 in our experiments), and $\tau$ is the active-unit threshold. The factor $3N$ is therefore the total number of atom-coordinate latent dimensions.

\subsection{Case Study for Long Trajectory Generation}
Our framework naturally extends simulations via continuous iterative sampling. As shown in the figure at \url{https://anonymous.4open.science/r/GPCRLMD/assets/6Z10_1us.png}, extending the trajectory for 6Z10 captures a large-scale transition (TM2-TM6 distance shifting from $17.5$ \AA to $13.8$ \AA).

\subsection{Case Study for Pharmacologically Relevant Conformational Ensembles}

To connect the statistical improvements with pharmacologically relevant dynamics, we analyzed two complexes containing the same receptor but ligands with distinct pharmacological functions. Their time-dependent TM2–TM6 distances are shown in our anonymous repository:
\url{https://anonymous.4open.science/r/GPCRLMD/assets/pharmacological_relevant_traj.png}.
The TM2–TM6 distance is an activation-related conformational coordinate for GPCRs. For the agonist-bound complex, GPCRLMD generates an ensemble with higher occupancy of large-distance, active-like conformations. In contrast, the antagonist-bound complex predominantly remains in small-distance, inactive-like conformations. Interestingly, for 8IYX, which starts from an intermediate conformation, the generated trajectory shifts toward the inactive-like state.

These results suggest that the improvements in flexibility and distributional metrics correspond to better recovery of ligand-dependent conformational preferences rather than generic structural fluctuations alone. We will add this analysis and clarify that it supports the recovery of pharmacologically relevant conformational ensembles.

\section{Model Architecture}

\subsection{Energy-Guided \ourVAE for Long Trajectory Prediction}\label{appsec:Energy_HPVAE}

We extend the \ourVAE to an energy-guided formulation to enhance the physical fidelity of long-term trajectory predictions.\paragraph{Energy-Augmented Posterior.}Standard generative models capture global conformational variability but often violate low-entropy geometric constraints, such as chemically valid side-chain bond lengths and angles. These constraints are naturally expressed as coordinate-space energies rather than through a Gaussian decoder. To address this, we incorporate a physics-based potential $E_{\mathrm{sc}}(\mathbf{x})$ directly into the inference process. We define a corrected conditional distribution (energy-augmented posterior):

\begin{equation}\tilde{p}(\mathbf{x} \mid \mathbf{z}) \propto \mathcal{N}(\mathbf{x}; g(\mathbf{z}), \sigma^2 I) \exp(-\beta E_{\mathrm{sc}}(\mathbf{x})),\end{equation}

where $g(\mathbf{z})$ is the raw decoder output. Finding the most likely structure $\mathbf{x}^*$ under this distribution corresponds to a Maximum A Posteriori (MAP) estimate:

\begin{equation}
\mathbf{x}^* = \arg\min_{\mathbf{x}} \left( \frac{1}{2\sigma^2} |\mathbf{x} - g(\mathbf{z})|^2 + \beta E_{\mathrm{sc}}(\mathbf{x}) \right).
\end{equation}

This formulation acts as a proximal operator, $\operatorname{prox}_{\mu E}(g(\mathbf{z}))$ with $\mu=\sigma^2/\beta$. It effectively applies a likelihood correction that enforces physical plausibility while preserving the learned manifold of the generative model.\paragraph{Test-Time Optimization.}In our specific implementation, we employ a local structural potential $E_{\mathrm{sc}}(\mathbf{x})$ that regularizes side-chain bond stretching and bond angles, using the initial frame $\mathbf{x}_0$ as a reference topology to define the equilibrium geometry. We integrate this minimization directly into the sampling process. After generating the trajectory via flow matching:

\begin{align}
\hat{\mcal{R}}^0 &=[\bm{\epsilon}_1,\bm{\epsilon}_2,\cdots,\bm{\epsilon}_{T-1}]\\
    \hat{\mcal{R}}^{\tau +1}&= \hat{\mcal{R}}^\tau + v_\theta(\hat{\mcal{R}}^{\tau}, \mcal{C}, \tau) d\tau,\; 
\end{align}

we apply $K$ steps of gradient-based energy minimization to the final prediction $\hat{\mathcal{R}}^1$:
\begin{align}
\hat{\mcal{R}}^{1}_{k+1}&= \hat{\mcal{R}}^{1}_{k} - \alpha \nabla E_{\mathrm{sc}}(\hat{\mcal{R}}^{1}_{k}), k=0,2,...,K-1,
\end{align}
where $\alpha$ is the step size. This ensures the final trajectory adheres to local chemical constraints without requiring retraining of the flow network.
In practice, we define the side-chain geometric energy as:$$E_{\mathrm{sc}}(\hat{\mathbf{x}}_t) := \frac{2}{|\mathcal{E}_{\text{bond}}|} \sum_{(i,j) \in \mathcal{E}_{\text{bond}}} (\hat{d}^{t}_{ij}-d^0_{ij})^2 + \frac{1}{|\mathcal{E}_{\text{angle}}|}\sum_{(i,j,k)\in\mathcal{E}_{\text{angle}}} \left| \cos(\hat{\theta}^t_{ijk})-\cos(\theta^0_{ijk}) \right|,$$ where $d^0_{ij}$ and $\hat{d}^t_{ij}$ represent the bond lengths of side-chain bond $(i,j)$ in the initial frame ($t=0$) and the predicted frame $t$, respectively. Similarly, $\theta^0_{ijk}$ and $\hat{\theta}^t_{ijk}$ denote the bond angles defined by side-chain atoms $(i,j,k)$ in frame $0$ and frame $t$, respectively. The cosine terms are calculated following Eq.~\ref{eq:sc_angle_loss}.

\subsection{Weighted Rigid Aligned MSE}\label{appsec:weighedMSE} To resolve the global rotational and translational invariance of the generated structures while emphasizing the structural accuracy of the ligand, we follow AF3~\cite{abramson2024accurate} to employ a weighted Mean Squared Error (MSE) loss with rigid alignment. 
Let $\mathbf{x} = \{\mathbf{x}_i\}_{i=1}^{N}$ and $\hat{\mathbf{x}} = \{\hat{\mathbf{x}}_i\}_{i=1}^{N}$ denote the ground-truth and predicted coordinates for a complex with $N$ atoms. We assign an importance weight $w_i$ to each atom to prioritize the ligand geometry:
\begin{equation}
    w_i = 
    \begin{cases} 
    1 + \lambda_{\text{ligand}} & \text{if atom } i \in \text{Ligand} \\
    1 & \text{if atom } i \in \text{Protein}
    \end{cases}
\end{equation}
where $\lambda_{\text{ligand}}$ is a hyperparameter balancing the contribution of ligand atoms (we use  $\lambda_{\text{ligand}}=10$ in training \ourM).
Before computing the error, we compute the optimal rigid transformation $(R^*, \mathbf{t}^*)$ that aligns the ground truth to the prediction by minimizing the weighted RMSD:
\begin{align}
    (R^*, \mathbf{t}^*) &= \operatorname*{argmin}_{R \in SO(3), \mathbf{t} \in \mathbb{R}^3} \sum_{i=1}^{N} w_i \left\| \hat{\mathbf{x}}_i - (R \mathbf{x}_i + \mathbf{t}) \right\|^2,\\
    \mathbf{x}^{\text{GT-aligned}}_i &= R^* \mathbf{x}_i + \mathbf{t}^*,
\end{align}
where $\mathbf{x}^{\text{GT-aligned}}_i$ represents the aligned ground-truth coordinates. The weighted rigid alignment loss is then defined as the mean squared discrepancy between the predicted and aligned coordinates:
\begin{equation}
    \mathcal{L}_{\text{align-mse}} = \frac{1}{\sum w_i} \sum_{i=1}^{N} w_i \left\| \hat{\mathbf{x}}_i - \mathbf{x}^{\text{GT-aligned}}_i \right\|^2.
\end{equation}
This formulation ensures that the model learns the precise internal geometry of the ligand and its relative pose within the binding pocket, decoupled from the global frame of reference.

\section{Auxiliary Losses}\label{appsec:auxiliary}

By enforcing local geometric constraints within the decoder, the model learns a compact latent representation that filters out redundant local fluctuations and retains the essential conformational features of the trajectory.

\paragraph{Ligand geometric center loss.}
To stabilize the global placement of the ligand and prevent spurious rigid translations, we align the predicted and reference geometric centers of ligand atoms. Let $\mathbf{x}_t^\ell=\{\mathbf{x}_{t}^{\ell,i}\}_{i=1}^{N_\ell}$ and $\hat{\mathbf{x}}_t^\ell=\{\hat{\mathbf{x}}_{t}^{\ell,i}\}_{i=1}^{N_\ell}$ denote ground-truth and predicted ligand coordinates at step $t$. The geometric center is
\[
C(\mathbf{x}_t^\ell)=\frac{1}{N_\ell}\sum_{i=1}^{N_\ell} \mathbf{x}_{t}^{\ell,i},\qquad
C(\hat{\mathbf{x}}_t^{\ell,i})=\frac{1}{N_\ell}\sum_{i=1}^{N_\ell} \hat{\mathbf{x}}_{t}^{\ell,i},
\]
and the loss is the mean-squared discrepancy
\[
\mathcal{L}_{\mathrm{center}}
= \left\| C(\hat{\mathbf{x}}_t^\ell)-C(\mathbf{x}_t^\ell) \right\|_2^2.
\]
This term softly anchors the ligand’s global position while remaining agnostic to its internal geometry.

\paragraph{Collision loss.}  
To penalize steric clashes, we define a collision loss between protein–ligand atoms and within ligand atoms.  
Let $\x_t^\ell$ and $\x_t^\mcal{P}$ denote ligand and protein atom coordinates at step $t$, and $\hat{\x}_t^\ell$, $\hat{\x}_t^\mcal P$ their predictions.  
We compute predicted distances  
\[
d^{PL}_{ij} = \|\hat{\x}_t^{\mcal P,i} - \hat{\x}_t^{\ell,j}\|_2, 
\quad 
d^{L}_{ij} = \|\hat{\x}_t^{\ell,i} - \hat{\x}_t^{\ell,j}\|_2,
\]
and corresponding ground-truth minimal distances 
\[
d^{PL,gt}_{ij} = \min_t \|\x_t^{\mcal P,i} - \x_t^{\ell,j}\|_2, 
\quad 
d^{LL,gt}_{ij} = \min_t \|\x_t^{\ell,i} - \x_t^{\ell,j}\|_2.
\]

Protein–ligand and ligand–ligand thresholds are set as
\[
\zeta^{PL}_{ij} = \min\!\left(0.9\, d^{PL,gt}_{ij}, \; \zeta_{pl}\right)\, 
\quad
\zeta^{LL}_{ij} = \min\!\left(0.9\, d^{LL,gt}_{ij}, \; \zeta_{ll}\right),
\]
where $\zeta_{pl}=3.0\,$Å and $\zeta_{ll}=2.0\,$Å.  

The collision loss is then defined as
\[
\mathcal{L}_{\text{collision}} = 
\sum_{i,j} \mathbf{1}\!\left(d^{PL}_{ij}<\zeta^{PL}_{ij}\right)\,(\zeta^{PL}_{ij}-d^{PL}_{ij})^2 
+ \sum_{i\neq j} \mathbf{1}\!\left(d^{LL}_{ij}<\zeta^{LL}_{ij}\right)\,(1-b_{ij})\,(\zeta^{LL}_{ij}-d^{LL}_{ij})^2,
\]
where $\mathbf{1}(\cdot)$ represents the indicator function and $b_{ij}$ is the ligand bond mask to exclude bonded pairs.

\paragraph{Ligand bond loss.}  
To preserve ligand bond lengths, we penalize deviations between predicted and ground-truth bonded atom distances.  
Let $\mathcal{B}$ denote the set of bonded atom pairs according to the ligand bond mask.  
For each bond $(i,j) \in \mathcal{B}$, we compute the predicted and ground-truth distances
\[
d^{\ell}_{ij} = \|\hat{\x}_t^{\ell,i} - \hat{\x}_t^{\ell,j}\|_2, 
\quad 
d^{\ell,gt}_{ij} = \|\x_t^{\ell,i} - \x_t^{\ell,j}\|_2.
\]
The bond loss is then defined as the mean squared deviation:
\[
\mathcal{L}_{\text{bond}} 
= \frac{1}{|\mathcal{B}|} \sum_{(i,j)\in\mathcal{B}} 
\left(d^{\ell}_{ij} - d^{\ell,gt}_{ij}\right)^2.
\]

\noindent \textbf{Side-chain bond length loss.} To ensure physical plausibility and robustness against outliers in the predicted structure, we constrain bond lengths using a robust error function.
Given the ground-truth coordinates $\mathbf{x}$ and predicted coordinates $\hat{\mathbf{x}}$, for each bond in the side-chain bond index set $\mathcal{E}_{\text{bond}}$, we compute the Euclidean bond lengths $d_{ij} = \| \mathbf{x}_i - \mathbf{x}_j \|_2$ and $\hat{d}_{ij} = \| \hat{\mathbf{x}}_i - \hat{\mathbf{x}}_j \|_2$.
The loss is minimized using the Huber loss with a threshold $\delta = 0.05$:
\begin{equation}
    \mathcal{L}_{\text{sc-bond}} = \frac{1}{|\mathcal{E}_{\text{bond}}|} \sum_{(i,j) \in \mathcal{E}_{\text{bond}}} H_\delta(\hat{d}_{ij} - d_{ij}),
\end{equation}
where $H_\delta(\cdot)$ applies a quadratic penalty for small errors and a linear penalty for large errors:
\begin{equation}
    H_\delta(e) = 
    \begin{cases} 
    0.5 e^2 & \text{if } |e| \le \delta \\
    \delta (|e| - 0.5 \delta) & \text{otherwise}.
    \end{cases}
\end{equation}

\noindent \textbf{Side-chain bond angle loss.} To maintain correct local stereochemistry while avoiding numerical instabilities associated with angular periodicity, we define the loss directly on the cosine of the bond angles.
For each angle triplet $(i, j, k) \in \mathcal{E}_{\text{angle}}$, we compute the cosine via the dot product of normalized bond vectors $\mathbf{v}_{ji}$ and $\mathbf{v}_{jk}$:
\begin{equation}
    \cos \theta_{ijk} = \frac{\mathbf{v}_{ji}}{\|\mathbf{v}_{ji}\|} \cdot \frac{\mathbf{v}_{jk}}{\|\mathbf{v}_{jk}\|}.
\end{equation}
We minimize the $L_1$ distance between the predicted and ground-truth cosine values:
\begin{equation}\label{eq:sc_angle_loss}
    \mathcal{L}_{\text{sc-angle}} = \frac{1}{|\mathcal{E}_{\text{angle}}|} \sum_{(i,j,k) \in \mathcal{E}_{\text{angle}}} \left| \widehat{\cos \theta}_{ijk} - \cos \theta_{ijk} \right|.
\end{equation}
\paragraph{Sidechain torsion loss} To ensure the model captures correct side-chain rotameric states while circumventing the periodicity issues inherent in direct angular regression, we optimize the cosine of the torsion angles.
Let $\mathcal{T}$ be the set of side-chain torsion angles, where each angle is defined by an ordered sequence of four atoms $(i, j, k, l)$.
We define the bond vectors $\mathbf{v}_1 = \mathbf{x}_j - \mathbf{x}_i$, $\mathbf{v}_2 = \mathbf{x}_k - \mathbf{x}_j$, and $\mathbf{v}_3 = \mathbf{x}_l - \mathbf{x}_k$.
The normal vectors to the planes defined by these bonds are given by $\mathbf{n}_1 = \mathbf{v}_1 \times \mathbf{v}_2$ and $\mathbf{n}_2 = \mathbf{v}_2 \times \mathbf{v}_3$.
The cosine of the torsion angle $\phi$ corresponds to the dot product of the normalized plane normals:
\begin{equation}
    \cos \phi_{ijkl} = \frac{\mathbf{n}_1}{\|\mathbf{n}_1\|} \cdot \frac{\mathbf{n}_2}{\|\mathbf{n}_2\|}.
\end{equation}
The loss is defined as the mean absolute difference ($L_1$ loss) between the predicted and ground-truth cosine values:
\begin{equation}
    \mathcal{L}_{\text{sc-torsion}} = \frac{1}{|\mathcal{T}|} \sum_{(i,j,k,l) \in \mathcal{T}} \left| \widehat{\cos \phi}_{ijkl} - \cos \phi_{ijkl} \right|.
\end{equation}

\paragraph{Backbone torsion loss} To ensure the recovery of accurate secondary structure elements (such as $\alpha$-helices and $\beta$-sheets), we enforce geometric constraints on the backbone torsion angles $\phi$ and $\psi$.
Similar to the side-chain formulation, we mitigate angular periodicity issues by operating directly on the cosine of the angles.
Let $\mathcal{T}_{\text{bb}}$ be the set of backbone torsion angles defined by the consecutive backbone atoms (e.g., $C_{i-1}-N_i-C_{\alpha,i}-C_i$ for $\phi$ and $N_i-C_{\alpha,i}-C_i-N_{i+1}$ for $\psi$).
We minimize the $L_1$ distance between the predicted and ground-truth cosine values:
\begin{equation}
    \mathcal{L}_{\text{bb-torsion}} = \frac{1}{|\mathcal{T}_{\text{bb}}|} \sum_{\theta \in \mathcal{T}_{\text{bb}}} \left| \widehat{\cos \theta} - \cos \theta \right|.
\end{equation}
The cosine values are computed via the dot product of the normalized normal vectors of the planes defined by the atom quadruplets.

For both the backbone and side-chain torsion losses, we exclusively use the cosine of the angles rather than computing both the sine and cosine. Empirically, we found that relying solely on the cosine facilitates easier and more stable training convergence. This simplification does not introduce adverse effects, as the torsion loss serves as an auxiliary objective while the primary MSE loss drives the overall parameter optimization.

\paragraph{Smooth LDDT loss.} To directly optimize the local structural accuracy while maintaining differentiability, we employ a smoothed version of the Local Distance Difference Test (LDDT) score, as suggested in AlphaFold 3~\cite{abramson2024accurate}. Standard LDDT relies on a non-differentiable step function, and the smooth version approximates it using a sigmoid function.

Let $d_{ij}$ and $\hat{d}_{ij}$ be the ground-truth and predicted Euclidean distances between atoms $i$ and $j$. We define a binary mask $m_{ij}$ that is 1 if the pair exists in the ground truth and the distance $d_{ij}$ is within the inclusion radius $R_{\text{inc}}$, and 0 otherwise. Following AlphaFold 3, we set $R_{\text{inc}} = 15\text{\AA}$.

For each valid pair, we compute the absolute distance error $e_{ij} = |\hat{d}_{ij} - d_{ij}|$. The smooth score contribution is the average of sigmoid activations over four thresholds $\mathcal{T} = \{0.5, 1.0, 2.0, 4.0\}\text{\AA}$:
\begin{equation}
    s_{ij} = \frac{1}{4} \sum_{\tau \in \mathcal{T}} \sigma(\tau - e_{ij}),
\end{equation}
where $\sigma(\cdot)$ is the sigmoid function. This term smoothly approximates the fraction of thresholds satisfied by the prediction. The final loss is defined as:
\begin{equation}
    \mathcal{L}_{\text{smooth-LDDT}} = 1 - \frac{\sum_{i,j} m_{ij} s_{ij}}{\sum_{i,j} m_{ij} + \epsilon}.
\end{equation}

\paragraph{Final Objective.}
\begin{align}
    \mcal{L}_{aux} = 0.5 \cdot \mcal{L}_{\text{center}} + \mcal{L}_{\text{bond}} +  \mcal{L}_{\text{collision}} + 6 \cdot \mcal{L}_{\text{sc-angle}} + 6 \cdot \mcal{L}_{\text{sc-torsion}}  +  \mcal{L}_{\text{sc-bond}} + 1.5 \cdot \mcal{L}_{\text{bb-torsion}} + 3\cdot \mcal{L}_{\text{smooth-LDDT}}
\end{align}

Finally, the AAT-VAE loss is
\begin{align}
    \mcal{L}_{vae} =\mcal{L}_{\text{aligned-mse}}(\lambda_{\text{ligand}}=10) + 0.5 \cdot \text{min}(1.0, \text{step}/500)\cdot \text{KL}(\|) + \mcal{L}_{aux} 
\end{align}

For the flow matching loss, we also consider applying a higher weight to the ligand.
\begin{align}
\scalemath{0.999}{
    \mathcal{L}_{fm} = \frac{1}{N(T-1)}\sum_{t=1}^{T-1}\sum_{i=1}^N \omega_i\| \mathbf{r}_\theta(\mathbf{r}_{t,i}^\tau, \mathbf{c}_i, \tau) - \mathbf{r}_{t,i}^{1} \|_2^2}
\end{align}
where $\omega_i = 25 $ if atom $i$ belongs to ligand, otherwise $\omega_i =1$.

\section{Training and Inference Algorithms}
The main training and inference procedure is shown in \autoref{alg:main_loop}.

\begin{algorithm2e}[ht]\LinesNumbered
\caption{Main Loop of \ourM}\label{alg:main_loop}

\SetKwProg{KwIn}{\textcolor{green!50!black}{Input: }}{:}{}
\SetKw{Return}{\textcolor{green!50!black}{\textbf{return}}}

\SetKwIF{If}{ElseIf}{Else}{\textcolor{green!50!black}{\textbf{if}}}{\textcolor{green!50!black}{:}}{\textcolor{green!50!black}{\textbf{elif}}}{\textcolor{green!50!black}{\textbf{else:}}}{}

\KwIn{$\{\mathbf{f}^*\}$ , $\{\vec{\rvx}_{0,l}\}$, $N_{\text{cycle}} = 4$, $c_s = 384$, $c_z = 128$}
{
    \textcolor{brown}{\# Construct the token embeddings, including GPCR-ligand specific features}\;
    $\{\rvs_i^{\mathrm{inputs}}\} \gets \text{\color{teal}InputFeatureEmbedder}(\{\mathbf{f}^*\})$\;
    $\rvs_i^{\mathrm{init}} \gets \text{LinearNoBias}(\rvs_i^{\mathrm{inputs}})$\;
    $\rvz_{ij}^{\mathrm{init}} \gets \text{LinearNoBias}(\rvs_i^{\mathrm{inputs}}) + \text{LinearNoBias}(\rvs_j^{\mathrm{inputs}})$\;

    $\{\rvz_{ij}\}, \{\rvs_i\} \gets 0, 0$\;
    \ForEach{$c \in \{1, \ldots, N_{\text{cycle}}\}$}{
        $\rvz_{ij} \gets \rvz_{ij}^{\mathrm{init}} + \text{LinearNoBias}(\text{LayerNorm}(\rvz_{ij}))$\;
        $\{\rvz_{ij}\}, \{\rvs_i\}  \gets \text{\color{teal}GraphTransformer}(\{\vec{\rvx}_{0,l}\},\{\rvs_i\}, \{\rvz_{ij}\}, \{\rvs_i^{\mathrm{inputs}}\})$\;
        $\rvs_i \gets \rvs_i^{\mathrm{init}} + \text{LinearNoBias}(\text{LayerNorm}(\rvs_i))$\;
    }

    \uIf{TrainAATVAE}{
        $\mathcal{L}_{vae}\gets\text{\color{teal}TrainAATVAE}(\{\vec{\rvx}_{l}\}$, $\rvt$, $\{\mathbf{f}^*\}$, $\{\rvs_i^{\mathrm{inputs}}\}$, $\{\rvs_i\}$, $\{\rvz_{ij}\})$\;
        \Return{$\mathcal{L}_{vae}$}
    }
    \uElseIf{TrainLatentFlow}{
        $\mathcal{L}_{fm}\gets\text{\color{teal}TrainLatentFlow}(\{\vec{\rvx}_{l}\}$, $\rvt$, $\{\mathbf{f}^*\}$, $\{\rvs_i^{\mathrm{inputs}}\}$, $\{\rvs_i\}$, $\{\rvz_{ij}\})$\;
        \Return{$\mathcal{L}_{fm}$}
    }
    \Else{
        \textcolor{brown}{\# Inference}\;
        traj\_list=[$\{\vec{\rvx}_{0,l}\}$] \;
        $\{\vec{\mu}_l,\vec{\sigma}_l\},\{\rvs_i\},\{\rvz_{ij}\}, \{\rva_i^{\mathrm{skip}}\}, \{\rvc_l^{\mathrm{skip}}\}, \{\rvp_{lm}^{\mathrm{skip}}\} \gets \text{\color{teal}Encoder}(\{\vec{\rvx}_{0,l}\}$, $\rvt$, $\{\mathbf{f}^*\}$, $\{\rvs_i^{\mathrm{inputs}}\}$, $\{\rvs_i\}$, $\{\rvz_{ij}\})$\;
        
        $\{\vec{\rvz}_{t,l}^{\mathrm{pred}}\} \gets \text{\color{teal}SampleLatentFlow}(\{\vec{\rvx}_{0,l}\},\{\rva_i^{\mathrm{skip}}\}, \{\rvs_i\}, \{\rvc_l^{\mathrm{skip}}\}, \{\rvp_{lm}^{\mathrm{skip}}\})$\;
        
        $\text{traj\_list} \gets \text{Decoder}(\{\vec{\rvz}_{t,l}^{\mathrm{pred}}\},\{\rvs_i\}, \{\rvz_{ij}\}, \{\rvc_l^{\mathrm{skip}}\}, \{\rvp_{lm}^{\mathrm{skip}}\},\beta_{ij}=0 )$\;
        \Return{$\text{traj\_list}$}
    }
}
\end{algorithm2e}

\begin{algorithm2e}[ht]\LinesNumbered
\caption{TrainAATVAE}\label{alg:train_VAE}
\SetKwProg{KwIn}{\textcolor{green!50!black}{Input: }}{:}{}
\SetKw{Return}{\textcolor{green!50!black}{\textbf{return: }}}
\KwIn{$\{\vec{\rvx}_l\}$, $\rvt$, $\{\mathbf{f}^*\}$, $\{\rvs_i^{\mathrm{inputs}}\}$, $\{\rvs_i^{\mathrm{trunk}}\}$, $\{\rvz_{ij}^{\mathrm{trunk}}\}$}
{

\textcolor{brown}{\# Data Augmentation for exempting SE(3) Equivalent Model design (Algorithm 19 in AF3) }\;
    $\{\vec{\rvx}_l\} \leftarrow \text{CentreRandomAugmentation}(\{\vec{\rvx}_l\})$\;

    $(\{\vec{\mu}_l,\vec{\sigma}_l\},\{\rvs_i\}, \{\rvz_{ij}\},\{\rva_i^{\mathrm{skip}}\}, \{\rvc_l^{\mathrm{skip}}\}, \{\rvp_{lm}^{\mathrm{skip}}\} ) \leftarrow \text{\color{teal}Encoder}(\{\vec{\rvx}_l\}, \tau, \{\mathbf{f}^*\}, \{\rvs_i^{\mathrm{inputs}}\}, \{\rvs_i^{\mathrm{trunk}}\}, \{\rvz_{ij}^{\mathrm{trunk}}\})$\;
     $ \{{\rvq}_l\} \leftarrow \text{\color{teal}Reparameterization} (\{\vec{\mu}_l,\vec{\sigma}_l\})$\;

$\{\vec{\hat{\x}}_l\}  \gets \text{\color{teal}Decoder}( \{\rvq_l\}, \{\rvs_i\}, \{\rvz_{ij}\}, \{\rvc_{l}^{\mathrm{skip}}\}, \{\rvp_{lm}^{\mathrm{skip}}\},\beta_{ij}=0) $\;
    $\mcal{L}_{vae} = \text{MSE}(\{\vec{\hat{\x}}_l\},\{\vec{\rvx}_l\}) + \{\text{KL}(\mcal{N}(z; \vec{\mu}_l,\vec{\sigma}_l) \Vert \mcal{N}(\x_l, \sigma \mbf{I})))\}$\;
    $\mcal L_{aux}=\text{GeometricConstraint}(\{\vec{\hat{\x}}_l\},\{\vec{\rvx}_l\})$\;
\Return{$\mcal{L}_{vae}+\mcal L_{aux}$}
}
\end{algorithm2e}

\begin{algorithm2e}[ht]\LinesNumbered
\caption{TrainLatentFlow}\label{alg:train_flow}
\SetKwProg{KwIn}{\textcolor{green!50!black}{Input: }}{:}{}
\SetKw{Return}{\textcolor{green!50!black}{\textbf{return: }}}
\KwIn{$\{\vec{\rvx}_l\}$, $\{\mathbf{f}^*\}$, $\{\rvs_i^{\mathrm{inputs}}\}$, $\{\rvs_i^{\mathrm{trunk}}\}$, $\{\rvz_{ij}^{\mathrm{trunk}}\}$}
{
\textcolor{brown}{\# Data Augmentation for exempting SE(3) Equivalent Model design (Algorithm 19 in AF3) }\;
    $\{\vec{\rvx}_l\} \leftarrow \text{CentreRandomAugmentation}(\{\vec{\rvx}_l\})$\;
    $(\{\vec{\mu}_l,\vec{\sigma}_l\},\{\rvs_i\}, \{\rvz_{ij}\},\{\rva_i^{\mathrm{skip}}\}, \{\rvc_l^{\mathrm{skip}}\}, \{\rvp_{lm}^{\mathrm{skip}}\} ) \leftarrow \text{\color{teal}Encoder}(\{\vec{\rvx}_l^{\tau}\}, \tau, \{\mathbf{f}^*\}, \{\rvs_i^{\mathrm{inputs}}\}, \{\rvs_i^{\mathrm{trunk}}\}, \{\rvz_{ij}^{\mathrm{trunk}}\})$\;
     $ {\rvq}_l \leftarrow \vec{\mu}_l$\;
    $\{\rvq_l\mathrm{
    .stop\_grad()}\}, \{\rvs_i\mathrm{
    .stop\_grad()}\}, \{\rvz_{ij}\mathrm{
    .stop\_grad()}\}$\;
    $\{\rva_i^{\mathrm{skip}}\mathrm{
    .stop\_grad()}\}, \{\rvc_l^{\mathrm{skip}}\mathrm{
    .stop\_grad()}\}, \{\rvp_{lm}^{\mathrm{skip}}\mathrm{
    .stop\_grad()}\}$\;
$\mcal{L}_{fm}  \gets \text{\color{teal}LatentFlowMatching}( \{\rvq_l\}, \{\vec{\rvx}_{0,l}\}, 
\{\rva_i^{\mathrm{skip}}\}, \{\rvs_i\}, \{\rvc_{l}^{\mathrm{skip}}\}, \{\rvp_{lm}^{\mathrm{skip}}\}$\;

\Return{$\mcal{L}_{fm}$}
}
\end{algorithm2e}

\begin{algorithm2e}[h]\LinesNumbered
\caption{LatentFlowMatching}\label{alg:LatentFlowMatching}
\SetKwProg{KwIn}{\textcolor{green!50!black}{Input: }}{:}{}
\SetKw{Return}{\textcolor{green!50!black}{\textbf{return: }}}
\KwIn{$\{\vec{\rvz}_l\}$,$\{\vec{\rvx}_{0,l}\}$, $\{\rva_i^{\mathrm{skip}}\}$, $\{\rvs_i^{\mathrm{trunk}}\}$,$\{\rvc_{l}^{\mathrm{skip}}\}, \{\rvp_{lm}^{\mathrm{skip}}\}$}
{
\textcolor{brown}{\# Independent noise levels }\;
$\tau \sim (\mathcal{U}(0, 1),\mathcal{U}(0, 1),\cdots,\mathcal{U}(0, 1))$\;

$\{\vec{\rvr}^0_l\} \sim \mathcal{N}(\vec{0}, \mathbf{I}_3)$ \;
\textcolor{brown}{\# Calculate the residual vector}\;
$\vec{{\rvr}}_l = \vec{{\rvz}}_l - \vec{\rvx}_{0,l}$ \;
    $\{\vec{{\rvr}}_l^{\tau}\} = 
    \tau \{\vec{\rvr}_l \}+ (1-\tau)\{\vec{\rvr}^0_l\}$\;
    
    $\{\vec{\hat{\rvr}}_l^1\} \leftarrow \text{\color{teal}LatentVelocityNetwork}(\{\vec{{\rvr}}_l^{\tau}\}, \tau, \{\rva_i^{\mathrm{skip}}\}, \{\rvs_i^{\mathrm{trunk}}\},\{\rvc_{l}^{\mathrm{skip}}\}, \{\rvp_{lm}^{\mathrm{skip}}\})$\;
    $\mcal{L}_{flow} = \text{MSE}(\{\vec{{\rvr}}_l\},\{\vec{\hat{\rvr}}_l^1 \})$\;

\Return{$\mcal{L}_{flow}$}
}
\end{algorithm2e}

\begin{algorithm2e}[h]\LinesNumbered
\caption{SampleLatentFlow}\label{alg:sample_flow}
\SetKwProg{KwIn}{\textcolor{green!50!black}{Input: }}{:}{}
\SetKw{Return}{\textcolor{green!50!black}{\textbf{return: }}}
\KwIn{$\{\vec{\rvx}_{0,l}\}$, $\{\mathbf{a}^{\mathrm{skip}}_i\}$, $\{\rvs_i^{\mathrm{trunk}}\}$, $\{\rvc_l^{\mathrm{skip}}\}, \{\rvp_{lm}^{\mathrm{skip}}\}$}
{
$\vec{\rvr}_l^{0} \sim   \mathcal{N}(\vec{0}, \mathbf{I}_3)$\;
$d\tau=0.1$\;
\ForEach{ $\tau$ in $\{0, 0.1,0.2,...,0.9\}$}{
    $\{\vec{\hat{\rvr}}_l^1\} \leftarrow \text{\color{teal}LatentVelocityNetwork}(\{\vec{\rvr}_l^{\tau}\}, \tau,\{\mathbf{a}^{\mathrm{skip}}_i\}, \{\rvs_i^{\mathrm{trunk}}\},\{\rvc_l^{\mathrm{skip}}\}, \{\rvp_{lm}^{\mathrm{skip}}\})$\;
    $\vec{\rvr}_l^{\tau+d\tau} \leftarrow \vec{\rvr}_l^{\tau} +  d\tau \cdot (\frac{\vec{\hat{\rvr}}_l^1-\vec{\hat{\rvr}}_l^{\tau}}{1-\tau})$\;
}

\Return{$\{\vec{\rvr}_l^{1} + \vec{\rvx}_{0,l}\}$ }
}
\end{algorithm2e}

\begin{algorithm2e}[h]\LinesNumbered
\caption{InputFeatureEmbedder}\label{alg:construct_embedding}
\SetKwProg{KwIn}{\textcolor{green!50!black}{Input: }}{:}{}
\SetKw{Return}{\textcolor{green!50!black}{\textbf{return: }}}
\KwIn{\text{}$\{\mathbf{f}^*\}$}{
        \textcolor{brown}{\# Construct the GPCR-ligand complex specific features. $c = 8 + 8$}\;
    $\mathbf{a}_i^{\mathrm{gpcr}} \gets \text{\color{teal}{concat}}(\mathbf{f}_i^{\mathrm{TM\_index}}, \mathbf{f}_i^{\mathrm{lig\_func}})$\;
    \textcolor{brown}{\# Concatenate the per-token features.}\;
    $\mathbf{s}_i \gets \text{concat}(\mathbf{a}_i^{\mathrm{gpcr}}, \mathbf{f}_i^{\mathrm{restype}})$\;
    
    \Return{$\{\mathbf{s}_i\}$}
}
\end{algorithm2e}

\begin{algorithm2e}[h]\LinesNumbered
\caption{Encoder}\label{alg:encoder}
\SetKwProg{KwIn}{\textcolor{green!50!black}{Input: }}{:}{}
\SetKw{Return}{\textcolor{green!50!black}{\textbf{return: }}}
\KwIn{$\{\vec{\rvx}_l\}$, $\rvt$, $\{\mathbf{f}^*\}$, $\{\rvs_i^{\mathrm{inputs}}\}$, $\{\rvs_i^{\mathrm{trunk}}\}$, $\{\rvz_{ij}^{\mathrm{trunk}}\}$,
$\sigma_{\mathrm{data}} = 16$, $c_{\mathrm{atom}} = 128$, $c_{\mathrm{atompair}} = 16$, $c_{\mathrm{token}} = 768$}
{
$\{\rvs_i\}, \{\rvz_{ij}\} \gets \text{\color{teal}VAEConditioning}(\{\mathbf{f}^*\}, \{\rvs_i^{\mathrm{inputs}}\}, \{\rvs_i^{\mathrm{trunk}}\}, \{\rvz_{ij}^{\mathrm{trunk}}\}, \sigma_{\mathrm{data}})$\;

\textcolor{brown}{\# Sequence-local Atom Attention and aggregation to coarse-grained tokens (following Ag. 5 in AF3), and we also add frame-index ($\rvt$) embeddings before AtomTransformer into $\rvq_l$ }\;
$\{\rva_i^{\mathrm{skip}}\},\{\rvq_l\}, \{\rvc_l^{\mathrm{skip}}\}, \{\rvp_{lm}^{\mathrm{skip}}\} \gets \text{\color{black} AtomAttentionEncoder}(\{\mathbf{f}^*\}, \{\vec{\rvx}_l\}, \{\rvs_i\}, \rvt, \{\rvz_{ij}\}, c_{\mathrm{atom}}, c_{\mathrm{atompair}}, c_{\mathrm{token}})$\;
\textcolor{brown}{\#  Linear heads to get the mean and variane of VAE with atom-level structure-relevant variable $\rvq_l$}\;
$\{\vec{\mu}_l,\vec{\sigma}_l\} \gets \text{MeanVarianceHeads}(\{\rvq_l\})$\;
\Return{$\{\vec{\mu}_l,\vec{\sigma}_l\}, \{\rvs_i\}, \{\rvz_{ij}\}, \{\rva_i^{\mathrm{skip}}\}, \{\rvc_l^{\mathrm{skip}}\}, \{\rvp_{lm}^{\mathrm{skip}}\}$} 
}
\end{algorithm2e}

\begin{algorithm2e}[h]\LinesNumbered
\caption{VAEConditioning}\label{alg:diffusion_conditioning}
\SetKwProg{KwIn}{\textcolor{green!50!black}{Input: }}{:}{}
\SetKw{Return}{\textcolor{green!50!black}{\textbf{return: }}}
\KwIn{$\{\mathbf{f}^*\}$, $\{\rvs_i^{\mathrm{inputs}}\}$, $\{\rvs_i^{\mathrm{trunk}}\}$, $\{\rvz_{ij}^{\mathrm{trunk}}\}$, $\sigma_{\mathrm{data}}$, $c_z = 128$, $c_s = 384$}
{
\textcolor{brown}{\# Pair conditioning (AF3)} \;
$\rvz_{ij} \gets \text{concat}([\; \rvz_{ij}^{\mathrm{trunk}},\ \text{RelativePositionEncoding}(\{\mathbf{f}^*\}) \;])$\;
$\rvz_{ij} \gets \text{LinearNoBias}(\text{LayerNorm}(\rvz_{ij}))$\;
\ForEach{$b \in \{1, 2\}$}{
    $\rvz_{ij} \pluseq \text{Transition}(\rvz_{ij},\ n=2)$\;
}

\textcolor{brown}{\# Single conditioning}\;
$\rvs_i \gets \text{concat}([\; \rvs_i^{\mathrm{trunk}},\ \rvs_i^{\mathrm{inputs}} \;])$\;
$\rvs_i \gets \text{LinearNoBias}(\text{LayerNorm}(\rvs_i))$\;
\ForEach{$b \in \{1, 2\}$}{
    $\rvs_i \pluseq \text{Transition}(\rvs_i,\ n=2)$\;
}

\Return{$\{\rvs_i\}, \{\rvz_{ij}\}$}
}
\end{algorithm2e}

\begin{algorithm2e}[h]\LinesNumbered
\caption{Decoder}\label{alg:Decoder}
\SetKwProg{KwIn}{\textcolor{green!50!black}{Input: }}{:}{}
\SetKw{Return}{\textcolor{green!50!black}{\textbf{return: }}}
\KwIn{$\{\rvq_l\}$, $\{\rvs_i\}$, $\{\rvz_{ij}\}$, $ \{\rvc_{l}^{\mathrm{skip}}\}, \{\rvp_{lm}^{\mathrm{skip}}\}$, $\{\beta_{ij}\}$, $N_{\mathrm{block}} = 3$, $N_{\mathrm{head}} = 4$}{

$\{\rvq_l\} \gets \text{Linear}(\{\rvq_l\})$\;

\textcolor{brown}{\# Aggregate Latent atom representations to token representation.}\;
$\rva_i \gets \text{\color{teal}{AggregateAtomToToken}}(\{\rvq_l\})$\;

\textcolor{brown}{\# Full self-attention on token level}\;
$\rva_i += \text{LinearNoBias}(\text{LayerNorm}(\rva_i))$\;
$\{\rva_i\} \gets \text{\color{teal}TemporalDiffusionTransformer}(\{\rva_i\}, \{\rvs_i\}, \{\rvz_{ij}\}, \beta_{ij} = 0, N_{\mathrm{block}} = 6, N_{\mathrm{head}} = 16)$\;
$\rva_i \gets \text{LayerNorm}(\rva_i)$\;
\textcolor{brown}{\# Broadcast token activations to atoms and run Sequence-local Atom Attention}\;
$\{\vec{\rvx_l}^{\mathrm{out}}\} = \text{\color{teal}AtomTemporalAttentionDecoder}(\{\rva_i\}, \{\rvq_l\}, \{\rvc_{l}^{\mathrm{skip}}\}, \{\rvp_{lm}^{\mathrm{skip}}\})$ \;

\Return{$\{\vec{\rvx_l}^{\mathrm{out}}\}$}
}
\end{algorithm2e}

\begin{algorithm2e}[h]
\caption{LatentVelocityNetwork}\label{alg:latent_velocity_network}
\SetKwProg{KwIn}{\textcolor{green!50!black}{Input: }}{:}{}
\SetKw{KwRet}{\textcolor{green!50!black}{\textbf{return: }}}
\KwIn{$\{\rvr_i^{\tau}\}$,$\tau$,$\{\rva_i\}$, $\{\rvs_i\}$, $\{\rvc^{skip}_l
\}, \{\rvp_{lm}^{\mathrm{skip}}\}$, $N_{\mathrm{block}} = 3$, $N_{\mathrm{head}} = 4$}
{

\textcolor{brown}{\# Stochastic Token Propagator is used to transfer the structural context from the initial frame tokens ($\mbf{a}_0$) into the temporal domain }\;
$\mathbf{\epsilon} \sim \mcal N(\mathbf{0},\mathbf{1})$\;
$\{\rva_{t>0,i}\} \gets \{\tau \cdot \rva_{0,i} + 0.1 \cdot (1-\tau) \cdot \mathbf{\epsilon} \}$\;
$\rva_i \gets \rva_{\{t\},i}$\;
$\rva_i \gets \text{LayerNorm}(\rva_i + \text{LinearNoBias}(\text{LayerNorm}(\rvs_i)))$\;

$\rvr_l = \text{LinearNoBias}(\rvr_l^{\tau})$\;
$\rvr_l \gets 0.1\cdot\text{LinearNoBias}   (\rva_{\mathrm{tok\_idx(l)}}) + \rvr_l$\;
\textcolor{brown}{\# Cross attention transformer from AF3.}\;
$\{\rvr_l\} = \text{AtomTransformer}(\{\rvr_l\}, \{\rvc^{skip}_l
\}, \{\rvp_{lm}^{\mathrm{skip}}\}, N_{\mathrm{block}} = 3, N_{\mathrm{head}} = 4) $\;
\textcolor{brown}{\# Map to positions update}\;
$\rvr_l^{\mathrm{update}}= \text{LinearNoBias} (\text{LayerNorm}(\rvr_l))
$\;

\KwRet $\{\mathbf{r}_l^{\mathrm{update}}\}$
}
\end{algorithm2e}

\begin{algorithm2e}[h]\LinesNumbered
\caption{AtomTemporalTransformer}\label{alg:atom_transformer}

\SetKwProg{Fn}{\textcolor{green!50!black}{Input: }}{:}{}
\SetKw{KwRet}{\textcolor{green!50!black}{\textbf{return: }}}

\Fn{$\{\mathbf{q}_l\}, \{\mathbf{c}_l\}, \{\mathbf{p}_{lm}\}, N_{\mathrm{block}}=3, N_{\mathrm{head}}, N_{\mathrm{queries}}=32, N_{\mathrm{keys}}=128, \mathcal{S}_{\mathrm{subset\ centres}} = \{15.5, 47.5, 79.5, \dots\}$}{

    $\{\mathbf{q}_l\} = \textcolor{teal}{\text{TemporalDiffusionTransformer}}(\{\mathbf{q}_l\}, \{\mathbf{c}_l\}, \{\mathbf{p}_{lm}\}, \{\beta_{ij}=0\}, N_{\mathrm{block}}, N_{\mathrm{head}})$\;
    
    \KwRet $\{\mathbf{q}_l\}$
}
\end{algorithm2e}

\begin{algorithm2e}[h]\LinesNumbered
\caption{AtomTemporalAttentionDecoder}\label{alg:atom_attention_decoder}

\SetKwProg{Fn}{\textcolor{green!50!black}{Input: }}{:}{}
\SetKw{KwRet}{\textcolor{green!50!black}{\textbf{return: }}}

\Fn{$\{\mathbf{a}_i\}, \{\mathbf{q}_l^{\mathrm{skip}}\}, \{\mathbf{c}_l^{\mathrm{skip}}\}, \{\mathbf{p}_{lm}^{\mathrm{skip}}\}$}{

    \textcolor{brown}{\# Broadcast per-token activiations to per-atom activations and add the skip connection}\;
    $\mathbf{q}_l = \text{LinearNoBias}(\mathbf{a}_{\text{tok\_idx}(l)}) + \mathbf{q}_l^{\mathrm{skip}}$\;
    
    \textcolor{brown}{\# Cross attention temporal transformer.}\;
    $\{\mathbf{q}_l\} = \textcolor{teal}{\text{AtomTemporalTransformer}}(\{\mathbf{q}_l\}, \{\mathbf{c}_l^{\mathrm{skip}}\}, \{\mathbf{p}_{lm}^{\mathrm{skip}}\}, N_{\mathrm{block}}=3, N_{\mathrm{head}}=4)$\;
    
    \textcolor{brown}{\# Map to positions update.}\;
    $\mathbf{x}_l = \text{LinearNoBias}(\text{LayerNorm}(\mathbf{q}_l))$\;
    
    \KwRet $\{\mathbf{x}_l\}$
}
\end{algorithm2e}

\begin{algorithm2e}[h]\LinesNumbered
\caption{TemporalDiffusionTransformer}\label{alg:diffusion_transformer}

\SetKwProg{KwIn}{\textcolor{green!50!black}{Input}}{:}{}
\SetKw{KwRet}{\textcolor{green!50!black}{\textbf{return}}}

\KwIn{$ \{\mathbf{a}_i\}, \{\mathbf{s}_i\}, \{\mathbf{z}_{ij}\}, \{\beta_{ij}\}, N_{\mathrm{block}} = 6, N_{\mathrm{head}} = 16$}{
    
\For{$n \in [1, \ldots, N_{\mathrm{block}}]$}{
    \text{\color{brown} \# Algorithm 24 in AF3}\;
    $\{\rvb_i\} \gets \text{AttentionPairBias}(\{\rva_i\}, \{\rvs_i\}, \{\rvz_{ij}\}, \{\beta_{ij}\}, N_{\mathrm{head}})$\;
    \text{\color{brown} \#  Multihead attention on time-level, following ~\citet{feng2025biomd}}\;
    $\{\rvb_i\} \gets \text{\color{black}TemporalAttention}(\{\rva_i + \rvb_i\})$\;
     \text{\color{brown} \# Algorithm 25 in AF3}\;
    $\rva_i \gets \rvb_i + \text{ConditionedTransitionBlock}(\rva_i, \rvs_i)$\;
}
    
    \KwRet $\{\mathbf{a}_i\}$
}
\end{algorithm2e}

\begin{algorithm2e}[!t]
\LinesNumbered
\caption{{TemporalAttention}}\label{alg:temporal_attention}
\SetKwProg{KwIn}{\textcolor{green!50!black}{Input: }}{:}{}
\SetKw{Return}{\textcolor{green!50!black}{\textbf{return: }}}
\KwIn{Single representation $c_s$ of shape ($T,B, N, c_s$), where $T$ is time, $B$ is batch size, and $N$ is residues.}
{

\textcolor{brown}{\# Permute dimensions to make time the sequence axis for attention}\;
$c_s' \gets \text{Permute}(c_s, \text{dims}=(2, 1, 0, 3))$\;

\textcolor{brown}{\# Project to Query, Key, Value for each residue independently}\;
$Q \gets \text{Linear}_{Q}(c_s')$\;
$K \gets \text{Linear}_{K}(c_s')$\;
$V \gets \text{Linear}_{V}(c_s')$\;

\textcolor{brown}{\# Calculate scaled dot-product attention scores across time}\;
$d_k \gets \text{dimension of } K$\;
$\text{logits} \gets (Q \cdot K^T) / \sqrt{d_k}$\;
$\text{weights} \gets \text{Softmax}(\text{logits}, \text{dim}=-1)$\;

\textcolor{brown}{\# Apply attention weights to values}\;
$\text{output}' \gets \text{weights} \cdot V$\;

\textcolor{brown}{\# Permute back to the original dimension order}\;
$\rvo \gets \text{Permute}(\text{output}', \text{dims}=(2, 1, 0, 3))$\;

\Return{$\rvo$}\;
}
\end{algorithm2e}

\section{Evaluation}
\subsection{Main Quantitative  Metrics}\label{appsec:main_metrics}
We adopt the evaluation criteria proposed by ~\citet{jingAlphaFoldMeetsFlow2024}, which can be grouped as follows:  
\begin{itemize}

\item  \textbf{Flexibility correlation (↑):} Pearson correlation coefficient $r$ computed for pairwise RMSD, global RMSF, and per-target RMSF.  
\item  \textbf{Distributional accuracy:} Root mean of the 2-Wasserstein distance ($W_{2}$-dist) along with its translation and variance components (↓), molecular dynamics (MD) PCA $W_{2}$-dist (↓), joint PCA $W_{2}$-dist (↓), and the proportion of samples achieving PC-sim $> 0.5$ (↑).     
\item  \textbf{Ensemble observables (↑):} Jaccard similarity $J$ for weak contacts, transient contacts, and exposed residues, together with the Spearman correlation $\rho$ of the exposed mutual information (MI) matrix. 
\end{itemize}
\paragraph{JS Divergence.}

We use the JS Divergence to measure the distribution distinction in this paper.
The Jensen-Shannon Divergence between two probability vectors $p$ and $q$ is defined as,

$${\frac{D(p \parallel m) + D(q \parallel m)}{2}}$$

where $m$ is the pointwise mean of and $p$ and $q$, and $D$ is the Kullback-Leibler divergence.

This routine will normalize $p$ and $q$ if they do not sum to 1.0.

We use $JSD$ or $JS^2$ to represent the JS Divergence throughout this paper.

\subsection{TICA Analysis} To evaluate the kinetic fidelity of the generated ensembles, we employ Time-lagged Independent Component Analysis (TICA)~\cite{jing2024generative}. TICA identifies collective variables (slow modes) by maximizing their time-lagged autocorrelation:

\begin{equation} \text{maximize } \rho_i = \text{corr}\big(y_i(t), y_i(t+\tau)\big). \end{equation} 


\subsection{Fidelity  Metrics}
We use dihedral angle coupling distributions to verify the physical Fidelity of the conformational ensemble.
\begin{itemize}
    \item Side chain torsional angles $\chi_1,\chi_2$. 
    \item Backbone torsional angle $\phi$ and $\psi$.
\end{itemize}




\subsection{Dual-cutoff Contact Map calculation}\label{appsec:dual_contactmap_calculate}
To characterize stable ligand-protein interactions and mitigate high-frequency noise arising from thermal fluctuations at the cutoff boundary, we employ a dual-cutoff hysteresis scheme for contact definition.
Let $d_{ij}^{(t)}$ denote the Euclidean distance between a protein heavy atom $i$ and a ligand heavy atom $j$ at frame $t$. The binary atomic contact state $s_{ij}^{(t)}$ is determined recursively to enforce stability:
\begin{equation}
    s_{ij}^{(t)} = 
    \begin{cases} 
    1 & \text{if } d_{ij}^{(t)} \le r_{\text{on}} \\
    0 & \text{if } d_{ij}^{(t)} \ge r_{\text{off}} \\
    s_{ij}^{(t-1)} & \text{otherwise (hysteresis region)},
    \end{cases}
\end{equation}
with thresholds set to $r_{\text{on}} = 3.5\text{\AA}$ and $r_{\text{off}} = 5.0\text{\AA}$ and $s_{ij}^{(0)} = 0$.
We aggregate these atomic states to the residue level: a residue $R_k$ is considered in contact with ligand atom $j$ at frame $t$ if any of its constituent heavy atoms are in contact:
\begin{equation}
    c_{kj}^{(t)} = \bigvee_{i \in R_k} s_{ij}^{(t)}.
\end{equation}
The final contact map represents the interaction frequency over the trajectory of length $T$:
\begin{equation}
    P_{kj} = \frac{1}{T} \sum_{t=1}^{T} c_{kj}^{(t)}.
\end{equation}

Similarly, the calculation of dual-cutoff contact maps between amino acids can be performed simply by replacing the ligand with amino acids.


\begin{figure*}
    \centering
    \includegraphics[width=0.999\linewidth]{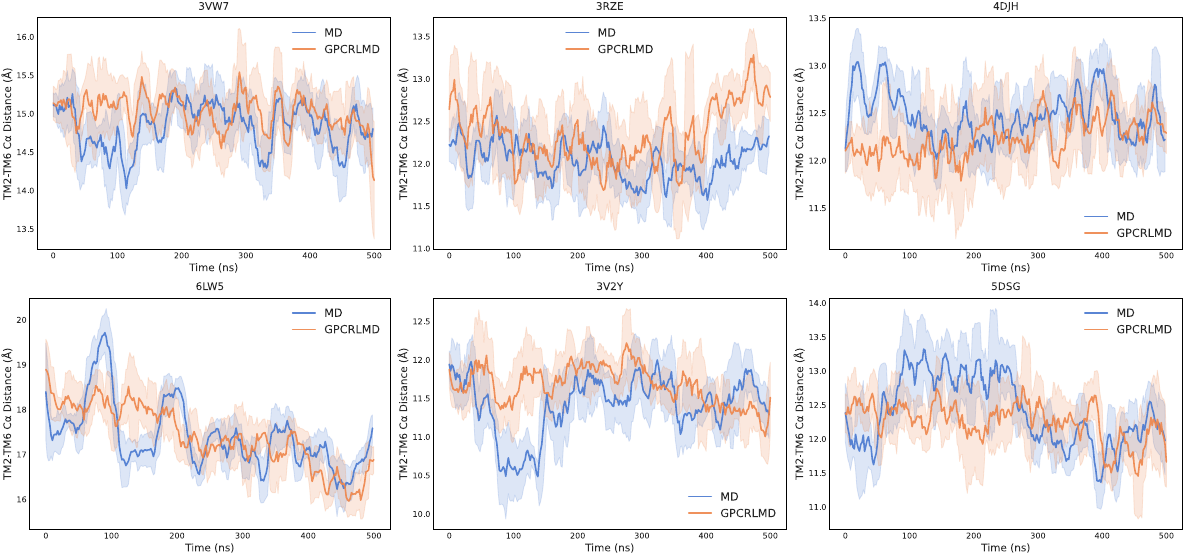}
    \caption{TM2-TM6 distance. For GPCR class A, B1, and C, we calculate the distance between residue $2\times 46$ and $6\times 37$. For class F, we calculate distance $2\times 46 - 6\times 31$ \cite{aranda2025large}.
    Our model keeps a similar movement tendency to the MD reference.
    }
    \label{fig:tm_distance}
\end{figure*}

\begin{figure*}
    \centering
    \includegraphics[width=0.9995\linewidth]{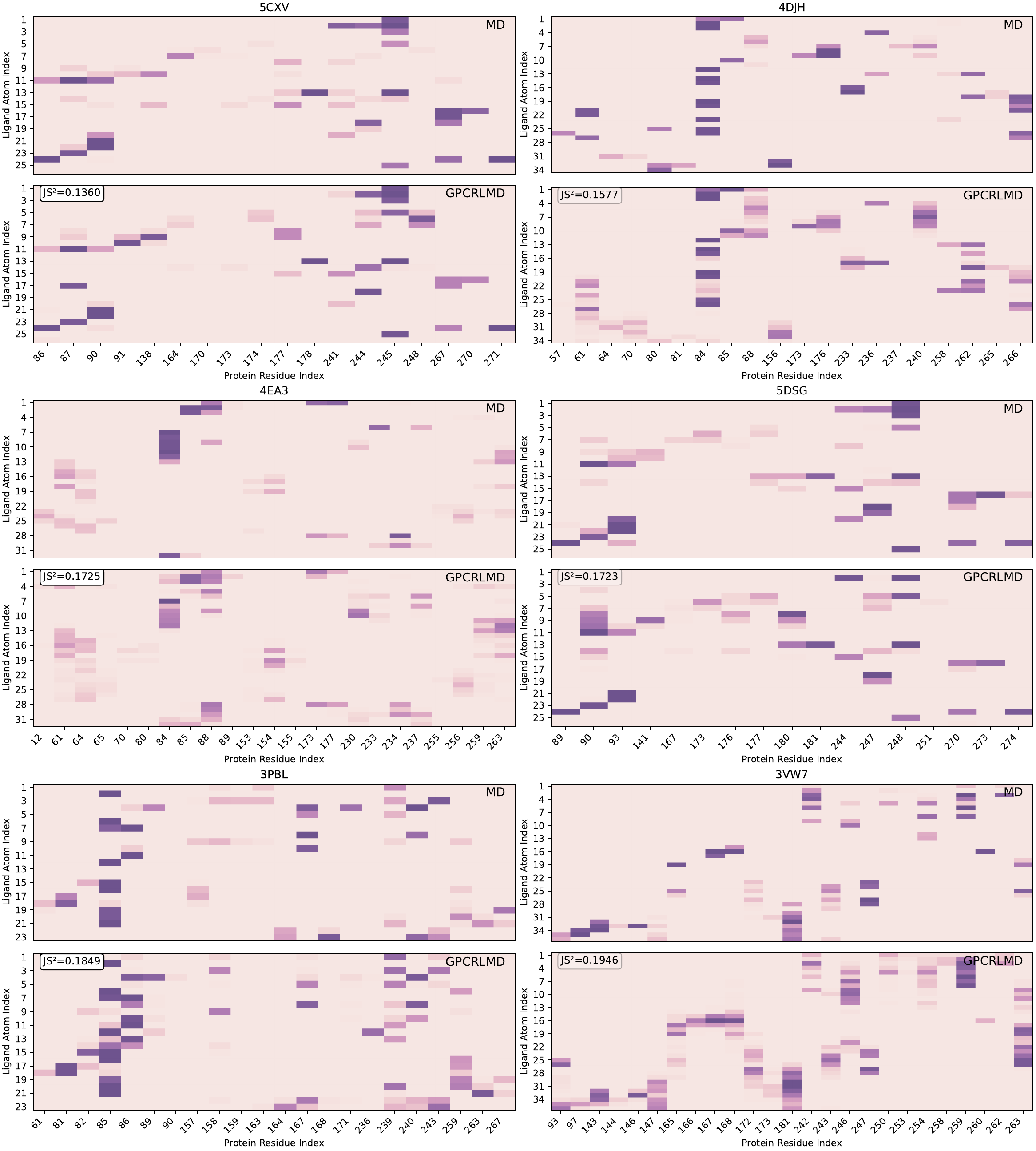}
    \caption{The contact map between ligand and receptor. We compare the 100ns trajectory between the reference MD and \ourM prediction, and the results indicate that our approach is highly consistent with the ligand-receptor interactions presented by Ground Truth MD.}
    \label{fig:contactmap}
\end{figure*}

\begin{figure*}
    \vspace{-2mm}
    \centering
    
    \includegraphics[width=0.93\linewidth]{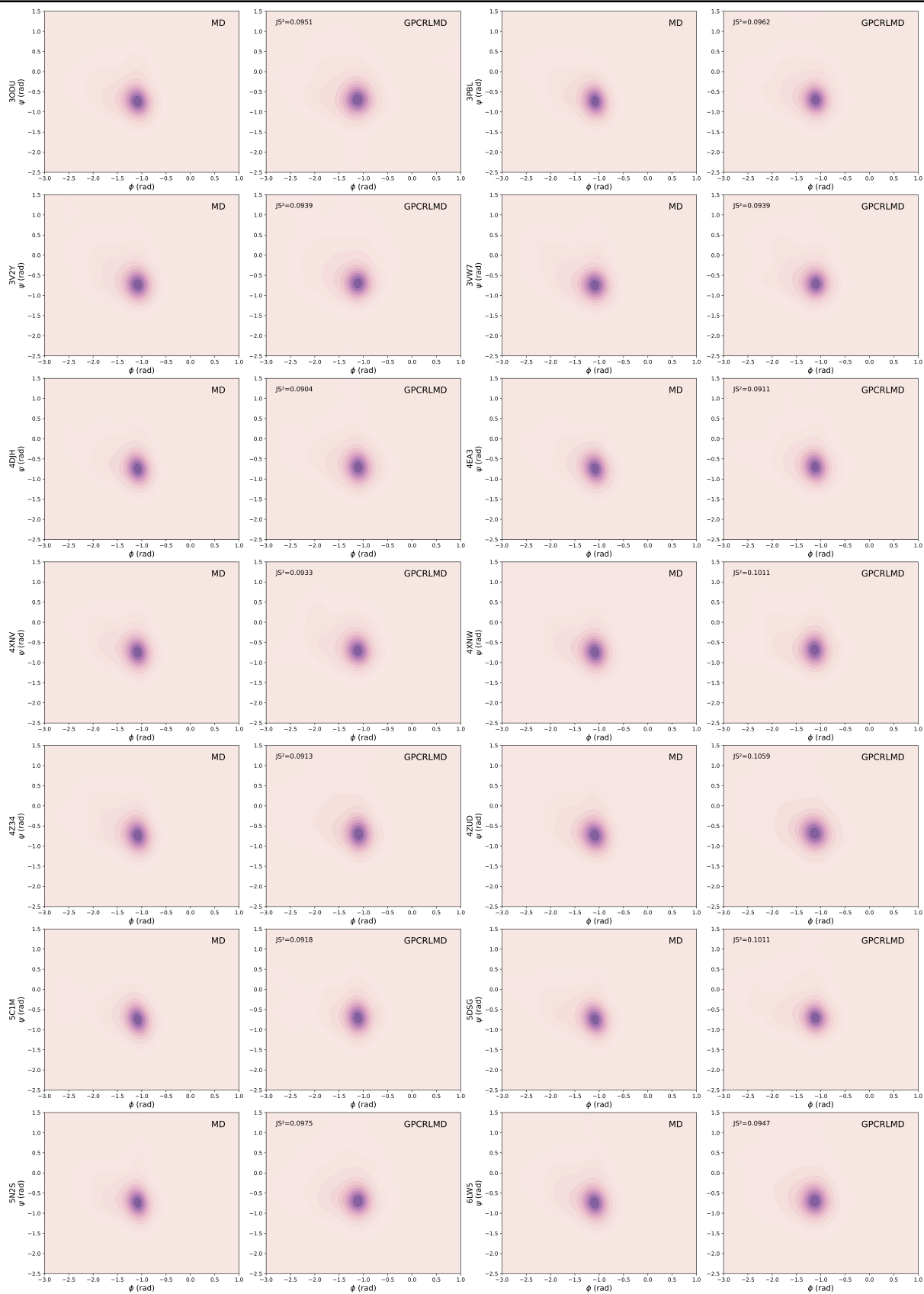}
    \caption{Backbone dihedral angles $\psi, \phi$ coupling distribution visualization. Our method generates a receptor ensemble with a high degree of similarity in backbone angular coupling compared to the reference MD. }
    \vspace{-2mm}
    \label{fig:phi_psi_distri}
\end{figure*}

\begin{figure*}
    \centering
    \includegraphics[width=0.958\linewidth]{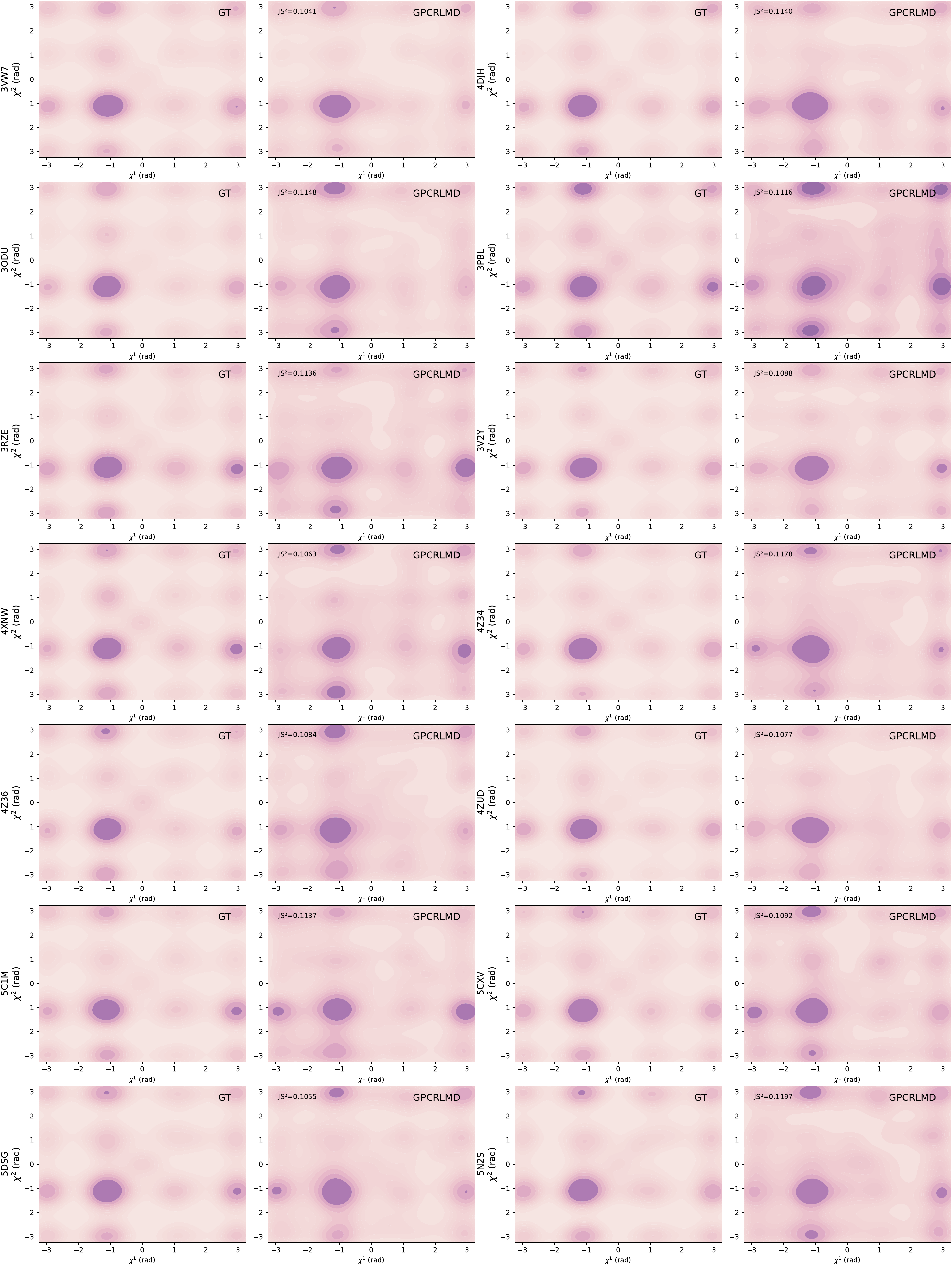}
    \caption{Side chain torsional angles $\chi^1, \chi^2$ distribution visualization. Our method generates a receptor ensemble with a high degree of similarity in side-chain angular coupling compared to the reference MD. }
    \label{fig:chi1_chi2_distri}
\end{figure*}
\begin{figure*}
    \centering
    \includegraphics[width=1\linewidth]{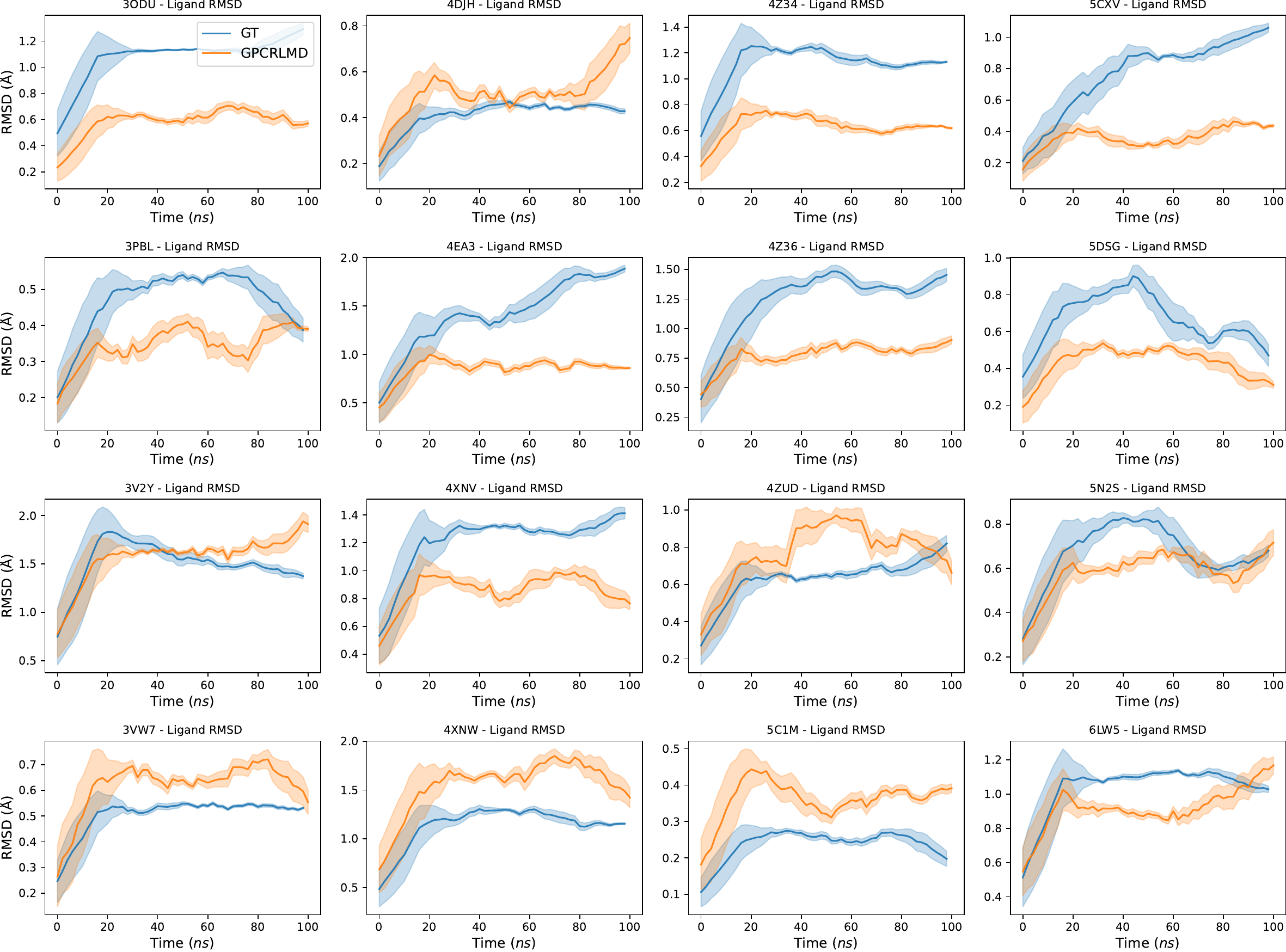}
    \caption{The ligand RMSD change curve over time. We calculate the  RMSD between the first frame (initial ligand conformation) and $t$-th frame in the MD and prediction trajectory, respectively. The results indicate that our method can maintain the same pattern of reference MD in about one-third of the samples and also shows higher stability than reference MD in the remaining samples. }
    \label{fig:ligand_rmsd}
\end{figure*}
\begin{figure*}
    \centering
    \includegraphics[width=0.895\linewidth]{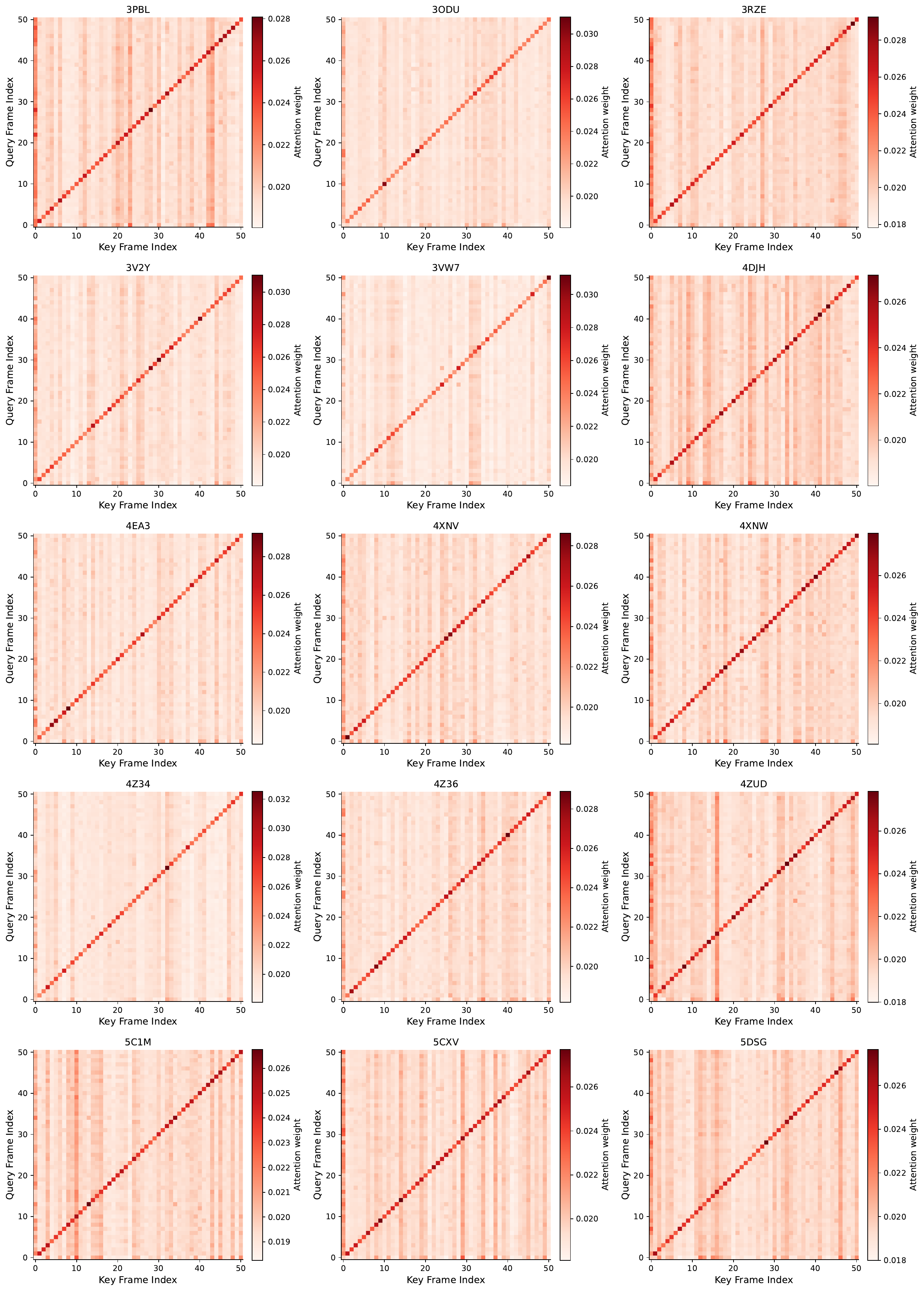}
    \caption{The attention map in the token temporal attention decoder. We focus on the last-layer temporal attention, which captures the most task-relevant and long-range dependencies across frames. }
    \label{fig:attnmap}
\end{figure*}

\end{document}